\documentstyle[preprint,epsfig,tighten,floats,aps,amsfonts,amssymb]{revtex}
\renewcommand{\S}{{\cal S}}
\newcommand{\semi}{{\rtimes}}

\newcommand{\su}{{\frak su}}

\newcommand{\R}{{\Bbb R}}
\newcommand{\C}{{\Bbb C}}
\newcommand{\Z}{{\Bbb Z}}
\renewcommand{\Re}{{\rm Re}}
\newtheorem{thm}{Theorem}
\newcommand{\et}{\hspace{-0.08in}{\bf .}\hspace{0.1in}}
\newcommand{\BOX}{\hbox {$\sqcap$ \kern -1em $\sqcup$}}
\newcommand{\qed}{\hskip 3em \hbox{\BOX} \vskip 2ex}
\def\ut#1{\rlap{\lower1ex\hbox{$\sim$}}{#1}}

\def\U{{\rm U}}
\def\SU{{\rm SU}}

\def\SO{{\rm SO}}
\def\D{{\cal D}}
\def\E{\Sigma}
\def\g{{\frak g}}
\def\k{\kappa}
\def\A{{\cal A}}
\def\G{{\cal G}}

\def\P{{\cal P}}   
\def\H{{\cal H}}
\def\X{{\cal X}}
\def\Xg{{\cal X}_{\rm grav}}

\def\Ab{\overline\A}

\def\cyl{{\rm Cyl}}
\def\phys{{\rm phys}}
\def\bh{{\rm bh}}

\def\ag{{{}^\gamma\!A}}
\def\Eg{{{}^\gamma \E}}
\def\Xg{{{}^\gamma\!{\cal X}}}
\def\u{\underline}

\def\Ad{{\rm Ad}}
\def\maps{\colon}
\def\mod{{\rm mod}}

\def\bfA{{\bf A}}
\def\bfF{{\bf F}}
\def\bfE{{\bf E}}
\def\bfpi{{\bf \Pi}}

\def\l{\ell_{P}}
\def\Diff{{\rm Diff}}
\def\T{T}

\def\ba{\begin{eqnarray}}
\def\ea{\end{eqnarray}}
\def\be{\begin{equation}}
\def\ee{\end{equation}}
\newcommand{\ban}{\begin{eqnarray*}}
\newcommand{\ean}{\end{eqnarray*}}
\newcommand{\barr}{\begin{array}}
\newcommand{\earr}{\end{array}}
\newcounter{letter} 
\newenvironment{alphalist}{\begin{list}{{\normalshape(\alph{letter})}}
{\usecounter{letter}} }{\end{list}}
 
\newcommand{\iso}{\cong}
\newcommand{\Tr}{{\rm Tr}}
\newcommand{\we}{\wedge}

\preprint{\vbox{\baselineskip=12pt
\rightline{NSF-ITP-99-153}}}

\begin{document}
\draft
\title{Quantum Geometry of Isolated Horizons and Black Hole Entropy}
\author {Abhay\ Ashtekar${}^{1,2}$, 
%\thanks{E-mail address: ashtekar@phys.psu.edu}
John\ C.\ Baez${}^{3,1}$
%\thanks{E-mail address: baez@math.ucr.edu}
and Kirill\ Krasnov${}^{2,4}$
%\thanks{E-mail address: krasnov@cosmic.physics.ucsb.edu}
}
\address{1. Center for Gravitational Physics and Geometry \\
Department of Physics, The Pennsylvania State University \\
University Park, PA 16802, USA}

\address{2. Institute of Theoretical Physics\\
University of California, Santa Barbara, CA 93106, USA}
  
\address{3. Department of Mathematics,  University of California \\
      Riverside, CA 92521, USA }

\address{4.  Department of Physics, University of California \\
      Santa Barbara, CA 93106, USA}
 
\maketitle

\begin{abstract}
Using the classical Hamiltonian framework of \cite{ack} as the point
of departure, we carry out a non-perturbative quantization of the
sector of general relativity, coupled to matter, admitting
non-rotating isolated horizons as inner boundaries.  The emphasis is
on the quantum geometry of the horizon.  Polymer excitations of the
bulk quantum geometry pierce the horizon endowing it with area.  The
intrinsic geometry of the horizon is then described by the quantum
Chern-Simons theory of a $\U(1)$ connection on a punctured 2-sphere,
the horizon.  Subtle mathematical features of the quantum Chern-Simons
theory turn out to be important for the existence of a coherent
quantum theory of the horizon geometry.  Heuristically, the intrinsic
geometry is flat everywhere except at the punctures.  The
distributional curvature of the $\U(1)$ connection at the punctures
gives rise to quantized deficit angles which account for the overall
curvature.  For macroscopic black holes, the logarithm of the number
of these horizon microstates is proportional to the area, irrespective
of the values of (non-gravitational) charges.  Thus, the black hole
entropy can be accounted for entirely by the quantum states of the
horizon geometry.  Our analysis is applicable to all non-rotating
black holes, including the astrophysically interesting ones which are
very far from extremality. Furthermore, cosmological horizons (to
which statistical mechanical considerations are known to apply) are
naturally incorporated.

An effort has been made to make the paper self-contained by including
short reviews of the background material.

\end{abstract}
\pacs{04070B,0420}

\section{Introduction}
\label{s1}

Isolated horizons are a generalization of the event horizons of stationary
black holes to physically more realistic situations \cite{ack,abf}.
The generalization is in two directions.  First, while one needs the
entire spacetime history to locate an event horizon, isolated horizons
are defined using \textit{local} spacetime structures.  Second,
spacetimes with isolated horizons need not admit \textit{any} Killing
field.  Thus, although the horizon itself is stationary, the outside
spacetime can contain non-stationary fields and admit radiation.  This
feature mirrors the physical expectation that, as in statistical
mechanics of ordinary systems, a discussion of the equilibrium
properties of black holes should only require the black hole to be in
equilibrium and not the whole universe.  These generalizations are
also mathematically significant. For example, while the space of
stationary solutions to the Einstein-Maxwell equations admitting
\textit{event} horizons is three dimensional, the space of solutions
admitting \textit{isolated} horizons is infinite dimensional.  Yet,
the structure available on isolated horizons is sufficiently rich to
allow a natural extension of the standard laws of black hole mechanics
\cite{abf,other}. Finally, cosmological horizons to which
thermodynamic considerations also apply \cite{gh} are special cases of
isolated horizons.

It is then natural to ask if one can analyze the \textit{quantum}
geometry of isolated horizons in detail and account for entropy from
statistical mechanical considerations.  We will see that the answer to
both questions is in the affirmative. (A summary of these results
appeared in \cite{abck}. For early work, see \cite{K,Rov}.)

The first paper in this series \cite{ack} introduced the notion of an
undistorted, non-rotating, isolated horizon and examined 4-dimensional
spacetimes which are asymptotically flat and admit such an isolated
horizon as inner boundary.  If one chooses a partial Cauchy surface
$M$ in such a spacetime, the intersection of $M$ with the horizon will
be a 2-sphere $S$, as shown in Figure 1. In \cite{ack,abf}, the
Hamiltonian framework adapted to such 3-manifolds with internal
boundaries was constructed starting from a suitable action principle.
This framework has the novel feature that the expression for the
gravitational symplectic structure contains, in addition to the
familiar volume integral over $M$, a surface term given as an integral
over the boundary $S$.  Furthermore, this surface term coincides with
the standard symplectic structure of the $\U(1)$ Chern-Simons theory,
where the $\U(1)$ connection $W$ is simply the gravitational
spin-connection on the horizon 2-sphere.%
\footnote{In \cite{ack} this $\U(1)$ connection was called $V$ instead
of $W$.  In this paper $V$ is generally used as a suffix to label
structures associated with volume fields defined on $M$ and $S$ as a
suffix which labels structures associated with surface fields defined
on $S$.}
This framework offers a natural point of departure for
non-perturbative quantization.

The purpose of this paper is to carry out this quantization in detail,
examine the resulting quantum geometry of isolated horizons and use
the associated quantum states to calculate the statistical mechanical
entropy of these horizons in the context of non-perturbative quantum
gravity.  Conceptually, perhaps the most striking feature of this
description is that ideas from three distinct sources ---the analysis
of isolated horizons in classical general relativity \cite{ack}, the
quantum theory of geometry
\cite{AI,AL1,B1,AL2,AL3,MM,P,RS2,B2,B3,ALMMT,RS3,AL4,AL5,T1,BS,LT},
and the $\U(1)$ Chern-Simons theory--- fuse together seamlessly to
provide a coherent description of the quantum states of isolated
horizons.  At certain points there is even a delicate matching between
numerical coefficients calculated independently within these areas,
suggesting that the underlying unity is potentially deep; its
ramifications are yet to be fully understood. {}From a practical
viewpoint, the most notable feature is that the framework can
incorporate, in a single stroke, a wide variety of horizons, without
any restriction on near-extremality, on the ratio of the horizon
radius to the cosmological radius, etc.\ made in other approaches.

%Pl label ${\Cal I}^\pm$ in the figure.@@
\begin{figure}
\centerline{\hbox{\epsfig{figure=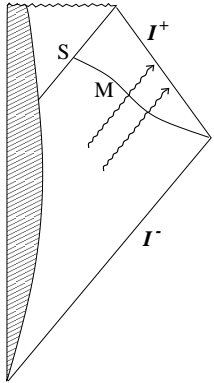,height=2in}}}
\medskip
\caption{A typical spacetime of interest depicting a gravitational
collapse. The horizon geometry becomes time-independent at late times,
say to the future of a cross-section $S$. However, there is
gravitational radiation crossing any (partial) Cauchy surface $M$ in
the exterior region.}
\label{fig1}
\end{figure}
 
We have attempted to make this paper self-contained.  In particular,
for the convenience of readers who may not be interested in the
intricacies of global issues in classical general relativity discussed
in \cite{ack,abf}, in Section \ref{s2} we review the background
material needed here.  This section is divided into two parts.  In the
first, we recall from \cite{ack} the Hamiltonian framework which
serves as the point of departure for quantization.  As usual, the
phase space consists of certain smooth fields on a 3-manifold $M$.%
\footnote{As in \cite{ack,abf}, here we will be interested primarily
in non-rotating isolated horizons.  The classical theory needed in the
treatment of more general cases is now well-understood \cite{other}
and the quantum theory will be discussed in subsequent papers.  For
brevity, in the rest of this paper, $S$ will be referred to simply as
the `horizon'.}
The horizon boundary conditions imply that the only independent degree
of freedom on $S$ is a $\U(1)$ connection $W$.  Note however that in
the classical theory $W$ does {\it not} represent a new degree of
freedom; it is determined by the limiting value of the connection in
the bulk.  But, as mentioned above, the symplectic structure is
somewhat unusual: in addition to the familiar volume term, it contains
also a Chern-Simons term for $W$ on the internal boundary $S$.  In the
second part of this section, we recall the quantum theory of geometry
\cite{AI,AL1,B1,AL2,AL3,MM,P,RS2,B2,B3,ALMMT,RS3,AL4,AL5,T1,BS,LT}
which has been developed on manifolds without boundary. In this
theory, the fundamental excitations are the Wilson lines of an
$\SU(2)$ connection $A$.  Physically, these Wilson lines can be
thought of as `flux lines of area': heuristically, they endow each
surface which they intersect with a quantum of area.  Because of the
one-dimensional nature of these excitations, the resulting quantum
geometry is often referred to as `polymer geometry'.

\begin{figure}
\centerline{\hbox{\epsfig{figure=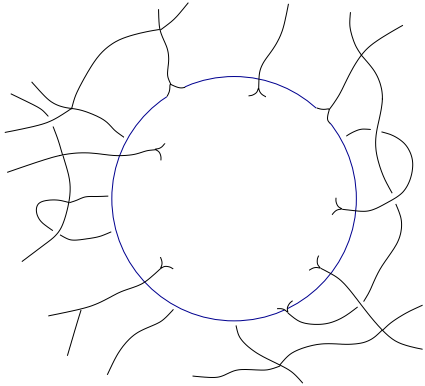,height=2in}}}
\medskip
\caption{Quantum Horizon. Polymer excitations in the bulk puncture the
horizon, endowing it with quantized area. Intrinsically, the horizon
is flat except at punctures where it acquires a quantized deficit
angle. These angles add up to $4\pi$.}
\label{fig2}
\end{figure}

The first goal of this paper is to extend this quantum geometry to
situations in which there is an inner boundary $S$ representing the
horizon.  The overall strategy is summarized in section \ref{s3}.
Section \ref{s4} discusses the kinematical Hilbert space.  Recall that
in quantum field theories with local degrees of freedom, quantum
states are functions of {\it generalized} fields which need not be
continuous.  Therefore, we are led to consider generalized connections
\cite{AL2,AL3} whose behavior on the boundary $S$ can be quite
independent of their behavior in the bulk.  Thus, there is an
interesting and important departure from the situation in the
classical theory.  As a result, surface states are no longer
determined by the volume states.  Rather, the total Hilbert space is a
subspace of the tensor product $\H_V \otimes \H_S$ of a volume Hilbert
space $\H_V$ with a surface Hilbert space $\H_S$.  States in the
volume Hilbert space $\H_V$ represent polymer geometries as before.
However, now the one-dimensional excitations can end on $S$ where they
make punctures. (See Figure 2.) At each puncture, they induce a
specific distributional curvature for the $\U(1)$ connection $W$.
Furthermore, the space of $\U(1)$ connections on $S$ naturally
inherits the Chern-Simons symplectic structure from the classical
Hamiltonian framework.  Therefore, the two pieces of information
needed in the quantization of Chern-Simons theory on a punctured
surface are now at hand, one supplied by the quantum theory of
geometry, and the other by the classical theory of isolated horizons;
the three theories are naturally intertwined. The surface Hilbert
space $\H_S$ is the space of states of this Chern-Simons theory.

\begin{figure}
\centerline{\hbox{\epsfig{figure=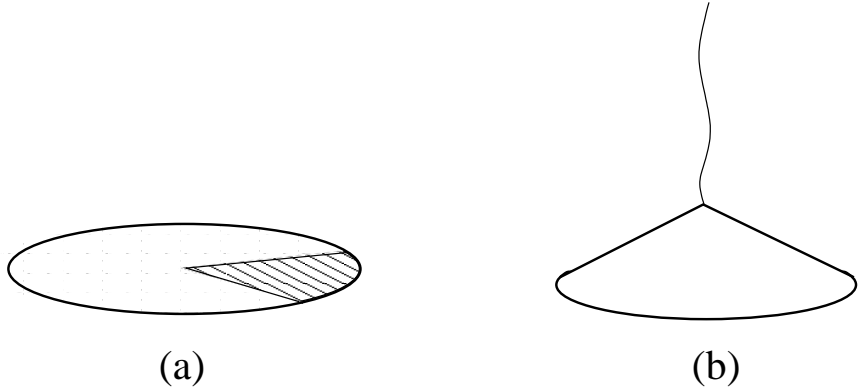,height=2in}}}
\medskip
\caption{(a) Deficit angle in the intrinsic horizon 2-geometry.
(b) A 3-dimensional perspective: a bulk polymer excitation `exerts a 
tug' on the horizon causing a deficit angle.} 
\label{fig3}
\end{figure}
 
Section \ref{s5} extracts physical states by imposing the quantum
version of the phase space boundary conditions and constraints on the
space of kinematical states.  It turns out that the quantum boundary
conditions ensure that the volume and the surface states are coupled
just in the correct way to ensure gauge invariance of the total state.
This is another illustration of the unexpected and delicate matching
between the isolated horizon boundary conditions from classical
general relativity, quantum geometry and the quantum Chern-Simons
theory.  The quantum geometry of the horizon emerges from this
interplay.  The holonomy of the $\U(1)$ connection $W$ around a loop
in $S$ is non-trivial if and only if the loop encloses a puncture.
The holonomy around each puncture endows it with a deficit angle, as
depicted in Figure \ref{fig3}.  All these angles are quantized and add
up, in a suitable sense, to $4\pi$.  Thus, heuristically, one can say
that the quantum horizon is flat except at the punctures.

The second goal of the paper is to use this quantum horizon geometry
to account for the entropy of isolated horizons.  This task is carried
out in Section \ref{s6}.  Note that the physical states considered so
far include information about gravitational and electromagnetic
radiation far away from the black hole which is obviously irrelevant
to the calculation of black hole entropy.  What is relevant are the
states directly associated with the horizon of a given area, say
$a_0$.  One is therefore led to trace over the volume degrees of
freedom and construct a density matrix $\rho_\bh$ describing a maximum
entropy mixture of surface states for which the area of the horizon
$a$ lies in the range $a_0-\delta \le a \le a_0+\delta$ for some small
$\delta$ ($\sim \l^2$), and for which the electric, magnetic and
dilaton charges lie in similar small intervals containing the
classical value.  The statistical mechanical entropy is then given by
$S_\bh = -{\rm Tr}(\rho_\bh\, \ln\rho_\bh$).  As usual, this number
can be calculated simply by counting states: $S_\bh = \ln N_\bh$ where
$N_\bh$ is the number of Chern-Simons surface states satisfying the
area and charge constraints.  We find that, irrespective to the values
of charges, for $a_0 \gg \l^2$, the (leading term in the expression of
the) entropy is proportional to the area.  Thus, the entropy can be
traced back directly to the quantum states of the geometry of the
horizon, states which can interact with the other degrees of freedom
in the physical, curved spacetime geometry around the black hole.
There are no corresponding surface states for Maxwell or dilaton
fields. Section \ref{s7} summarizes the results and compares and
contrasts our approach to other approaches available in the
literature.

As discussed in detail in \cite{ack,abf}, isolated horizons encompass
a wider class of situations than black holes.  In particular, they
also include `cosmological horizons' of the type encountered in the
deSitter spacetime which has no black hole at all.  It has been known
for some time \cite{gh} that thermodynamic considerations apply also
to cosmological horizons and recent work \cite{abf,other} has shown
that the laws of black hole mechanics extends to all isolated
horizons.  Therefore, one would expect a statistical mechanical
description to extend to this wider class of horizons.  Results of
this paper will show that this expectation is correct.  Furthermore,
there exists a \textit{single} statistical mechanical framework that
unifies these apparently distinct situations.  However, for simplicity
of presentation, we will focus on black holes in the main body of the paper
and briefly discuss the more general cases in Section \ref{s7}.

For convenience of the reader, we have organized the material such
that the background material and the main results are described in
Sections \ref{s2}, \ref{s3} and \ref{s6.A} and \ref{s6.C}. These
sections can be read independently of others which contain proofs and
technical subtleties. An index of notation is included at the end of the
paper.

\section{Preliminaries}
\label{s2}

This section is divided into two parts. In the first, we briefly
review the classical Hamiltonian framework of \cite{ack} for general
relativity \textit{in presence of an inner boundary} representing an
isolated horizon.  In the second, we sketch the quantum theory of
Riemannian geometry on manifolds \textit{without} boundary developed
in \cite{AL1,B1,AL2,AL3,MM,P,RS2,B2,B3,ALMMT,RS3,AL4,AL5,T1,BS,LT}. In
sections \ref{s3}, \ref{s4} and \ref{s5}, these two ingredients will be
combined and further developed to obtain the quantum geometry of
isolated horizons.

\subsection{Classical Hamiltonian framework}
\label{s2.A}

The non-perturbative quantization used in this paper is based on a
Hamiltonian framework.  Therefore, the starting point is the classical
phase space.  The arena is a 3-manifold $M$, the complement of the
unit open ball in $\R^3$.  The boundary of $M$ is a 2-sphere, which we
denote by $S$.  We think of $M$ as the partial Cauchy surface exterior
to the black hole, and think of $S$ as the intersection of an isolated
horizon with $M$ (see Figure \ref{fig1} for a prototype situation).
For simplicity we will refer to $S$ simply as the `horizon' and use
under-bars to denote the pullback of fields from $M$ to $S$.

Let $P$ be the trivial $\SU(2)$ bundle over $M$.  The fields of our
classical system are a connection $A$ on $P$ and an $\Ad P$-valued
2-form $\E$ on $M$.  We denote the curvature of $A$ by $F$.  We may
identify $A$ with an $\su(2)$-valued 1-form and identify $\E$ and $F$
with $\su(2)$-valued 2-forms.  We use lower-case Roman letters
$a,b,c,\dots$ for spatial tensor indices and lower-case Roman letters
$i,j,k,\dots$ for `internal' indices running over a basis of $\su(2)$.
Thus in component notation we write $A$ as $A_a^i$ and $\E$ as
$\E_{ab}^i$. The internal indices are raised and lowered using the
Cartan-Killing form on $\su(2)$.  However, we often suppress the
internal indices% 
\footnote{In the previous paper \cite{ack} which dealt with the
classical theory, it was more convenient to work in the
spin-$\frac{1}{2}$ representation of $\SU(2)$. There, the basic fields
were denoted by $A_{aA}{}^{B}$ and $\Sigma_{abA}{}^{B}$ in the
component notation. The relation between these fields and those in
this paper is given by: $A_{aA}{}^B = - \frac{i}{2}A_a^i\,
\tau_{iA}{}^B$ and $\Sigma_{abA}{}^B = - \frac{i}{2}\Sigma_{ab}^i\,
\tau_{iA}{}^B$, where $\tau_{iA}{}^{B}$ are the $2\times 2$ traceless,
Hermitian matrices satisfying $\tau_i\tau_j = i \epsilon_{ijk}\tau_k
+\delta_{ij}1$.}.
The connections may be thought of as the configuration variables and
$\E$ as their canonical momenta.  Thus the kinematical phase space is
the same as in $\SU(2)$ Yang-Mills theory, the duals of the 2-forms $\E$
playing the role of the Yang-Mills electric fields. However, the
interpretation of these fields is quite different.  Roughly, the
connections can be thought of as `gravitational spin connections' and
their momenta as `spatial triads'.  However, there are a number of
technical subtleties \cite{A1,I}.  The final picture can be summarized
as follows.  

For each positive number $\gamma$, there exists a phase space $\Xg$
consisting of pairs $(\ag, \Eg)$ of smooth fields on $M$ satisfying
certain boundary conditions.  $\gamma$ is called the Barbero-Immirzi
parameter and in some ways is analogous to the $\theta$-parameter in
Yang-Mills theories.  These phase spaces are naturally isomorphic to
one another.  One can write Einstein's equations using any value of
$\gamma$.  They have the same general form and the same physical
content at the classical level; the difference lies only in certain
relative numerical factors involving $\gamma$ between various terms.
Therefore, at the classical level, without loss of generality one can
just work with a fixed value of $\gamma$ and, if so desired, pass to
another $\gamma$-sector via a canonical transformation.  However, in
the quantum theory, these canonical transformations fail to be
unitarily implementable.  Hence distinct $\Xg$ lead to unitarily
inequivalent quantum theories.  These theories are physically distinct
because, as we explain in Section \ref{s2.B}, the spectra of geometric
operators, such as those defining surface-areas, depend on $\gamma$
and are distinct in different $\gamma$-sectors.  As with the $\theta$
parameter in quantum chromodynamics, the value of $\gamma$ can be
determined only experimentally, or via \textit{new} theoretical
inputs.  Therefore, for our purposes it is important to explore all
$\gamma$ sectors.

The geometrical meaning of the fields $(\ag,\Eg)$ is as follows.  
Starting from the 2-forms $\Eg$ one can define an orthonormal triad $E$ 
of density weight one on $M$ as follows:
\be \label{geom1} 
E^a_i := \gamma\, \eta^{abc}\, (\Eg_{bci}) \, , \ee
where $\eta^{abc}$ is the metric-independent Levi-Civita 3-form of
density weight one.  Thus, the 3-metric $q_{ab}$ on $M$ is given by
\be\label{metric}
 E^a_i E^{bi} = {\bf q }\, q^{ab}
\ee
where $\bf{q}$ is the determinant of the metric $q_{ab}$. The triad
determines an unique torsion-free derivative operator $D$ which acts
on both, tensor and internal indices. Denote by $\Gamma_{a}^i$ the
corresponding $\SU(2)$ connection on $P$ and by $K_a^i$ the $\Ad
P$-valued 1-form constructed from the extrinsic curvature $K_{ab}$ of
$M$ via $K_a^i = (1/\sqrt{\bf q}) \, K_{ab}E^{bi}$.  In the spacetime
picture, the connection $\Gamma$ is constructed from spatial
derivatives of the triad $E$ while the 1-forms $K$ determine their
time derivatives. In terms of these fields, our phase space variable\,
$\ag$\, is given by:
\be \label{geom2}
\ag_a := \Gamma_a - \gamma K_a \, ;
\ee
it depends on spatial and temporal derivatives of the triads.
\textit{For notational simplicity, from now on we will drop the
suffix} $\gamma$.  Thus, unless otherwise stated, from now on $(A,
\E)$ will uniformly denote $(\ag, \Eg)$.

The boundary conditions at infinity ensure that the geometry is
asymptotically flat at spatial infinity \cite{ack,abf}.  Since they
will not play an essential role in the present paper, we will not
state them explicitly.  The structure at the horizon, on the other
hand, {\it will} play a crucial role.  To spell it out, let us first
note that the geometrical interpretation (\ref{geom1}) of $\E$
provides a formula for area of any 2-surface $\T$ in $M$. 
\be \label{area}
A_\T = \gamma \int_\T \left(\tilde{\E}^i \tilde{\E}^j k_{ij} 
\right)^{1\over 2}\, d^2x\, ,
\ee
where $\tilde\E^i = \eta^{ab}\E^i_{ab}$ with $\eta^{ab}$ the
(metric-independent) Levi-Civita density on $\T$, and where, as before
$k_{ij}$ is the Cartan-Killing metric on $\su(2)$. The integral is
well-defined because the integrand is a density of weight one; it
gives the area because the integrand equals the square root of the
determinant of the 2-metric induced on $\T$.

We can now state the horizon boundary conditions. First, only those
2-forms $\E$ are admissible for which the horizon area has a fixed
value, $a_0$.  Second, the pullback $\underline{A}$ of $A$ to $S$ is
completely determined by a $\U(1)$ connection $W$ on $S$ and the
constant $a_0$.  The third and final boundary condition is that the
pullback $\u{\E}$ of $\E$ to $S$ is completely determined by the
curvature $F = dW$ of $W$.

To specify these restrictions explicitly, let us first fix a smooth
function $r \maps S \to \su(2)$ with $|r| = 1$ which has degree $1$
when viewed as a map from a 2-sphere to itself.  Clearly, $r$ is fixed
by a $\U(1)$ subgroup of the gauge group at each point of $S$.  It
thus picks out a $\U(1)$ sub-bundle $Q$ of the $\SU(2)$ bundle $P|_S$
given by restricting $P$ to $S$.  (Intrinsically, $Q$ is just the spin
bundle of $S$.)  Then $W$ is a connection on $Q$, defined in terms of
the $\SU(2)$ spin connection $\Gamma$ on $P$ via
\be \label{W}
W_a := - \frac{1}{\sqrt{2}}\,\, \underline{\Gamma}_a^i r_i,
\ee
and its curvature $F$ is related to $\E$ via
\be \label{bc}
 F_{ab} = -{2 \pi \gamma \over a_0} \underline{\E}_{ab}^i\, r_i \,.
\ee

The phase space $\X$ consists of pairs $(A, \E )$ of asymptotically
flat, smooth fields on $M$ satisfying the horizon boundary conditions
just stated.  $\X$ is an infinite-dimensional smooth manifold ---
technically, a Frech\'et submanifold of the vector space of smooth
$(A,\E)$ pairs equipped with a suitable topology defined using the
Cartesian components of $A$ and $\E$.  The symplectic structure on
$\X$ is given by
\be \label{ss} \Omega_{\rm grav}((\delta A, \delta E), (\delta A',
\delta E')) = {1\over 8\pi G}\left[ \int_M {\rm Tr} \left( \delta A
\wedge\delta^\prime \E - \delta^\prime A \wedge\delta\E \right) +
\frac{a_0}{\gamma\pi} \oint_S \delta W\wedge\delta^\prime W\right] \ee
for any tangent vectors $(\delta A, \delta \E)$ and $(\delta A',\delta
E')$ at a point of $\X$.  Note that, in addition to the familiar
volume term, this symplectic structure has a surface term which
coincides with the symplectic structure of the $\U(1)$ Chern-Simons
theory.  The symplectic structure $\Omega_{\rm grav}$ is weakly
nondegenerate --- that is, a tangent vector to $\X$ whose pairing
with any other tangent vector vanishes must be zero.

The phase space $\X$ serves as the arena for the Hamiltonian
formulation of general relativity. Points of $\X$ represent
`kinematical states'; they are not all physically realized.  This is
because some of Einstein's equations contain no time derivatives and
thus constrain the physically admissible states to lie in a
submanifold $\tilde\X$ of $\X$.  The constraint submanifold
$\tilde\X$ is defined by three sets of restrictions, called the
Gauss, diffeomorphism and Hamiltonian constraints.  They are of first
class in Dirac's terminology: the pullback of the symplectic structure
to the $\tilde\X$ is degenerate and the degenerate directions
correspond precisely to the Hamiltonian vector fields generated by the
constraint functions.  Therefore, motions generated by these vector
fields can be interpreted as gauge transformations in the phase space.
A careful analysis \cite{ack} shows that they correspond to: i)
$\SU(2)$ internal rotations on $(A, \E)$ that reduce to $\U(1)$
rotations preserving $r$ on the boundary $S$; ii) spatial
diffeomorphisms generated by vector fields which are tangential to
$S$; and, iii) `bubble-time evolution' in spacetime where the
3-surfaces $M$ are kept fixed both at the horizon and at infinity.  It
turns out that the boundary condition (\ref{bc}) plays an important
role in ensuring full invariance with respect to the $\SU(2)$ internal
rotations on the boundary.  Without it, as in the case of the scalar
constraint, only those internal rotations whose generators vanish on
$S$ could be regarded as gauge.

Finally, note that we can define a `volume phase space' $\X_V$ and a
`surface phase space' $\X_S$ as follows.  $\X_V$ is isomorphic with
$\X$ as a manifold but it is equipped with a symplectic structure
$\Omega_V$ given just by the volume term in (\ref{ss}).  $\X_S$ is
the space of $\U(1)$ connections $W$ on $S$ equipped with the
Chern-Simons symplectic structure --- i.e., the surface term in
(\ref{ss}).  Then, we have natural maps
\[ p_V \maps \X \to \X_V , \qquad p_S \maps \X \to \X_S \]
given by
\[ p_V (A, \E) = (A, \E) \qquad p_S (A, \E) = W  \]
such that 
\[  \Omega_{\rm grav} = p_V^\ast \Omega_V + p_S^\ast \Omega_S .\]
We will see that this structure on the kinematical space of classical
states is faithfully mirrored in the kinematical space of quantum
states.

\subsection{Quantum geometry}
\label{s2.B}

In general relativity, geometry is treated as a \textit{physical}
entity with degrees of freedom of its own.  Indeed, some of the most
dramatic predictions of the theory ---black holes and gravitational
waves--- center around purely geometric notions .  This viewpoint plays
a central role in non-perturbative quantum gravity.  Thus, unlike in
other approaches such as string theory, one avoids the introduction of
a classical, background geometry on which quantum matter is to live.
Rather, matter and geometry are \textit{both} treated quantum
mechanically from the very beginning.  Quantum gravity is regarded as
a theory of quantum geometry interacting with quantum matter.  In this
section, we will review the relevant features of quantum geometry on a
spatial 3-manifold $M$ \textit{without boundary}.  Modifications
required by the presence of the internal boundary at the horizon will
be discussed in Section \ref{s4}.

Recall that in quantum field theory states are often described by
square-integrable wave functions, not on the classical configuration
space of smooth fields, but on certain completion of this which also
contains more singular fields.  The space of smooth fields is usually a
set of measure zero with respect to the natural measure on this larger
space.  As we saw in Section \ref{s2.A}, the classical configuration
space $\A$ of general relativity can be taken to consist of smooth
$\SU(2)$ connections on the spatial 3-manifold $M$.  Its completion
$\Ab$ consists of `generalized' $\SU(2)$ connections \cite{AI,AL1}. 
Like a connection, a generalized connection $A$ describes parallel
transport along paths: it assigns to each path $\eta$ a holonomy
$A(\eta)$, which we may think of as an element of $\SU(2)$.  However,
with a generalized connection there is no requirement that this holonomy
vary smoothly with the path.    Consequently, the space $\Ab$ is  very
large.  However, in the natural topology, the classical configuration
space $\A$ is densely embedded in $\Ab$.  $\Ab$ is sometimes referred to
as the `quantum configuration space'.    (For precise definitions, see
Section \ref{s4.A}.)

There are several distinct ways of constructing the space $\Ab$.  To do
geometry and measure theory on $\Ab$, it is convenient to construct it
as the projective limit of configuration spaces $\A_\g$ of $\SU(2)$
lattice gauge theories associated with graphs $\g$ embedded in $M$
\cite{B1,AL2,AL3}.%  
\footnote{Note that, in spite of the similarity of an intermediate
stage of our construction with lattice gauge theory, we do
\textit{not} follow the traditional strategy of first working with
lattices and then taking the continuum limit by letting the edge
length go to zero.  Rather, we work with the continuum theory from the
very beginning; lattice configuration spaces $\A_\g$ only serve to
provide useful approximations to the full continuum quantum
configuration space $\Ab$.}
More precisely, we can proceed as follows.  First consider any graph
$\g$ with finitely many vertices and edges analytically embedded in
the 3-manifold $M$.  Define a `connection on $\g$' to be a map $A$
assigning an element $A(e) \in \SU(2)$ to each edge $e$ of $\g$
equipped with an orientation, and such that: i) If $e_1, e_2, e_1 
e_2$ are all analytic, then $A(e_1)A(e_2) = A(e_2 e_1)$; and,
ii) $A(e^{-1}) = A(e)^{-1}$ when $e^{-1}$ is $e$ equipped with the
opposite orientation.  The group element $A(e)$ may be thought of 
as the holonomy of $A$ along the edge $e$. Let $\A_\g$ be the space 
of connections on the graph $\g$.  $\A_\g$ is naturally endowed with 
the structure of a finite-dimensional compact manifold, diffeomorphic 
to a product of copies of $\SU(2)$, one for each edge of $\g$.  It can 
be thought of as the configuration space of the $\SU(2)$ lattice gauge 
theory associated with the graph $\g$.

Next, note that the set of graphs in $M$ is naturally equipped with a 
partial ordering where $\g \le \g'$ if $\g'$ is obtained from $\g$
by subdividing the edges of $\g$ and/or adding new edges.  Moreover,
when $\g \le \g'$ there is an obvious projection 
\[     p_{\g,\g'} \maps \A_{\g'} \to \A_\g  .\]
Since there is no `largest' graph in $M$, the family of spaces $\A_\g$
does not contain a largest space from which one can project to any other
space in the family.  However, as is well known, one can construct such 
a space, called the `projective limit' of the spaces $\A_\g$.   This
projective limit is precisely our quantum configuration space $\Ab$. 
By definition of the projective limit, there are natural projections
\[      p_\g \maps \Ab \to \A_\g   \]
for every graph $\g$.  Moreover, these are consistent with the family of
projections $p_{\g,\g'}$, in the sense that 
\[
 p_\g = p_{\g,\g\prime} \, p_{\g^\prime}
\] 
for all $\g,\g'$ with $\g \le \g'$.  

As noted above, the space $\Ab$ of generalized connections is very
large.  At first it may even seem mathematically uncontrollable.  
However, the projective limit construction endows it with a rich
structure \cite{AL3}.  The basic idea is to first note that each space
$\A_\g$ is a equipped with an interesting topology, measure and
geometry, and then to induce these structures on $\Ab$ through the
projection maps $p_\g$.  More precisely, there is a one-to-one
correspondence between `consistent sets' of structures on the family of
finite-dimensional manifolds $\A_\g$ and structures on $\Ab$.  

Let us illustrate this relation through some examples.  Note first that
the projection map $p_\g$ enables us to pull back functions from any 
space $\A_\g$ to $\Ab$.   Denote by $\cyl$ the space of  functions on
$\Ab$ obtained by pulling back $C^\infty$ functions on $\A_\g$ for all
graphs $\g$.  We will refer to the elements of $\cyl$ as `cylinder
functions'.  They play the role of smooth functions on $\Ab$.  Since
$\A_\g$ is a finite-dimensional manifold, the space of $C^\infty$
functions on any one $\A_\g$ is relatively small and fully manageable. 
Yet, since in the construction of $\cyl$ we allow $\g$ to vary over
all possible graphs in $M$, the space $\cyl$ is very large.

A second useful example is provided by measures. Since $\A_\g$ is
diffeomorphic to a product of copies of $\SU(2)$, one for each edge of
$\g$, the Haar measure on $\SU(2)$ naturally gives rise to a
probability measure $\mu_\g$ on each space $\A_\g$, The measures
$\mu_\g$ form a consistent set in the sense that if $\g \le \g'$, then
$\mu_\g$ is the pushforward of $\mu_{\g'}$ under the projection from
$\A_{\g'}$ to $\A_\g$:
\[    \mu_\g = (p_{\g,\g'})_\ast \, \mu_{\g^\prime}  \]
Therefore, there exists \cite{AL1,B1,AL2} a unique probability 
measure $\mu$ on $\Ab$ that projects down to all the measures
$\mu_\g$:
\[      \mu_\g = (p_\g)_\ast \mu .  \]
This measure $\mu$ is called the `uniform measure'.  By construction,
all functions in $\cyl$ are integrable with respect $\mu$.  Moreover,
since we did not introduce any extra structure such as a background
metric in its construction, $\mu$ is invariant under the natural
action of gauge transformations and diffeomorphisms on $\Ab$.  As in
more traditional quantum field theories, although the classical
configuration space $\A$ is topologically dense in the quantum
configuration space $\Ab$, measure-theoretically it is sparse: $\A$ is
contained in a measurable subset of $\Ab$ with measure zero \cite{MM}.

Using the uniform measure $\mu$, we can now define the Hilbert space
$\H$ of kinematical states for quantum gravity:
\[            \H = L^2(\Ab).    \]
This state space is the quantum analog of the kinematical phase space
$\X$.  In both cases, one has to impose the Einstein constraint equations
to extract physical states.  In the classical theory, the space of
kinematical states provide an arena for a precise formulation of the
constraint equations which play a central role in the dynamics of the
Einstein's theory.  Moreover, much of our physical intuition regarding
geometric quantities such as the spatial metric, areas and volumes,
extrinsic curvature, holonomies and their relation to local curvature
comes from this kinematical arena.  One expects, with appropriate
caveats, the situation to be similar in quantum theory.

It turns out that the space $\cyl$ of cylinder functions is dense in
$\H$.  This fact turns out to be very useful in quantum geometry.  In
particular, often one can first define interesting operators with $\cyl$
as their domain and then show that they are essentially self-adjoint, 
thus determining self-adjoint operators on $\H$.  (Thus $\cyl$ plays a role
in quantum gravity rather similar to that played by $C^\infty$ functions
of compact support in non-relativistic quantum mechanics.)  This
strategy greatly simplifies the task of defining geometrical operators
on $\H$.  For, since elements of $\cyl$ are pullbacks of smooth
functions on the spaces $\A_\g$, geometrical operators can now be
defined as consistent families of operators on the Hilbert spaces 
\[          \H_\g= L^2(\A_\g).   \]
Thus, technically, a field-theoretic problem is now reduced to a set of
quantum-mechanical problems.  A further simplification arises because
one can introduce a convenient basis in each $\H_\g$ using a suitable
generalization of Penrose's spin networks \cite{P,RS2,B2,B3,AL4,BS}, 
as described in Section \ref{s4.B}.  This further reduces the
quantum-mechanical problem of defining operators on $\H_\g$ to that of
defining operators on finite-dimensional Hilbert spaces, characteristic
of spin systems.

We can now introduce the geometrical operators.  On the classical
phase space, information about the Riemannian geometry of the 3-manifold
is coded in the $\Ad P$-valued 2-forms $\E$.  Specifically, orthonormal
triads $E$ of density weight one which determine the Riemannian
3-metric $q_{ab}$ via equation (\ref{metric}) are given by:
\be E^a_i = \gamma\, \eta^{abc}\, \E_{bci} \, .\ee
(Recall that for simplicity we are suppressing the superscript
$\gamma$ on $\E$.) Therefore, any Riemannian geometric quantity, such
as the area of a surface or volume of a region, can be constructed
from these 2-forms $\E$.  The basic object in quantum Riemannian
geometry are the corresponding operators $\hat{\E}$.  Since $\E$ is a
2-form on $M$, from geometric considerations one would expect
$\hat{\E}$ to be a 2-dimensional operator-valued distribution on $\H$.
In other words, one expects to have an operator $\hat{\E}_{\T,f}$
given a surface $\T$ in $M$ and an $\Ad P$-valued test field $f$ on
this surface.  This expectation is borne out.

We begin by describing some useful operators on the Hilbert spaces
$\H_\g$.  Suppose that $v$ is a vertex of $\g$ and that $e$ is an edge
of $\g$ having $v$ as one of its endpoints.  Since $\A_\g$ is a
product of copies of $\SU(2)$, one for each edge of $\g$, the left-
and right-invariant vector fields $L^i, R^i$ on $\SU(2)$ give
operators $L^i(e),R^i(e)$ on the $C^\infty$ functions on $\A_\g$.
Using these we define
\[
J^i(e,v) = \cases{\, iL^i(e)  & {if the edge $e$ is 
oriented to be outgoing at $v$,}\cr %\hfil &\hfil\cr
iR^i(e) & {if the edge $e$ is oriented to be ingoing at $v$}}
\]
and introduce `vertex operators'
\[ J^i(v) = \sum_e\, J^i(e,v) \]
where the sum extends over all edges having $v$ as an endpoint.  
These angular momentum-like operators $J^i(v)$ have a natural
geometrical interpretation: they are the three generators of the
$\SU(2)$ gauge rotations at $v$.  

We are now ready to introduce the triad operators $\hat{\E}_{T,f}$.
Let $\psi$ be a $C^\infty$ function on $\A_\g$. For the purposes of
this paper, it will suffice to assume that all edges of the graph $\g$
lie `above' $T$, although some may intersect $S$ from above. (The
notion of `above' and `below' refer to the orientation of $M$ and
$T$.) By subdividing the graph if necessary, we can assume without
loss of generality that each intersection point between $\g$ and and
$T$ is a vertex of $\g$.  Then, given an $\Ad P$-valued function $f$
on $\T$, the smeared triad operator is given by:
\be \label{Eop}
\hat{\E}_{\T,f} [p_\g^* \psi] = p_\g^*\left[ 
{8\pi\gamma\l^2}\,\sum_{v}  \, f_i(v) \,J^i(v) \,\, \psi\right] \, ,
\ee
where the sum is taken over vertices $v$ where $\g$ intersects $\T$.
The general definition, without any restriction on $\g$, is given in
\cite{AL4}.  One can show that $\hat{\E}_{\T,f}$ is a well-defined
operator on $\cyl$, which moreover is essentially self-adjoint.  Note
that the action of the smeared triad operator $\hat{\E}_{\T,f}$ is
localized at intersections of the smearing surface $\T$ and the graph
$\g$ associated with the state $\psi$.  Furthermore, the action is
very natural, defined by the right- and left-invariant vector fields
on $\SU(2)$.  As with the construction of the Hilbert space of states, 
the definition of $\hat{\E}_{\T,f}$ does \textit{not} refer to any 
background structure.  The operator is therefore covariant under gauge 
transformations and diffeomorphisms of space.

Classically, geometric quantities such as areas of surfaces and volumes
of regions are functions of the triads.  Therefore, it is natural to
construct the corresponding quantum operators by first expressing the
classical quantities in terms of triads and then replacing the triads
$\E$ by operators $\hat\E$.  However, the functional form of these
quantities can be quite complicated.  Indeed, already formula
(\ref{area}) for the area of a surface is non-polynomial in the triads
$\E$.  Nonetheless, somewhat surprisingly, the corresponding quantum
operators can be constructed via suitable regularization \cite{AL3}. 
The resulting length \cite{T1}, area \cite{RS3,AL4} and volume
\cite{RS3,AL5} operators are again covariant with respect to spatial
diffeomorphisms.

In this paper, we will need only the area operator.  Let us therefore
focus on that case.  As we saw in Section \ref{s2.A}, given an
oriented surface $\T$ embedded in $M$, its area (\ref{area}) defines
a function $A_\T$ on the classical phase space $\X$.  The
action of the corresponding quantum operator $\hat{A}_\T$ 
is given by:
\be \label{areaop} 
\hat{A}_\T\, [p_\g^* \, \psi] = p_\g^*\left[{8\pi \gamma \l^2}\,\, 
\sum_v \, \sqrt{J^i(v) J^j(v) k_{ij} }\,\, \psi \right] \, .  
\ee
where, for simplicity, we have again restricted ourselves to the case
in which all edges of $\g$ lie above $T$.  (The square-root is
well-defined because the operator in the parenthesis is a positive
definite, essentially self-adjoint operator on $\H_\g$.) 

With $\cyl$ as its domain, $\hat{A}_\T$ is an essentially
self-adjoint, positive definite operator on $\H$.  Furthermore, it has
some physically striking properties.  First, all its eigenvalues are
known in a closed form \cite{AL3} and they are all discrete multiples
of $\l^2$, the square of the Planck length.  Thus, at the Planck
scale, the continuum picture breaks down in a precise sense.  For any
surface $\T$, the smallest eigenvalue of $\hat{A}_\T$ is of course
zero.  It turns out that the area gap ---the value of the first
non-zero eigenvalue--- depends on the topology of $\T$.  Finally, the
`level spacing' ---i.e., the difference between the consecutive
eigenvalues--- goes to zero rapidly, as the exponential of the
square-root of the area (in Planck units).  Consequently, the
continuum limit is reached \textit{very} rapidly.

Since $\cyl$ is dense in $\H$, heuristically a `typical' state $\psi$
is associated with a graph $\g$.  Note that such functions $\psi(A)$
depend only on the action of the generalized connection $A$ on the
edges of the graph $\g$.  In particular, in such states the area
assigned to a surface not intersecting $\g$ must vanish identically.
More generally, these states represent excitations of geometry only
along the graph $\g$.  In this sense, typical excitations of quantum
geometry are one-dimensional, like a polymer.  Therefore one says that
in non-perturbative quantum gravity, space has a `polymer geometry'.

Finally, as mentioned above, elements of $\H$ only represent
`kinematical states'.  Nonetheless, just as the full phase space $\X$
plays an important role in the classical Hamiltonian framework, the
structure provided by $\H$ is important to the quantum theory.  In
particular, this structure provides the tools that are needed to
define the quantum constraint operators.  Physical states of
non-perturbative quantum gravity are annihilated by these operators.
Because of the geometrical nature of the Gauss and diffeomorphism
constraints, the corresponding operators are essentially unambiguous
\cite{ALMMT,BS,LT}.  For the Hamiltonian constraint, such geometrical
guidelines are not yet known.  Nonetheless, specific proposals for the
corresponding operators have been made (see in particular
\cite{T2,MoreHam}) and are currently being analyzed from various
angles.  Fortunately, the analysis of this paper depends only on some
general assumptions on the solutions to the quantum Hamiltonian
constraint (since the lapse function smearing this constraint has to
vanish on the horizon).  Therefore, our results are largely
insensitive to the detailed form that the final `correct' constraint
operator will have.

\section{The Quantization Strategy}  
\label{s3}

The quantization procedure used to handle the boundary conditions at the
horizon and the relevant parts of the quantum Einstein equations is
technically quite subtle.  In particular, it requires a careful
treatment of a number of delicate features of the $\U(1)$ Chern-Simons
theory.  These issues are discussed in detail in the next two sections. 
In this section, we will sketch the general strategy to orient the
reader and to provide the overall logic behind the detailed, technical
treatment that follows.

Recall from Section \ref{s2.A} that the volume and surface degrees of
freedom cannot be separated in the classical theory: Since the horizon
$S$ is the inner boundary of the spatial 3-manifold $M$, all fields on
$S$ are determined by fields in the interior of $M$ by continuity.
However, as we saw in Section \ref{s2.B}, in the quantum theory the
fields describing geometry become discontinuous in a certain precise
sense, so the fields on $S$ are no longer determined by fields in $M$;
in this case there are independent degrees of freedom living on the
boundary.  These surface degrees of freedom dictate the quantum
geometry of the horizon and account for black hole entropy in our
approach.

More precisely, since the holonomies of generalized connections along
paths that lie on $S$ are quite independent from their holonomies along
paths in the interior of $M$, the quantum configuration space $\Ab$ is
now given by a product $\Ab = \Ab_V\times \Ab_S$ of the quantum
configuration space associated with the interior of $M$ and that
associated with the horizon $S$.  Therefore, it is natural to begin with
the product $\H_V\otimes \H_S$ of Hilbert spaces consisting of suitable
functions on $\Ab_V$ and $\Ab_S$.  The construction of volume states
closely follows the procedure outlined in Section \ref{s2.B}. Thus,
there is a uniform measure $\mu_V$ on $\Ab_V$ which enables one to
construct $L^2(\Ab_V)$.  We could take this space as the volume Hilbert
space.  However, to avoid a proliferation of spaces and symbols, it is
more convenient to impose at this stage the technically trivial part of
the quantum Gauss constraint so that we can focus only on the nontrivial
part later.  Let $\G_V$ denote the space of all (not necessarily
continuous) gauge transformations of $P$ \textit{that are the identity
on} $S$.  We will take the volume Hilbert space $\H_V$ to be the
subspace of $L^2(\Ab_V)$ consisting of vectors that are invariant
under $\G_V$.
 
The form of the symplectic structure (\ref{ss}) suggests that the
surface Hilbert space $\H_S$ should be closely related to the Hilbert
space of quantum states of Chern-Simons theory.  To construct this
space, let us first recall that, \textit{even at the kinematical level},
fields $(A,\E)$ in the phase space $\X$ have to satisfy the boundary
condition (\ref{bc}) which arises because $S$ is the intersection of an
isolated horizon with the spatial manifold $M$.  This boundary condition
must be incorporated also in the quantum theory.  The heuristic idea is
to quantize equation (\ref{bc}) by replacing $F$ and $\underline\E\cdot
r$ by the corresponding operators.  By imposing this condition as an
operator equation, physically we are allowing the triad as well as the
curvature to fluctuate at the horizon but asking that they do so in
tandem.

Now, since $F$ is the curvature of the surface connection $W$, one
expects the field operator $\hat{F}$ to act on the surface Hilbert space
$\H_S$ while, as we saw in Section \ref{s2.B}, the operator $\hat\E$
acts on $\H_V$.  Thus the quantum version of the horizon boundary
condition (\ref{bc}) imposes a relation between the surface and volume
states.  One would naively expect it to constrain states $\Psi$ in
$\H_V \otimes \H_S$ via
\be
\label{qbc1}
(1 \otimes \hat{F}) \, \Psi = \, (-{2 \pi \gamma\over a_0} 
\hat{\underline{\E}} \cdot r \otimes 1)\, \Psi .
\ee
However, it turns out that, because of certain subtleties associated
with the quantum Chern-Simons theory (discussed in Sections
\ref{s4.C.2} and \ref{s5.A.2}), it is only the `exponentiated version'
$\exp(i\hat{F})$ of $\hat{F}$ that is
well-defined on $\H_S$.  Therefore, in place of (\ref{qbc1}), we are
led to impose
\be
\label{qbc2}
(1 \otimes \exp(i \hat{F})) \, \Psi = 
(\exp( -i {2 \pi \gamma\over a_0}
\hat{\underline{\E}} \cdot r) \otimes 1)\, \Psi
\ee
We call this the `quantum boundary condition'. Only those elements of
$\H_V \otimes \H_S$ that satisfy this boundary condition can qualify as
kinematical quantum states of our system.

The structure of this equation implies that we can obtain a basis
$\Psi_V \otimes \Psi_S$ of solutions such that $\Psi_V$ and $\Psi_S$
are eigenstates of $\hat{\underline{\E}}\cdot r$ and
$\exp (i\hat{F})$ respectively, with
\be 
\label{qbc3}
\Psi_V \otimes \exp(i \hat{F}) \Psi_S = 
\exp(-i {2 \pi \gamma\over a_0} \hat{\underline{\E}} \cdot r)
\Psi_V \otimes \Psi_S
\ee
Now, all the eigenstates of ${\hat{\underline\E}}\cdot r$ are known
\cite{AL4}.  They satisfy
\be \label{ev}
(\hat{\underline{\E}} \cdot r) \Psi_V = 
8\pi \l^2 \sum_{i = 1}^n \, m_i\, \delta^2(x,p_i)\, \eta \,\,\Psi_V
\ee 
where
\[ \P=\{p_1,\ldots, p_n,\} \]     
is some finite set of points on $S$, $m_i$ are spins (i.e.,
half-integers) labelling these points, $\delta^2$ is the delta
distribution on $S$, and $\eta$ the Levi-Civita density on $S$.  This
fact allows a useful decomposition of the volume Hilbert space $\H_V$,
and also leads to a precise construction of the surface Hilbert space
$\H_S$.

Let us first consider $\H_V$. {}From our discussion in Section \ref{s2.B}
it follows that the states in $\H_V$ satisfying equation (\ref{ev}) are
cylinder functions based on graphs in $M$ whose edges have ends at the
horizon at the points $\P$.  If we let $\H_V^{\P,m}$ be the space
of all states satisfying equation (\ref{ev}), then 
\[       \H_V = \bigoplus_{\P,m} \H_V^{\P,m} ,  \]
where $\P$ ranges over all finite sets of points on $S$ and $m$ ranges
over all ways of labelling these points with nonzero spins.  
This decomposition will be useful for solving the quantum boundary
condition.

Let us now turn to the construction of the surface Hilbert space
$\H_S$.  Since classically the pullback of the connection $A$ to $S$
is determined by the $\U(1)$ connection $W$, it is natural to write
the surface states $\Psi_S$ as a function of a {\it generalized}
$\U(1)$ connection on $S$.  Then (\ref{qbc3}) implies that $\Psi_S$
has support only on generalized $\U(1)$ connections that are flat
everywhere except at finitely many points $p_i$ where the polymer
geometry excitations in the bulk puncture $S$.  Let us fix a set of
points $\P$ and denote the space of such generalized connections by
$\A^\P$ (for a precise definition of $\A^\P$, see Section
\ref{s4.C.1}).  As with the volume Hilbert space, it is convenient to
incorporate the technically trivial part of quantum Einstein's
equations already in the definition of $\H_S$.  Therefore, let us
quotient $\A^\P$ by the action of the group $\G^\P$ of $\U(1)$ gauge
transformations that are identity at the punctures $p_i$ and of the
group $\D^\P$ consisting of diffeomorphisms of $S$ that fix each of
the punctures $p_i$ together with certain structure needed for
quantization. (For precise definitions of these groups, see Section
\ref{s4.C.1}.) One can show that the resulting space
\[  \X^\P = \A^\P/(\G^\P \semi \D^\P)  \] 
is equipped with the Chern-Simons symplectic structure coming from the
surface term in equation (\ref{ss}).  Therefore, $\X^\P$ can be thought of 
as the phase space of the surface degrees of freedom associated with
the set $\P$ of punctures.  Denote by $\H^\P_S$ the Hilbert space
obtained by geometric quantization of $\X^\P$.  Then, the total surface
Hilbert space $\H_S$ can be constructed as a direct limit of the 
spaces $\H^\P_S$ as the punctures $\P$ range over all finite subsets of $S$. 

One can show that $\X^\P$ is isomorphic with a $2(n-1)$-dimensional
torus: 
\[              \X^\P = \C^{n-1}/\Lambda \] 
for the lattice $\Lambda = (2\pi\Z)^{2(n-1)}$.  Therefore, to quantize
$\X^\P$, one can begin with the space of holomorphic functions on
$\C^{n-1}$ as in the Bargmann-Segal representation of the canonical
commutation relations \cite{Bargmann,Segal} and then select only those
states that are invariant under the discrete subgroup defined by
$\Lambda$.  These are the so-called `theta functions', studied
extensively since the 1800s.  Now, as is well known, the translation
group on $\C^n$ acts only projectively in the Bargmann-Segal
representation because, if we set $z=q+ip$, the translations in the $q$
directions fail to commute with the translations in the $p$ directions.
As a result, nontrivial states that are invariant under the action of
$\Lambda$ exist if and only if the constant
\be  \label{level}
k = \frac{a_0}{4\pi \gamma \l^2} 
\ee
is an integer.  (The combination on the right comes from the coefficient
in front of the surface term in the symplectic structure (\ref{ss})
which determines the phase factor in the projective representation.) 
This is the familiar `prequantization condition', and $k$ is called the
`level' of the Chern-Simons theory.  Next, associated with any small
loop $\eta_i$ winding once around the puncture $p_i$ there is an operator 
$\hat{h}_i$ on $\H_S^\P$ which measures the holonomy around this loop.
It turns out that eigenstates of these operators provide a basis
$\psi_{\P,a}$  of the surface Hilbert space, where $\P$ ranges over all
finite sets of points on the horizon and $a$ ranges over all ways of
labelling these points by nonzero elements $a_i$ of $\Z_k$ (the group of
integers modulo $k$), satisfying:
\be
\label{sum}
  a_1 + \cdots + a_n =0  .
\ee
We have
\be
\label{holonomy}
\hat{h}_i\, \Psi_{\P,a} = e^{\frac{2 \pi i a_i}{k}}\, \Psi_{\P,a}\, ,  
\ee
so heuristically the state $\psi_{\P,a}$ describes a quantum geometry
of the horizon in which the $\U(1)$ connection is flat except at the
punctures $p_i$, with a holonomy of $\exp(2 \pi i a_i/k)$ around the
$i$th puncture.  In short, the horizon is flat except at the
punctures, where it has conical singularities with \textit{quantized
angle deficits}.  Note that this quantization is a direct consequence
of the prequantization condition on $k$.  Thus, there is an
interesting intertwining of quantum geometry and Chern-Simons theory.

The geometrical meaning of equation (\ref{sum}) is clarified if we work
with the $\SO(2)$ `Levi-Civita' connection instead of the $\U(1)$
connection.  In these terms, the angle deficit at the $i$th puncture is
$4\pi a_i/k$.  However, since $a_i$ is only defined modulo $k$, these
angle deficits are only defined modulo $4 \pi$, just as one would
expect, given that our approach to quantum geometry is based on
parallel transport of spinors.  Equation (\ref{sum}) says that the sum
of these angle deficits vanishes modulo $4 \pi$.  This is a quantum
analogue of the Gauss-Bonnet theorem, which says that for any metric on
the 2-sphere, the integral of the scalar curvature equals $4 \pi$.

So far, we have discussed the structure of the volume and surface
Hilbert spaces, $\H_V$ and $\H_S$, whose construction was motivated by
the quantum boundary condition (\ref{qbc3}).  Now that we have
specific Hilbert spaces on hand, we can impose this condition on $\H_V
\otimes \H_S$ in a precise fashion and explore its consequences.  Note
first that a state $\Psi_V \otimes \Psi_S$ can satisfy (\ref{qbc3}) if
and only if the eigenvalue of $\exp(-i \textstyle{2 \pi \gamma\over
a_0} \hat{\underline{\E}} \cdot r)$ on $\Psi_V$ equals the eigenvalue
of $\exp (i \hat{F})$ on $\Psi_S$.  Now, the first of these is an
operator on $\H_V$ defined in the quantum geometry framework, while
the second is an operator on $\H_S$ constructed using Chern-Simons
theory.  A priori it is not at all obvious that the spectra of these
two distinct operators have any overlap.  If they do not, (\ref{qbc3})
would have no solutions and this approach to the quantum geometry of
horizons and black hole entropy would not be viable.

Now, it follows from the above discussion that each operator has a 
nontrivial action only at the punctures.  By equation (\ref{ev}), at
each puncture $p_i$ the eigenvalues of $\exp(-i \textstyle{2 \pi
\gamma\over a_0} \hat{\underline{\E}} \cdot r)$ are of the form
\[    \exp (- {2 \pi i\gamma\over a_0}(8 \pi \l^2 m_i)) \]
where $m_i$ is an half-integer.  Similarly, equation (\ref{holonomy}) 
implies that the eigenvalues of $\exp(i\hat{F})$ at $p_i$ are of the 
form 
\[      \exp ({\frac{2\pi i a_i}{k}})      \]
where $a_i$ is any integer mod $k$.  (For details, see Sections
\ref{s4.C.2} and \ref{s5.A}.)  Simple algebra using equation
(\ref{level}) shows that these spectra in fact coincide!  The
eigenvalues match when $2m_i = -a_i$ mod $k$.  This is a striking
example of the unexpected, detailed matching between classical general
relativity (which dictates the horizon boundary condition), quantum
geometry (which determines the action of $\hat{\underline\E}\cdot r$)
and quantum Chern-Simons theory (which determines $\exp (i \hat{F})$).

With this background material at hand, we can now exhibit the space of
solutions to the quantum boundary conditions, i.e., the kinematical
Hilbert space: It is simply the subspace of $\H_V\otimes\H_S$ given by
\be\label{kinH}
\H^{\rm Kin} \,  = \bigoplus_{\P,m,a\, \colon\; 2m = -a \, \mod \, k} 
\H_V^{\P,m} \otimes \H_S^{\P,a}\, .
\ee
where $\H_S^{\P,a}$ is the one-dimensional Hilbert space spanned by
the surface state $\psi_{\P,a}$ introduced above.  As one might have
expected, the volume states are correlated with the surface states 
at the punctures in a specific way.

Finally, one has to impose the quantum versions of Einstein
constraints to extract physical states of our system.  As noted in
Section \ref{s2.A}, there are three sets of constraints: the Gauss,
diffeomorphism and Hamiltonian constraints.  Since the `lapse function'
smearing the Hamiltonian constraint must vanish at the horizon, this
constraint will not play an essential role in determining the quantum
geometry of the horizon.  Let us therefore concentrate on the other two. 

Let us begin with the Gauss constraint which demands that physical
states be gauge invariant.  By construction, all elements of $\H_V^\P
\otimes \H_S^\P$ are invariant under gauge transformations that are
identity at the punctures. Therefore, it only remains to ensure gauge
invariance \textit{at} the punctures.  Now, since the commutation
relations on the surface Hilbert space are dictated by the
Chern-Simons symplectic structure, gauge rotations on $\H_S^\P$ are
implemented \textit{precisely} by the surface operator in the quantum
boundary condition (\ref{qbc2}). Similarly, on the volume Hilbert
space, the operator $\hat{\E}\cdot r$ generates $\U(1)$ gauge
transformations on the horizon.  As a consequence, the volume operator
in (\ref{qbc3}) implements the action of $\U(1)$ gauge transformations
at the punctures on $\H_V^\P$.  Thus, the meaning of equation
(\ref{qbc3}) turns out to be rather simple: it ensures that the volume
and surface states are `coupled' in precisely the correct way so that
the total state is invariant under the allowed $\U(1)$ internal
rotations on the horizon.  Thus, the quantum Gauss constraint is
automatically satisfied on the Hilbert space $\H$.

Finally, let us consider the diffeomorphism constraint.  This
constraint simply implies that two states in the Hilbert space $\H$
should be regarded as equivalent if they are related by a
diffeomorphism of $M$ that maps $S$ to itself.  For the quantum
geometry of the horizon, only the action of the diffeomorphisms on $S$
is relevant. This action is rather subtle because the construction of
$\H^\P_S$ requires the introduction of an extra structure on
$S$. Nonetheless, the final result is transparent and easy to state.
Since any two sets $\P$ and $\P'$ with the same number of punctures
are related by a diffeomorphism, quantum geometries of the horizon
compatible with $\P$ are physically indistinguishable with those
compatible with $\P'$.  The `locations' of punctures are irrelevant;
what matters is only the \textit{number} of punctures.  This fact
plays an important role in entropy calculation of Section \ref{s6}.

\section{Kinematical Hilbert Spaces}
\label{s4}

Using the overall strategy presented in Section \ref{s3} as a
guideline we will now introduce the kinematical Hilbert spaces.  In
Section \ref{s4.A}, we extend the theory of generalized connections to
the case of a manifold with boundary.  We simply describe the
necessary results, because the proofs are similar to the case of
manifolds without boundary.  In Section \ref{s4.B}, we obtain the
volume Hilbert space again by a simple extension of the framework of
the case without boundary discussed in Section \ref{s2.B}. In Section
\ref{s4.C} we construct the surface Hilbert space using quantum
Chern-Simons theory. This discussion is more detailed.

\subsection{Generalized Connections}  \label{gen.conn} 
\label{s4.A}

In Section \ref{s2.B} we introduced generalized connections on a
manifold without boundary. The situation is a bit more complicated in
our treatment of the black hole, due to the boundary conditions at the
horizon $S$.  However, the theory of generalized connections still
forms the basis of our treatment and, as emphasized before, is in fact
crucial for understanding the separation of volume and surface degrees
of freedom that occurs. We shall show that the space $\overline \A$ of
generalized connections can be written as a product of two spaces,
$\overline \A_V$ and $\overline \A_S$, consisting of generalized connections 
`in the volume' and `on the surface' respectively.  Starting with these
spaces, in the next two sections we construct in detail the volume and
surface Hilbert spaces for the quantum black hole along the lines
outlined in Section \ref{s3}.

For the sake of possible future generalizations, we proceed quite
generally in this section, and specialize to the case at hand later.
Suppose $X$ is a real-analytic manifold with boundary $\partial X$.  We
define a `path in $X$' to be an equivalence class of analytic maps
$\eta \maps [0,1] \to X$ with $\eta'(t) \ne 0$, where we consider
two such maps the same if they differ by a reparametrization, that is,
an analytic orientation-preserving diffeomorphism of $[0,1]$. Given a
path $\eta \maps [0,1] \to X$, we call $\eta(0)$ its `source' and
$\eta(1)$ its `target'.   If $x \in X$ is the source or target of
$\eta$ we say $\eta$ is `incident' to $x$.  Given  $\eta_1$ and
$\eta_2$, if the target of $\eta_1$ equals the source of $\eta_2$ we
let the product $\eta_1 \eta_2$ be the path consisting of $\eta_1$
followed by $\eta_2$.  Similarly, we let $\eta^{-1}$ be the path
formed by reversing the orientation of the path $\eta$.    

Suppose $P$ is a smooth principal $G$-bundle over $X$, with $G$ a
compact connected Lie group.  We fix a trivialization of $P$ at each
point $x \in X$, so that we can think of parallel transport along any
path in $X$ as an element of $G$.  (We do not demand that this
trivialization vary continuously with $x$.)  We define a `generalized
connection on $P$' to be a map $A$ assigning to each path in $X$ an
element of $G$, satisfying the following properties:
\begin{enumerate}
\item  $A(\eta_1 \eta_2)$ = $A(\eta_2) A(\eta_1)$
\item  $A(\eta^{-1}) = A(\eta)^{-1}$
\end{enumerate}
We think of the group element $A(\eta)$ as describing parallel
transport from the source of $\eta$ to its target.  

Let $\A$ be the space of smooth connections on $P$ and let $\Ab$ be
the space of generalized connections on $P$. There is a one-to-one map
from $\A$ to $\Ab$ that assigns to each smooth connection the
generalized connection having the same holonomies along paths.  This
allows us to think of $\A$ as a subspace of $\Ab$.  There is a natural
topology on $\Ab$ for which this subspace is dense, so we may think of
$\Ab$ as a `completion' of the space of smooth connections.  A key
advantage of working with this completion is that it possesses a
natural measure, invariant under both diffeomorphisms and gauge
transformations, called the `uniform measure'.

The uniform measure $\mu$ on $\Ab$ is defined just as in the case of a
manifold without boundary \cite{AL1,AL2,B2}. In Section \ref{s2.B}, we
introduced it as the projective limit of the family of measures
$\mu_\g$ on configuration spaces $\A_\g$ associated with graphs
$\g$.  We will now give an different but equivalent characterization
which brings out how natural this measure is.  Define an `edge in $X$' to 
be a path in $X$ which restricts to an embedding of the interval $(0,1)$.
Given a finite set of paths $\eta_i$ in $X$, we can always find a
finite set of edges $e_j$ in $X$ such that:
\begin{alphalist}
\item Each path $\eta_i$ is a product of finitely many edges $e_j$
and their inverses.
\item Distinct edges $e_j$ intersect, if at all, only at their endpoints.
\end{alphalist}
A finite set of edges with property b) is called a `graph in $X$', and
we call the endpoints of these edges the `vertices' of the graph.  The
uniform measure on $\Ab$ is characterized by the following property:
for any graph $\g=\{e_j\}$ in $X$, the group elements $A(e_j)$ are
independent $G$-valued random variables, each distributed according to
normalized Haar measure on $G$.  We denote this measure by $\mu$.

For applications to black holes we need a way of separating a
generalized connection into two parts: the `volume' part, which
describes holonomies along paths having no subinterval lying in the
boundary $\partial X$, and the `surface' part, which describes
holonomies along paths lying completely in $\partial X$. To do this,
we define a `path in the volume' to be one which has no subinterval
lying in $\partial X$, and define a `generalized connection in the
volume' to be a map $A$ assigning to each path in the volume an
element $A(\eta) \in G$, satisfying the properties 1 and 2 listed
above.  Similarly, we define a `generalized connection on the surface'
to be a map $A$ assigning to each path in $\partial X$ an element
$A(\eta) \in G$ satisfying properties 1 and 2.

Let $\Ab_V$ denote the space of generalized connections
in the volume, and let $\Ab_S$ denote the space of
generalized connections on the surface.  A generalized connection
on $P$ determines generalized connections in the volume and on
the surface, so we have maps
\[          p_V \maps \Ab \to \Ab_V , \qquad 
            p_S \maps \Ab \to \Ab_S .\] 
One can check that in fact 
\[         \Ab =  \Ab_V \times \Ab_S, \]
with the above maps being the projections onto the two factors.  

We can define `uniform measures' on $\Ab_V$ and $\overline
\A_S$ by  pushing $\mu$ forwards along the projections from $\overline
\A$ to these two spaces:
\[      \mu_V = (p_V)_\ast \mu , \qquad \mu_S = (p_S)_\ast \mu  .\]
Moreover, one can check that, with their uniform measures, $\overline
\A$ is equal {\it as a measure space} to the product $\Ab_V
\times \Ab_S$.  This boils down to the fact that with respect to
the uniform measure, holonomies along collections of paths in the volume
are independent (as random variables) from holonomies along collections
of paths in the boundary $\partial X$.  

\subsection{The Volume Hilbert Space}   \label{vol.hilb} 
\label{s4.B}

Now let us focus attention on the case of physical interest: the
3-manifold $M$ with the 2-sphere $S$ as its interior boundary.  Here $P$
is the trivial $\SU(2)$ bundle over $M$, and we use a fixed
trivialization to think of holonomies along paths in $M$ as group
elements.

Using the results of the previous section we define the space
$\Ab_V$ of generalized connections in the volume, and equip
it with its uniform measure $\mu_V$.  Using this measure one can
define the Hilbert space $L^2(\Ab_V)$.  As in the case of
manifolds without boundary \cite{RS2,B2,B3,ALMMT,AL4}, one 
can introduce an explicit set of vectors spanning this space labelled by
triples $\psi = (\g,\rho,\nu)$ such that:
\begin{enumerate}
\item $\g$ is a graph in the volume,
\item $\rho$ assigns to each edge $e$ of $\g$ a nontrivial irreducible
representation $\rho_e$ of $\SU(2)$,
\item $\nu$ assigns to each vertex $v$ of $\g$ a vector $\nu_v$
in the tensor product of the representations $\rho_e$ labelling edges
$e$ incident to $v$. 
\end{enumerate}
Here we say a graph $\g$ in $M$ is `in the volume' if all its
edges have no subinterval lying in $S$.  

{}From any such triple we get a function in $L^2(\Ab_V)$, which
we also call $\psi$, as follows:
\[ \psi(A) = \langle \bigotimes_e \rho_e(A(e)), \bigotimes_v
\nu(v)\rangle .\]
Here the first tensor product is taken over the edges of $\g$, the
second is taken over the vertices of $\g$, and $\langle
\cdot,\cdot\rangle $ denotes the natural pairing given by contraction
of indices.  If we let $\g$ range over all graphs inside $M$, let
$\rho$ range over all labellings of edges by nonzero spins, and let
$\nu$ range over all labellings of vertices by vectors chosen from an
orthonormal basis, we obtain states $\psi$ forming an orthonormal
basis of $L^2(\Ab_V)$.

Let $\G_V$ denote the group of (not necessarily continuous) gauge
transformations of $P$ that equal the identity on $S$.  Since gauge
transformations act as unitary operators on $L^2(\Ab_V)$, we
may define the `volume Hilbert space', $\H_V$, to be the subspace of
$L^2(\Ab_V)$ consisting of vectors invariant under the action
of $\G_V$.  Starting from the  explicit description of $L^2(\overline
\A_V)$ in the previous paragraph, one can prove that the volume Hilbert
space is spanned by vectors $\psi$ corresponding to triples
$(\g,\rho,\iota)$ of the following form: 
\begin{enumerate}
\item $\g$ is a graph in the volume,
\item $\rho$ assigns to each edge $e$ of $\g$ a nontrivial irreducible
representation $\rho_e$ of $\SU(2)$,
\item $\iota$ assigns to each vertex $v$ of $\g$ a vector $\iota_v$
in the tensor product of the representations $\rho_e$ labelling edges
$e$ incident to $v$, and if $v$ lies in the interior of $M$, $\iota_v$ must
be invariant under the action of $\SU(2)$ on this tensor product.
\end{enumerate}
We call such a triple an `(open) spin network'.  We use the same
symbol $\psi$ to denote a spin network and the vector in the volume
Hilbert space that it determines.  If we let $\g$ range over all
graphs in $M$, let $\rho$ range over all labellings of edges by nonzero
spins, and let $\iota$ range over all labellings of vertices by vectors
chosen from an orthonormal basis, we obtain states forming an
orthonormal basis of $\H_V$.  

Any spin network $\psi = (\g,\rho,\iota)$ has a set of `ends',
namely the vertices of the graph $\g$ that lie in $S$.  For any
finite $\P = \{p_1,\dots, p_n\}$ of $S$, let $\H_V^\P$ be the subspace
of the volume Hilbert space spanned by all open spin networks whose
ends lie in the set $\P$.  Note that since every open spin network
lies in some space $\H_V^\P$, the volume Hilbert space is the closure
of the union of all these spaces $\H_V^\P$.  Also note that if $\P
\subseteq \P'$, then $\H_V^\P \subseteq \H_V^{\P'}$.  Technically,
these two facts together say that the Hilbert space volume Hilbert
space is the `direct limit' of the Hilbert spaces $\H_V^\P$:
\[        \H_V = \lim_{\P} \H_V^\P   \]
as $\P$ ranges over all finite subsets of $S$.  In the next section,
we shall give a similar description of the surface Hilbert space as a
direct limit.  These descriptions become important when we implement
the quantum boundary condition in Section \ref{s5.A}.

We conclude by describing an important operator on the volume Hilbert 
space, namely the horizon area operator. Construction of this operator 
follows the usual treatment in quantum geometry \cite{RS3,AL4}, as reviewed 
in Section \ref{s2.B}.  We need only adapt it slightly to take into account 
the special role of the horizon.

Let $\G_S$ denote the group of all (not necessarily continuous) gauge 
transformations of the bundle $P|_S$.  This group acts on $\overline\A_V$ 
in a measure-preserving way so it acts as unitary operators on 
$L^2(\overline\A_V)$.  Since these operators commute with the action of 
$\G_V$ on $L^2(\overline\A_V)$, they preserve the volume Hilbert space 
$\H_V$, which consists of the functions invariant under the action of 
$\G_V$.  We thus obtain a unitary representation of $\G_S$ on $\H_V$.

In particular, if we pick any point $p \in S$, we obtain a unitary
representation of $\SU(2)$ on $\H_V$ by considering gauge
transformations in $\G_S$ that are the identity at every point except
$p$.  Let $J^i(p)$ stand for the self-adjoint infinitesimal generators of
this representation, satisfying the usual angular momentum commutation
relations, and let
\[    J(p) \cdot J(p) = k_{ij} \, J^i(p) J^j(p)   \]
be the corresponding Casimir.   Then as described in equation (\ref{areaop}) 
there is a self-adjoint operator $\hat{A}_S$ on the volume Hilbert space that
measures the area of the horizon, given by
\[ \hat{A}_S = 8 \pi \gamma \l^2  \sum_{p \in S} \sqrt{J(p) \cdot J(p)} \;
. \]
Note that although the sum is uncountable, only countably many terms
give a nonzero contribution when this operator is applied to any
state.

Suppose $\P$ is a finite subset of points in $S$ and $j =
(j_1,\dots,j_n)$ is a way of labelling each point $p_i \in \P$  with a
nonzero spin (i.e. half-integers) $j_i$. Then there is a subspace 
$\H_V^{\P,j}$ of the volume Hilbert space consisting of all vectors 
$\psi$ such that
\[      (J(p_i) \cdot J(p_i)) \psi = j_i(j_i+1) \psi \]
for each point $p_i$ and 
\[      (J(p) \cdot J(p)) \psi = 0 \]
for all points $p$ other than the points $p_i$.  Clearly $\H_V^{\P,j}$
is a subspace of $\H_V^\P$, since gauge transformations at any point 
$p \in S$ act trivially on states corresponding to spin networks that
do not have $p$ as one of their ends.  Moreover, we have a direct
sum decomposition:
\be     
               \H_V = \bigoplus_{\P,j} \H_V^{\P,j}  
\label{H_V}
\ee
where $\P,j$ range over all finite sets of points in $S$ labelled
by nonzero spins.  This decomposition serves to diagonalize the area
operator $\hat{A}_S$, since any state $\psi \in \H_V^{\P,j}$ is an 
eigenstate of this operator with eigenvalue given by:
\be    
\hat A_S \psi = 8 \pi \gamma \l^2 \sum_i \sqrt{j_i (j_i + 1)} \psi ,
\label{area2} 
\ee
where the sum is taken over all the spins labelling points in $\P$.

\subsection{The Surface Hilbert Space}   \label{surf.hilb} 
\label{s4.C}

In this section we construct the surface Hilbert space $\H_S$ by
geometrically quantizing the space of generalized connections on the
2-sphere $S$ while taking into account the quantum boundary condition
(\ref{qbc2}) and the surface term in the symplectic structure
(\ref{ss}).  As discussed in Section \ref{s3}, the first task is to
construct, for each finite set $\P$ of points on $S$ the quantum phase
space $\X^\P$ consisting of generalized connections that are flat
everywhere on $S$ except at these points.  This will be carried out in
the first subsection.  In the second subsection, we will quantize this
phase space.  As noted in Section \ref{s3}, technical subtleties of this
quantization have an important effect on the quantum geometry of the
horizon.

\subsubsection{The phase space $\X^\P$} 
\label{s4.C.1}

Recall that the $\SU(2)$ bundle $P|_S$ has a $\U(1)$ sub-bundle $Q$
which is isomorphic to the spin bundle of the sphere $S$.   As in
Section \ref{gen.conn}, we define a `generalized connection on the
surface' to be a map $W$ assigning to each path $\eta$ in $S$ a
holonomy $W(\eta) \in \SU(2)$ in a consistent manner.   This
definition makes use of a trivialization of $P|_S$ over each point of
$S$.  If we wish to avoid this, we can alternatively think of $W(\eta)$ 
as a `transporter': a map from the fiber $P_{\eta(0)}$ to the fiber
$P_{\eta(1)}$.  We then call $W$ a `generalized $\U(1)$ connection' if
for each path $\eta$, $W(\eta)$ maps $Q_{\eta(0)}$ to
$Q_{\eta(1)}$.   We may also think of a generalized $\U(1)$ connection
as a generalized connection on the bundle $Q$.  

Next, let $\P = \{p_1,\dots, p_n\}$ be a finite set of points in $S$.  We
call these points `punctures'.   We say that a generalized $\U(1)$
connection $W$ is `flat except at the punctures' if it assigns the
same holonomies to paths as a connection $W_0$ on $Q$ with the 
following properties:
\begin{enumerate} 
\item $W_0$ is flat on $S - \P$.
\item For some neighborhood $U_i$ of each puncture $p_i$, some
smooth trivialization of $Q$ over $U_i$, and some analytic coordinate system 
$(x,y)$ on $U_i$ for which $p_i$ has the coordinates $(a,b)$, $W_0$ has
the following form:
\be  
W_0 = W_1 + c {(x - a)dy - (y - b)dx\over (x - a)^2 + (y - b)^2} 
\label{W_0}
\ee
on $U_i - \{p_i\}$, where $c \in \R$ and $W_1$ is a bounded smooth
1-form on $U_i - \{p_i\}$.
\end{enumerate}  
This definition requires some comment.  With condition 1 we demand
that $W$ be flat away from the punctures, and with condition 2 we
demand that at each puncture it has the usual sort of singularity
produced, say, in Maxwell theory by a magnetic flux line intersecting
$S$ transversely at that point.  Since the connection $W_0$ is
singular at the punctures we must actually check that it assigns a
well-defined holonomy to paths going {\it through} these points.  It
suffices to consider the case where $e \maps [0,1] \to S$ is an edge
lying completely within the neighborhood of $p_i$ where $W_0$ has the
stated form, and where $e(t) = p_i$ for $t = 0$ but for no other
values of $t$.  Using polar coordinates $(r,\theta)$ centered at the
point $p_i$, a naive calculation shows that
\[ \int_e W_0 = \int_e W_1 + \int_0^1 {d\theta(e(t))\over dt} dt .\]
The first integral on the right-hand side is well-defined because
$W_1$ is smooth and bounded on $U_i - \{p_i\}$.  Similarly, even
though $\theta$ is multi-valued and $\theta(e(t))$ is undefined at $t
= 0$, the second integral on the right-hand side is well-defined,
because using the fact that $e$ is analytic, one can show that
$d\theta(e(t))/dt$ is well-defined and continuous for $t > 0$ and has
a finite limit as $t \downarrow 0$.  We therefore use this formula to
{\it define} the holonomy of $W_0$ along the edge $e$. It is simple to
verify that the two conditions of Section \ref{s4.A}, required of
generalized connections, are satisfied.

Let $\A^\P$ be the space of generalized $\U(1)$ connections that are
flat except at the punctures.  Let $\G^\P$ be the group of (not
necessarily continuous) gauge transformations of $Q$ that equal the
identity at the punctures.  Fix a ray in the tangent space of each
puncture, and let $\D^\P$ be the identity component of the group of
orientation-preserving analytic diffeomorphisms that fix each puncture
together with these rays.  The semidirect product $\G^\P \semi \D^\P$ is
a subgroup of the group of all automorphisms of the bundle $Q$. Since
the space $\A^\P$ is defined in a gauge- and diffeomorphism-covariant
way, the group $\G^\P \semi \D^\P$ acts on $\A^\P$.  The quantum
Einstein equations require that we treat this action as gauge.  We are
therefore led to consider the quotient space $\X^\P = \A^\P/(\G^\P \semi
\D^\P)$.

Our next goal is to describe $\X^\P$. 

\begin{thm}\et \nonumber The space $\X^\P$
is diffeomorphic to a $2(n-1)$-dimensional torus.
\end{thm}

Proof - We begin by studying the consequences of conditions 1 and 2 above.  
In what follows, we shall not distinguish between a generalized $\U(1)$ 
connection $W$ that is flat except at the punctures and a distributional 
connection $W_0$ assigning the same holonomies to paths and satisfying 
conditions 1 and 2.  Also, if condition 2 holds for some real-analytic 
coordinate system in a neighborhood of the puncture $p_i$ and some smooth 
trivialization of $Q$ in this neighborhood, we say that $W$ has a 
`singularity of standard form' at $p_i$ with respect to this coordinate 
system and trivialization.

Suppose that $W$ has a singularity of standard form at $p$ with respect
to some local coordinate system $(x,y)$ and some local trivialization of
$Q$.  Then we can ask if this is also true with respect to some other
local coordinate system and/or trivialization.  If we change the
trivialization, this adds a smooth closed 1-form to $W$, so $W$ still
has a singularity of standard form at $p$ with respect to the new
trivialization.    If we change the coordinate system, $W$ may not have
a singularity of standard form with respect to the new coordinates. 
However, it will if the new coordinates $(x',y')$ satisfy 
% \[
$$      dx'^2 + dy'^2 = c(dx^2 + dy^2) $$ 
%  \]
at the point $p$, for some scale factor $c$.  This can be seen by an
explicit calculation. It follows that having a singularity of standard
form at $p$, which  {\it a priori} depends on a choice of a local
coordinate system and a local trivialization of $Q$, in fact depends
only on a choice of a conformal structure at $p$, i.e., a metric modulo
scale factor on the tangent space at $p$.

Now, choose an open disc $U$ containing all the punctures, a
trivialization of $Q$ over $U$, and an analytic coordinate system
$(x,y)$ on $U$ for which each puncture $p_i$ has the coordinates $(i,0)$
and the fixed ray in the tangent space of $p_i$ is spanned by the
tangent vector $\partial_y$.  Suppose that $W$ is a generalized $\U(1)$
connection flat except at the punctures.  Then it has singularities  of
standard form at each puncture $p_i$ with respect to some conformal
structure at $p_i$.  However, we can always find a transformation in
$\D^\P$ that maps all of these conformal structures at the punctures to
the conformal structure corresponding to the metric $dx^2 + dy^2$.   To
see this, first note that one can map any conformal structure at $p_i$
to any other while leaving a specified tangent vector fixed. Thus there
exists a smooth vector field $v$ generating a flow $F_t$ on  $S$ that
fixes the punctures together with the tangent vectors $\partial_y$  at
the punctures, and such that $F_1$ maps the conformal structures at the
punctures to the conformal structure corresponding to $dx^2 + dy^2$.
Choose a real-analytic vector field $w$ that agrees with $v$ to first
order at the punctures, and let $G_t$ be the flow generated by $w$.  The
diffeomorphism $G_1$ will then have all the desired properties.

It follows that for any point $[W] \in \X^\P$, we can find a representative
$W \in \A^\P$ that has singularities of standard form at the points
$p_i$, all with respect to the chosen coordinates and trivialization of
$Q$ over $U$.  Concretely, this means that on $U - \P$ we have
\be      W = W_1 + \sum_{i = 1}^n c_i 
{(x - i)dy - y dx\over (x - i)^2 + y^2}  \label{prelim} \ee
where $W_1$ is a bounded smooth 1-form on $U - \P$.  Since $W$ is  
flat except at the punctures, $W_1$ must be closed.   Note that the
constants $c_i$ are not independent: they must sum to zero modulo
$2\pi$, because the holonomy of $W$ around a loop enclosing all the 
punctures must be trivial.

\begin{figure}
\centerline{\hbox{\epsfig{figure=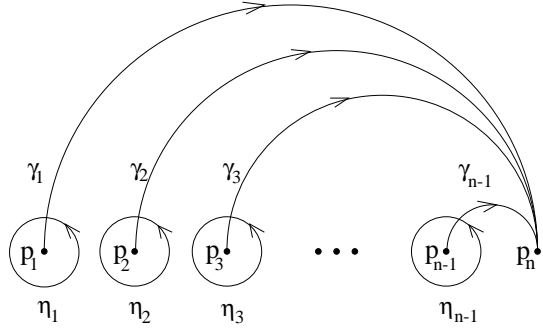,height=1.7in}}}
\caption{Choice of paths.}
\label{fig4}
\end{figure}
 
In fact, for any point in $\X^\P$ we can find a representative of a
very special form.  To describe this special form, we first choose
real-analytic paths $\gamma_1, \dots, \gamma_{n-1}$ and $\eta_1, \dots,
\eta_{n-1}$ as shown in Figure \ref{fig4}.  Note that $\gamma_i$ is a
semicircular arc in the upper half plane going from $p_i$ to $p_n$, while
$\eta_i$ is a circular loop going around $p_i$ once counterclockwise and
not containing any of the other punctures.   Next, for $1 \le i \le n-1$, 
we choose smooth 1-forms $X_i,Y_i$ on $S - \P$ with the following
properties:
\begin{alphalist}
\item $X_i$ satisfies
\[  \int_{\gamma_j} X_i = \delta_{ij}, \qquad
      \oint_{\eta_j} X_i = 0.   \]
\item $Y_i$ satisfies
\[  \oint_{\eta_j} Y_i = \delta_{ij}, \qquad
      \int_{\gamma_j} X_i = 0.   \]
\item $X_i$ and $Y_i$ satisfy
\[         \int_{S^2} X_i \we X_j = 0,  \qquad
           \int_{S^2} Y_i \we Y_j = 0,  \qquad
           \int_{S^2} X_i \we Y_j = \delta_{ij} . \]
\item For any real constants $x_i,y_i$ and any generalized $\U(1)$
connection $\tilde{W}$ that is flat except at the punctures and has
singularities of standard form at the punctures with respect to
the chosen coordinate system on $U$,
\be    
W = \tilde{W} + \sum_{i=1}^{n-1} (x_i X_i + y_i Y_i)  \label{special} 
\ee
defines a generalized $\U(1)$ connection that is flat except at the punctures
and has singularities of standard form.
\end{alphalist}
Note that property (d) implies the 1-forms $X_i,Y_i$ are closed.  

One can show that 1-forms with properties (a)-(d) exist using deRham
cohomology, but a more explicit construction is perhaps more illuminating.
We actually define $X_i$ and $Y_i$ on regions slightly larger than 
$S - \P$.  Define $X_i$ on all of $S$ by
\[     X_i = df_i  \]
where $f_i$ is a smooth real-valued function on $S$ with 
$f_i(p_j) = 1$ for $j \ne i$ and $f_i = 0$ in an open disc 
containing $p_i$ and the loop $\eta_i$.  To define $Y_i$, first set
\[   Y_i = {1\over 2\pi} {(x - i)dy - y dx\over (x - i)^2 + y^2} \]
in an open disc containing $p_i$ and the loop $\eta_i$, and
\[   Y_i = - {1\over 2\pi}{(x - n)dy - y dx\over (x - n)^2 + y^2}  \]
in an open disc containing $p_n$.  Then extend $Y_i$ smoothly to a
closed 1-form on all of $S - \{p_i,p_n\}$.  One can check that with 
these definitions, $X_i$ and $Y_i$ have properties (a)-(d).

One can show that by applying a suitable gauge transformation in $\G^\P$,
any connection of the form given by equation (\ref{prelim}) can be 
brought into the special form given by equation (\ref{special}).
Thus, given any point $[W] \in \X^\P$, we can choose a representative
$W \in \X^\P$ with the form given in equation (\ref{special}).  Define
\[         g_i = W(\gamma_i), \qquad h_i = W(\eta_i)  .\]
Using the chosen trivialization of $Q$ over the open disc $U$, we may
think of the holonomies $g_i$ and $h_i$ as elements of $\U(1)$.   Gauge
transformations in $\G^\P$ leave these holonomies unchanged, because
such gauge transformations equal the identity at the punctures.  With
more work, one can also show that all bundle automorphisms in $\G^\P
\semi \D^\P$ leave these holonomies unchanged.  (This is  the reason for
our somewhat complicated definition of $\D^\P$.)  It follows that these
holonomies depend only on the equivalence class $[W]$, so they define
functions
\[      g_i, h_i \maps \X^\P \to \U(1)  .\]
Taking all these functions together we obtain 
a map 
\[   \Phi \maps \X^\P \to  (\U(1) \times \U(1))^{n-1} \]
with
\[       \Phi([W]) =  (g_1,h_1,\dots,g_{n-1},h_{n-1}) .\]

To show that $\X^\P$ is a $2(n-1)$-dimensional torus, we just need
to show that $\Phi$ is a diffeomorphism.  It is clearly smooth.  
To see that $\Phi$ is onto, note that the generalized connection 
$W$ given in equation (\ref{special}) has
\[       g_i(W) = e^{ix_i} , \qquad h_i(W) = e^{iy_i}.  \]
To see that $\Phi$ is one-to-one, use deRham cohomology.    \qed

Since the map $\Phi$ constructed in the above proof is a 
diffeomorphism, we can use the functions $x_i$ and 
$y_i$ as `coordinates' on $\X^\P$.  Of course, these coordinates are only
defined modulo $2 \pi$.  More precisely, there is a covering
\ban       \R^{2(n-1)} &\to& \U(1)^{2(n-1)} \iso \X^\P   \\
           (x_i,y_i) &\mapsto& (e^{ix_i},e^{iy_i}), \ean
so we can identify $\X^\P$ with $\R^{2(n-1)}$ modulo the lattice
\[          \Lambda = (2\pi\Z)^{2(n-1)} .\]

In terms of these coordinates, it turns out that the physically
relevant symplectic structure on $\X^\P$ is given by
\be    
\omega = {k\over 2\pi} \sum_{i = 1}^{n-1} dx_i \wedge dy_i  
\label{surface.symplectic} 
\ee
To see this, note that the physically relevant symplectic 
structure on $\X^\P$ is given by 
\be  
\omega([\delta W],[\delta W']) = 
{k\over 2\pi} \int_{S^2} \delta W \we \delta W'  
\label{surface.symplectic.2}
\ee
where $[\delta W], [\delta W']$ are tangent vectors to $\X^\P$
corresponding to tangent vectors $\delta W,\delta W'$ at some point of
$\A^\P$.  This formula comes from the surface term in equation
(\ref{ss}), which was derived in the context of smooth connections. 
However, it makes sense even for generalized connections, as long as
$\delta W$ and $\delta W'$ are linear combinations of the 1-forms $X_i$
and $Y_i$, which we may assume without loss of generality.   
Using property (c) of the 1-forms $X_i$
and $Y_i$ (as described in the proof of the above theorem), we
may write
\[    \delta W = \sum_{i=1}^{n-1} (\delta x_i X_i + \delta y_i Y_i) ,
\qquad  \delta W' = \sum_{i=1}^{n-1} (\delta x'_i X_i + \delta y'_i Y_i). \]
This implies that
\[      \omega(\delta W,\delta W') = 
{k\over 2\pi} \int_{S^2} \Tr(\delta W \we \delta W') 
= {k\over 2\pi} \sum_{i=1}^{n-1} (\delta x_i \delta y'_i -
 \delta y_i \delta x'_i) , \]
proving equation (\ref{surface.symplectic}).

To summarize, in this section we gave a precise definition of
generalized connections which are flat everywhere except at the finite
set of points $\P$, proved that the resulting phase space $\X^\P$ is
diffeomorphic to a $2(n-1)$-dimensional torus, and showed that the
Chern-Simons surface term in the symplectic structure (\ref{ss}) of
the classical theory is well-defined on $\X^\P$ in spite of the fact
that the curvature of the connections now under consideration is
distributional.

Note that in the proof of Theorem 1, we needed to introduce some extra
structure on the surface $S$ to construct the map $\Phi$ that identifies
$\X^\P$ with the torus $(\U(1) \times \U(1))^{n-1}$.   In the next
section, our identification of $\X^\P$ with this torus plays a crucial
role in the geometric quantization.    By definition, the map $\Phi$
is invariant under the action of diffeomorphisms in $\D^\P$.  However,
$\Phi$ is not invariant under all diffeomorphisms of $S$ ---not even those 
that fix each puncture in $\P$--- because the extra structure needed
to define this map transforms nontrivially under such diffeomorphisms.     
Thus, in order to study the diffeomorphism constraint in Section 
\ref{s5.B.2}, it is important to know precisely what extra structure 
was needed to define the map $\Phi$.

First, of course, we needed to fix an ordering $p_1,\dots,p_n$ of the
punctures in $\P$.  Then we needed to fix a ray in the tangent space
of each puncture and a trivialization of $Q$ at each puncture.  Then
we needed to choose real-analytic paths $\gamma_i$ going from $p_i$ to
$p_n$, whose tangent vectors at both endpoints lie in the chosen rays.
We required that these paths look as in Figure \ref{fig4} with respect
to some analytic coordinate system.  However, we can continuously
deform the paths $\gamma_i$ without changing the holonomies along
these paths, as long as we keep their tangent vectors at at the
endpoints fixed throughout the deformation, and make sure their
interiors stay in $S-\P$ throughout the deformation.  We thus need
only a certain equivalence class of paths $\gamma_i$.  We shall use
the notation $\S$ to stand for a choice of the ordering, rays and
equivalence classes $\gamma_i$.  One can check that $\Phi$ is uniquely
defined given $\P$ and $\S$.
   
\subsubsection{Geometric quantization of $\X^\P$} 
\label{s4.C.2}

To geometrically quantize the phase space $\X^\P$, we first give it a
K\"ahler structure having the symplectic structure $\omega$ as its
imaginary part.  Then we find a holomorphic line bundle $L$ over $\X^\P$
equipped with a connection $\nabla$ whose curvature is $i\omega$.  This
`prequantum line bundle' only exists when the level $k$ is an integer.  
Then we construct the Hilbert space of holomorphic sections of $L$.  We
call this Hilbert space $\H_S^\P$, since states in this space describe
quantum geometries of the horizon $S$ which are flat except at the
punctures in the set $\P$.  Finally, we construct the surface Hilbert
space $\H_S$ by putting all these spaces $\H_S^\P$ together in a 
suitable way.  

To start, first note that we can introduce complex local coordinates
$z_i = x_i + iy_i$ on the space $\X^\P$.  In other words, we can
use the standard isomorphism $\R^{2(n-1)} \iso \C^{n-1}$ to make the
identification 
\[        \X^\P \iso \C^{n-1}/\Lambda , \]    
so that $\X^\P$ becomes a complex manifold.   In fact, it becomes a
K\"ahler manifold with a K\"ahler structure $g$ coming from ${k\over
2\pi}$ times the usual K\"ahler structure on $\C^{n-1}$.  This K\"ahler
structure is compatible with the symplectic structure given in equation
(\ref{surface.symplectic}), in the sense that ${\rm Im} g = \omega$.  

Next we construct a holomorphic complex line bundle $L$ over $\X^\P$
with a connection $\nabla$ whose curvature is $i\omega$.   Here we use
some ideas described in Mumford's book on theta functions
\cite{Mumford}. We first introduce $\nabla$  as a connection on the
trivial complex line bundle over $\C^{n-1}$, given  by:
\[   \nabla_{x_i} = \partial_{x_i}, \qquad 
     \nabla_{y_i} = \partial_{y_i} + {ik\over 2\pi} z_i .  \]
Note that
\[    [\nabla_{x_i},\nabla_{x_j}] = 0, \qquad
      [\nabla_{y_i},\nabla_{y_j}] = 0, \qquad
      [\nabla_{x_i},\nabla_{y_j}] = {ik\over 2\pi} \delta_{ij}, \]
so the curvature of this connection is ${ik\over 2\pi}$ times the standard
symplectic structure on $\C^{n-1}$.   Using this connection, we 
define parallel translation operators 
\[     U_i(t) = \exp(t \nabla_{x_i}), \qquad
       V_i(t) = \exp(t \nabla_{y_i})  \]
for all $t \in \R$.   More explicitly, we have:
\be 
(U_i(t)\psi)(z_1,\dots,z_{n-1}) =
\psi(z_1,\dots,z_i + t,\dots,z_{n-1}) ,
\label{U_i} 
\ee
\be (V_i(t)\psi)(z_1,\dots,z_{n-1}) = 
e^{{k\over 2\pi}(itz_i - {1\over 2}t^2)} \
\psi(z_1,\dots,z_i + it,\dots,z_{n-1}) .
\label{V_i} 
\ee
These parallel translation operators are clearly 1-parameter groups:
\[     U_i(s)U_i(t) = U_i(s+t), \qquad  V_i(s)V_i(t) = V_i(s+t), \]
and the above commutation relations imply
\[     U_i(s)U_j(t) = U_j(t)U_i(s), \qquad
       V_i(s)V_j(t) = V_j(t)V_i(s),  \quad
       U_i(s)V_j(t) = e^{{ik\over 2\pi}st} V_j(t)U_i(s). \]

The parallel translation operators give a representation of the
lattice $\Lambda$ on the space of holomorphic functions on $\C^{n-1}$. 
Given $w = u+iv$ with $u,v \in \R^{n-1}$, let 
\be  R(w) =  U_1(u_1) \cdots U_{n-1}(u_{n-1})
               V_1(v_1) \cdots V_{n-1}(v_{n-1})  .
\label{R}
\ee
Using the explicit formulas for the parallel translation operators it
is clear that $R(w)$ maps holomorphic functions to holomorphic 
functions.    We also have
\be  
        R(w+w') = e^{-{ik\over 2\pi}u' \cdot v} \, R(w)R(w')  
\label{R2}
\ee
when $w' = u'+iv'$ with $u',v' \in \R^{n-1}$.   Thus $R$ is a
projective representation of $\C^{n-1}$ on the space of holomorphic
functions.   If $w$ and $w'$ lie in the lattice $\Lambda$
then the above formula reduces to 
\[        R(w+w') = R(w)R(w') . \] 
Thus $R$ restricts to an honest representation of $\Lambda$.  

Now, the representation $R$ comes from an action of $\Lambda$ on the 
trivial line bundle over $\C^{n-1}$.   The quotient of this bundle by
this group action is a line bundle over $\C^{n-1}/\Lambda = \X^\P$,
which we denote by $L$.  The connection $\nabla$ gives rise to a
connection on $L$ which we also call $\nabla$.  The curvature of this
connection is $i \omega$, as desired.

Next let us describe the space of holomorphic sections of $L$.  A
holomorphic section of $L$ is the same as a holomorphic function on
$\C^{n-1}$ that is invariant under the operators $R(w)$ for all $w
\in \Lambda$.   Such functions are called `theta functions' 
\cite{Cartier,Mumford}.
We can obtain theta functions by the technique of group averaging.   To
make our job easier, suppose we start with a holomorphic function $f$ on
$\C^{n-1}$ that is invariant under $R(u)$ for real lattice vectors
$u$, that is, for $u \in (2\pi\Z)^{n-1}$.  Then we can try to average $f$
with respect to the imaginary lattice directions, forming the function
\[      \psi = \sum_{v \in (2\pi\Z)^{n-1}} R(iv) f .\]
If the sum converges uniformly on compact subsets of $\C^{n-1}$, it
defines a theta function.  

It is easy to find holomorphic functions that are invariant under
$R(u)$ for real lattice vectors $u$, since this condition simply
amounts to periodicity in the real lattice directions.  A basis of
such functions is given by
\[       f_a(z) = \exp(ia\cdot z)  \]
where $a \in \Z^{n-1}$.  If we apply the group averaging technique to 
such a function $f_a$, we obtain a theta function
\be  
\psi_a(z) = 
\sum_{v \in (2\pi\Z)^{n-1}} 
e^{{k\over 2\pi}(iv\cdot z - {1\over 2} v\cdot v)} e^{ia\cdot(z+iv)},
\label{psi_a}
\ee
since the sum indeed converges uniformly on compact subsets.  In fact
the functions $\psi_a$ form a basis of theta functions as we  let $a$
range over vectors with $a_i \in \{1,\dots,k\}$ for all $i$. The reason
for this restriction is that $\psi_a$ changes by only a scalar factor if
we add to $a$ a vector lying in $(k\Z)^{n-1}$.  For a full proof that
the functions $\psi_a$ form a basis of theta functions, see Mumford's
book \cite{Mumford}.  

Let $\H_S^\P$ denote the space of holomorphic sections of $L$, or
equivalently, the space of theta functions.   By the above remarks, the
dimension of this space is $k^{n-1}$.  We make this space into a Hilbert
space as follows: given theta functions $f$ and $g$, and defining $z =
x+iy$ for $x,y \in \R^{n-1}$, we define their inner product by
\be
\langle f,g\rangle = 
\int_{[0,2\pi]^{2(n-1)}} 
e^{-{k\over 2 \pi} y\cdot y}\; \overline{f}(z) g(z) 
\; d^{n-1}x \, d^{n-1}y . 
\label{inner}
\ee
This integral is easily seen to converge when $f,g$ lie in the
above basis of theta functions, so it converges for all theta
functions.

As we have just seen, the Hilbert space $\H_S^\P$ has a 
basis given by the states $\psi_a$ where $a = (a_1,\dots,a_{n-1})$ with
$a_i \in \{1,\dots,k\}$ for all $i$.  In fact, it is more convenient to
regard the $a_i$ as elements of $\Z_k$.  Also, to avoid treating the
point $p_n$ differently from the rest, we shall change our 
notation slightly and treat $a$ as an $n$-tuple $(a_1,\dots,a_n) \in 
\Z_k^n$, where we define $a_n$ by requiring that 
\[    a_1 + \cdots + a_n = 0. \] 
Taking advantage of this change of viewpoint, and adding a subscript
`$\P$' to denote the dependence on the set of punctures, it follows that
$\H_S^\P$ has a
basis $\psi_{\P,a}$ where $a$ ranges over all elements of $\Z_k^n$
satisfying the above equation.  Note that the dimension of
$\H_S^\P$ is $k^{n-1}$, just as one would predict using the
Bohr-Sommerfeld rule of `one quantum state per unit volume of phase
space', where phase space volume is measured in units of $h^{n-1}$.

We conclude by defining the `surface Hilbert space' $\H_S$, which
is built from all the Hilbert spaces $\H_S^\P$ as the set of punctures
$\P$ is allowed to vary.  We could do this using a direct limit
construction, but due to subtleties involving the extra structure
needed to quantize the phase spaces $\X^\P$, as described at the
end of Section \ref{s4.C.1}, it is more efficient to use the following
construction.  Let $\H_S^{\P,a}$ be the one-dimensional subspace of 
$\H_S^\P$ spanned by the state $\psi_{\P,a}$, and let
\be
\H_S = \bigoplus_{\P,a} \H_S^{\P,a}
\label{H_S}
\ee
where $\P,a$ ranges over all finite subsets of $S$ labelled with
{\it nonzero} elements $a_i \in \Z_k$ that sum to zero.  The reason
we demand that the $a_i$ be nonzero is as follows.  As we shall 
see in Section VA2, when $a_i = 0$ the vector $\psi_{\P,a}$ corresponds 
to a quantum state in which the $\U(1)$ connection is flat at the point 
$p_i$.   Such states already appear in the Hilbert space
$\H_S^{\P - \{p_i\}}$, so to avoid `double counting' these states, we 
exclude labellings $a$ where any $a_i$ equals zero.  (A direct limit 
construction would automatically avoid this `double counting'.)

It follows immediately that the surface Hilbert space has a basis of
states $\psi_{\P,a}$ corresponding to all ways of choosing finitely
many distinct points $p_1,\dots,p_n$ of $S$ and labelling these points
with nonzero numbers $a_1,\dots,a_n \in \Z_k$ satisfying
\[      a_1 + \cdots + a_n = 0 .\]
Thus, we have proved the assertions about the surface Hilbert space
$\H_S$ made in Section \ref{s3}.  In Section \ref{s5.A.2} we give a
geometrical interpretation of the basis states $\psi_{\P,a}$ and of
the above condition on the numbers $a_i$. These states will enable us
to impose the quantum boundary conditions (\ref{qbc3}) explicitly.

\section{Physical States}
\label{s5}

\subsection{Quantum Boundary Condition}
\label{s5.A}

Having constructed the volume and surface Hilbert spaces, we now wish
to impose the the quantum boundary condition, in order to pick out the
kinematical Hilbert space $\H$ as a subspace of $\H_V \otimes \H_S$.  
The naive version of this condition, equation (\ref{qbc1}), simply
states that kinematical states must be invariant under all $\U(1)$ 
gauge transformations on the horizon surface $S$.    This suggests
that we study the action of such gauge transformations on the volume 
and surface Hilbert spaces.  

In Section \ref{s5.A.1} we study the action of $\U(1)$ gauge
transformations on the volume Hilbert space.   In Section \ref{s5.A.2}
we carry out a similar study for the surface Hilbert space, but we
encounter an important subtlety: there is no operator $\hat F$
generating the action of $\U(1)$ gauge transformations on $\H_S$.
Instead, there is only a unitary operator corresponding to the
exponential $\exp(i\hat F)$.   Mathematically speaking, what this means
is that only gauge transformations taking values in a certain discrete
subgroup of $\U(1)$ act as unitary operators on $\H_S$.   This  subgroup
turns out to be $\Z_k \subset \U(1)$, the subgroup consisting of all
$k$th roots of unity, where $k$ is the level defined in equation
(\ref{level}).

For this reason, we need to impose the quantum boundary condition in 
its exponentiated form, equation (\ref{qbc2}), rather than the  naive
form given in equation (\ref{qbc1}).    In its exponentiated form, the
quantum boundary condition states that kinematical states must be
invariant under all $\Z_k$-valued gauge transformations on $S$. 
Starting from this fact, in Section (\ref{s5.A.3}) we explicitly
describe the kinematical Hilbert space $\H$ as a subspace of $H_V
\otimes \H_S$.

\subsubsection{Action of gauge transformations on $\H_V$} 
\label{s5.A.1}

First we describe operators on the volume Hilbert space that implement
$\U(1)$ gauge transformations at the horizon.   Most of our work is
already done, because in Section \ref{s4.B} we described how arbitrary
$\SU(2)$ gauge transformations at the horizon act on the volume Hilbert
space.  Here, however, we want a more explicit description of how the
subgroup consisting of $\U(1)$ gauge transformations acts on the volume
Hilbert space, so we can implement the quantum boundary condition.  

Recall now that $\G_S$ stands for the group of all (not necessarily
continuous) gauge transformations of the bundle $P|_S$. This has a
subgroup $\G_Q$ consisting of those gauge transformations that fix the
radial internal vector field $r \maps S \to \su(2)$. Alternatively, we
can think of $\G_Q$ as consisting of all gauge transformations of the
$\U(1)$ bundle $Q$, which abstractly is just the spin bundle of the
sphere $S$.  

Since $\G_S$ has a unitary representation on $\H_V$, so does the
subgroup $\G_Q$.   In particular, if we pick any point $p \in S$, we
obtain a unitary representation of $\U(1)$ on $\H_V$ by considering
gauge transformations in $\G_Q$ that are the identity at every point
except $p$.  In the notation of Section \ref{s4.B}, the self-adjoint
infinitesimal generator of $\U(1)$ action is the operator 
\[    J(p) \cdot r = J^i(p) r_i  .\]

Let $\P = \{p_1,\dots,p_n\}$ be any finite subset of $S$, and let $m$
stand for a way of labelling each point $p_i$ with a nonzero
half-integer $m_i$. Then there is a subspace $\H_V^{\P,m}$ of the
volume Hilbert space consisting of all vectors $\psi$ such that
\[      (J(p_i) \cdot r) \psi = m_i \psi \]
for each point $p_i$ and 
\[      (J(p)\cdot r) \psi = 0 \]
for all points $p$ other than the points $p_i$.  Clearly $\H_V^{\P,m}$
is a subspace of $\H_V^\P$, since gauge transformations at any point 
$p \in S$ act trivially on states corresponding to spin networks that
do not have $p$ as one of their ends.  Moreover, we have a direct
sum decomposition
\be
\H_V = \bigoplus_{\P,m} \H_V^{\P,m}  
\label{H_V2}
\ee
where $\P,m$ ranges over all finite sets of points in $S$ labelled by
nonzero half-integers.  This is analogous to the decomposition of the
surface Hilbert space into subspaces $\H_S^{\P,a}$ given in equation
(\ref{H_S}).  The reader may also wish to compare this direct sum
decomposition of the volume Hilbert space with the previous
decomposition, given in equation (\ref{H_V}), based on eigenspaces of 
the operator $J(p) \cdot J(p)$ which features in the expression of
quantum area. 

The present decomposition of the volume Hilbert space diagonalizes the
action of $\G_Q$, in the following sense.  Suppose $g \in \G_Q$ and
$\psi \in \H_V^{\P,m}$.  Then for each point $p_i \in \P$ we can think
of $g(p_i)$ as an element of $\U(1)$, and
\be g \psi = g(p_1)^{2 m_1}
\cdots g(p_n)^{2 m_n} \psi .
\label{V-action} 
\ee
The factors of 2 here come from the usual convention of using a
half-integer $m$ to label the eigenvalues of a specific component of
angular momentum.

\subsubsection{Action of gauge transformations on $\H_S$} 
\label{s5.A.2}

Next we describe operators on the surface Hilbert space that implement
gauge transformations at the punctures.  Classically the group of
allowed gauge transformations at each puncture is $\U(1)$.  However, not
all these gauge transformations can be implemented as operators on the
surface Hilbert space, but only those lying in the discrete subgroup
$\Z_k \subset \U(1)$, where $k$ is the level defined in equation
(\ref{level}).   Thus in a sense the group of gauge transformations must
itself be `quantized' as part of the quantization procedure.   

In fact, the `quantization of the gauge group' we see here is a simple
example of the relation between Chern-Simons theory and quantum groups
\cite{Atiyah}.  Recall that the surface Hilbert space is obtained by
quantizing the phase space for $\U(1)$ Chern-Simons theory on a sphere
with a fixed set of punctures, and then allowing the set of punctures 
to vary.  Since Chern-Simons theory can also be
described using quantum groups, one expects an alternative description
of the surface Hilbert space in terms of representations of `quantum
$\U(1)$'.  This quantum group is essentially just the group $\Z_k$.
More precisely, it is the usual Hopf algebra associated to $\Z_k$, but
equipped with a modified quasi-triangular structure \cite{Majid}.

However, we do not need the machinery of quantum groups in what
follows, since we can instead use well-known facts about about theta
functions.  As we saw in Section \ref{s4.C.2}, when there are $n$
punctures the classical phase space $\X^\P$ is a $2(n-1)$-dimensional
torus, and quantum states in $\H_S^\P$ are holomorphic sections of a
line bundle $L$ over this torus.  For each puncture we get an action
of $\U(1)$ as symplectic transformations of the torus. However, not
all elements of $\U(1)$ lift to holomorphic transformations of the
line bundle, but only those lying in the subgroup $\Z_k$, so only
these act as unitary operators on $\H_S^\P$.  This is a standard but
nontrivial result in the theory of theta functions \cite{Mumford2}, so
we shall not prove it here.  However, we will give an explicit
construction of the lift when it exists. It should explain why the
subgroup $\Z_k$ plays a special role.

As noted in Section \ref{s3}, this `quantization of the gauge group'
turns out to have very striking implications for the quantum geometry
of the horizon.  It means that the holonomy of the $\U(1)$ connection
around each puncture $p_i$ can only take on a discrete spectrum of
values: namely, those of the form $\exp(2 \pi i a_i/k)$ where $a_i \in
\Z_k$.  Heuristically, this means the horizon is flat except at the
punctures, where there are conical singularities with quantized angle
deficits.

To begin, let $\G_Q$ be the group of all gauge transformations of the
bundle $Q$. This group acts on $\A^\P$, and this action descends to an
action on the phase space $\X^\P$, where we have quotiented by the
action of the subgroup $\G^\P \semi \D^\P$.  However, the action of any
gauge transformation $g$ on $\X^\P$ depends only on its values $g(p_i)$,
so we obtain an action of $\U(1)^n$ on $\X^\P$.   Another way to see
this is to note that
 \[       \G_Q /\G^\P = \U(1)^n    .\]

The action of a gauge transformation at the point $p_n$ 
can be re-expressed in terms of gauge transformations at the other
points $p_i$, because constant gauge transformations act trivially
on $\X^\P$.  Thus we begin by only considering gauge transformations
that are the identity except at the first $n-1$ points.  We denote
the action of these gauge transformations on the phase space as
\[          T \maps \U(1)^{n-1} \times \X^\P \to \X^\P .\]
In terms of the holonomies $g_i,h_i$, we have the following explicit 
formula for $T$:
\[  T(e^{i \theta_1},\dots,e^{i \theta_{n-1}})(g_1,h_1,\dots,g_{n-1},h_{n-1})
=   (e^{-i\theta_1} g_1, h_1, \dots, e^{-i\theta_{n-1}} g_{n-1}, h_{n-1}) \]
In other words, $T$ is just the action of $\U(1)^{n-1}$ by translations
on $\X^\P \iso (\U(1) \times U(1))^{n-1}$.   

Given an element $g$ of $\U(1)^{n-1}$, we can now ask if $T(g)$ lifts
to a holomorphic transformation of the line bundle $L$. As mentioned
above, this is possible if and only if $g$ lies in the subgroup
$\Z_k^{n-1}$.  In this case, there is a natural way to choose a lift
that defines a unitary operator on the Hilbert space $\H_S^\P$.  To
see this, suppose that $g \in \U(1)^{n-1}$ and choose an element $w
\in \R^{n-1}$ that is sent to $g$ by the covering map
\[            \R^{n-1} \to (\R/2\pi\Z)^{n-1} = \U(1)^{n-1} .\]
Associated to $w$ there is an operator $R(w)$ on the space of
holomorphic functions on $\C^{n-1}$, given by equation (\ref{R}). This
operator preserves the space of theta functions if it commutes with
$R(w')$ for all $w' \in \Lambda$.   Using equation (\ref{R2}), one can
check that this holds precisely when $w$ lies in the lattice
$({2\pi\over k} \Z)^{n-1}$, or equivalently, when $g$ lies in
$\Z_k^{n-1}$.   Because $R(w)$ is defined in terms of parallel
translation with respect to the connection $\nabla$ and it preserves the
space of theta functions, it really does come from a lift of $T(g)$ to a
holomorphic map on bundle $L$, as desired.

As a unitary operator on the Hilbert space $\H_S^\P$, the operator
$R(w)$ is independent of the choice of $w$ mapping to $g \in
\Z_k^{n-1}$.  We shall thus write this operator simply as $R(g)$.
Using equations (\ref{R}) and (\ref{inner}), one can check that $R$ is
a unitary representation of $\Z_k^{n-1}$ on $\H_S^\P$.  Moreover, at
this point it is easy to go back and consider the action of gauge
transformations at the point $p_n$ as well as the points
$p_1,\dots,p_{n-1}$.  These give an action of $\U(1)^n$ on $\X^\P$,
which we again call $T$.  This is given explicitly as follows:
\be
\label{T} 
T(e^{i \theta_1},\dots,e^{i \theta_n}) (g_1,h_1,\dots,g_{n-1},h_{n-1})
=   (e^{i(\theta_n - \theta_1)} g_1, h_1, \dots, 
e^{i(\theta_n - \theta_{n-1})} g_{n-1}, h_{n-1}). 
\ee
By our previous results, $T(g)$ only lifts to a holomorphic map of the
line bundle $L$ when $g \in \Z_k^n$.  This gives us a unitary
representation of $\Z_k^n$ on $\H_S^\P$, which we again call $R$.  We
can think of $R$ as describing the action of the `quantized gauge
group' $\Z_k^n$ on geometries of the horizon for which the $\U(1)$
connection is flat except at the points $p_i$.

To describe $R$ more explicitly, we can use the basis $\psi_a$ of
$\H_S^\P$ described in Section \ref{s4.C.2}.  Recall that here $a =
(a_1,\dots,a_n)$ is any $n$-tuple of integers modulo $k$ that sum to
zero.  Suppose $(g_1,\dots,g_n) \in \Z_k^n$.  Then using the formulas
for the states $\psi_a$ and the representation $R$, one can show that
\be R(g_1,\dots,g_n) \psi_a = g_1^{a_1} \cdots g_n^{a_n} \psi_a
\label{R3} 
\ee
where we think of each $g_i$ as an element of $\U(1)$ using the
inclusion $\Z_k \subset \U(1)$.  Note that this formula makes sense
even though the integers $a_i$ are only defined modulo $k$, thanks to
the fact that the phases $g_i$ are $k$th roots of unity.

How can we understand the meaning of equation (\ref{R3})?  Naively, we
would expect the quantity ${2\pi\over k}F$ to generate $\U(1)$ gauge
transformations on the horizon, where $F$ is the curvature of the
$\U(1)$ connection $W$, thanks to the surface term in the symplectic
structure in equation \ref{surface.symplectic.2}.  This would suggest
that the curvature at a puncture is the infinitesimal generator of gauge
transformations at that puncture.  But the curvature at a puncture is
not a well-defined function on $\X^\P$, since this phase space is
defined in terms of generalized connections.  Instead, as a kind of
substitute, we have a function $h_i$ on $\X^\P$ measuring the holonomy
around the loop $\eta_i$ enclosing the $i$th puncture, as shown in
Figure \ref{fig4}.  Since this holonomy is essentially the exponential
of the curvature concentrated at the $i$th puncture, we expect the
corresponding operator in the quantized theory to be the {\it
exponential} of the generator of gauge transformations at $p_i$ ---
i.e., a unitary operator corresponding to a finite gauge transformation
at $p_i$.  The obvious candidate for this operator is
\[     \hat h_i = R(1, \cdots, 1,  e^{\frac{2\pi i}{k}}, 1, \dots, 1) ,\]
all the entries being $1$ except the $i$th. 

By equation (\ref{R3}), the states $\psi_a$ satisfy
\[     \hat h_i \psi_a = e^{\frac{2 \pi i a_i}{k}} \psi_a  .\]
The above heuristic argument thus suggests that $\psi_a$ represents a
quantum geometry for the horizon in which the connection $W$ is flat
except at the punctures and has holonomy $\exp(2\pi i a_i/k)$ around 
the $i$th puncture.   The condition that $a_1 + \cdots + a_n = 0$ then 
says that the product of all these holonomies is 1.  This is exactly
what one would expect, because the product of all these holonomies
equals the holonomy around a loop winding once around all the punctures,
and this sort of loop is contractible in $S - \P$.  

In the state $\psi_a$ we can visualize the horizon as having conical
singularities with specified angle deficits at the punctures, as shown
in Figure \ref{fig3}.  Of course, $W$ is a connection on the spin
bundle of the horizon; in terms of the $\SO(2)$ `Levi-Civita'
connection on the tangent bundle, the angle deficit at the $i$th
puncture is $4\pi a_i/k$.  Since $a_i$ is only defined modulo $k$,
this angle deficit is only defined modulo $4 \pi$.  The condition $a_1
+ \cdots + a_n = 0$ then says that the sum of these angle deficits
vanishes modulo $4 \pi$.  This is the quantum analogue of the
classical fact that for any metric on the 2-sphere, the integral of
the scalar curvature is $4 \pi$.

Our last order of business is to describe the action of gauge
transformations on the surface Hilbert space $\H_S$.  This is
straightforward given their action on the Hilbert spaces $\H_S^\P$.
To make precise the fact that only $\Z_k$-valued gauge transformations 
can be implemented in the quantum theory, we define $\hat \G_Q$, the group of
`quantized gauge transformations', to consist of all functions $g \maps
S \to \Z_k$.  We can think of this as a subgroup of the group $\G_Q$. 
The quantized gauge group has a unitary representation on $\H_S$,
defined as follows: for any group element $g \in \hat \G_Q$ and vector
$\psi \in \H_S^\P$ we have
\[  g \psi = R(g(p_1),\dots,g(p_n)) \psi \]
This implies that 
\[  g \psi_{\P,a} =  g(p_1)^{a_1} \cdots g(p_n)^{a_n} \psi_{\P,a}   \]
where $\psi_{\P,a}$ is the basis of the surface Hilbert space
described at the end of Section \ref{s4.C.2}.
Since we defined $\H_S^{\P,a}$ to be the subspace of $\H_S$ spanned 
by the state $\psi_{\P,a}$, we may also say this as follows:
\be
g \psi =  g(p_1)^{a_1} \cdots g(p_n)^{a_n} \psi
\label{S-action}
\ee
for any $g \in \hat \G_Q$ and $\psi \in \H_S^{\P,a}$.  This formula
is analogous to formula (\ref{V-action}), which describes the action
of gauge transformations on the volume Hilbert space.

\subsubsection{Implementing the quantum boundary condition} 
\label{s5.A.3}

In Section \ref{s5.A.1} we described how the group $\G_Q$ of $\U(1)$
gauge transformations on the horizon acts on the volume Hilbert space,
while in Section \ref{s5.A.2}, we described how the subgroup $\hat \G_Q$
consisting of $\Z_k$-valued gauge transformations acts on the surface
Hilbert space.  It is now a simple matter to implement the quantum
boundary condition, by picking out the subspace of $\H_V \otimes \H_S$
that is invariant under the action of $\hat \G_Q$.

Recall from equation (\ref{H_V2}) that the volume Hilbert space is a
direct sum of subspaces $\H_V^{\P,m}$, where $\P = \{p_1,\dots,p_n\}$
ranges over all finite sets of points in $S$ and $m$ ranges over all
ways of labelling these points by nonzero half-integers. By equation
(\ref{V-action}), any vector $\psi$ in $\H_V^{\P,m}$ transforms as
follows under the action of $g \in \G_Q$:
\[  g \psi = g(p_1)^{2 m_1} \cdots g(p_n)^{2 m_n}\,\, \psi   \]
Similarly, in equation (\ref{H_S}) we saw that the surface Hilbert
space is a direct sum of subspaces $\H_S^{\P,a}$, where $\P$ ranges
over all finite sets of points in $S$ and $a$ ranges over all ways
of labelling these points by nonzero elements of $\Z_k$.  By equation
(\ref{S-action}), any vector $\psi$ in $\H_S^{\P,a}$ transforms as
follows under the action of $g \in \hat \G_Q$:
\[  g \psi = g(p_1)^{a_1} \cdots g(p_n)^{a_n} \psi  . \]
It follows that the subspace of $\H_V \otimes \H_S$ consisting of
vectors invariant under $\hat \G_Q$ is
\be \H^{\rm Kin} = \bigoplus_{\P,m,a\, \colon\; 2m = -a\, \mod \, k}
\H_V^{\P,m} \otimes \H_S^{\P,a} 
\label{Hkin} \ee
where $\P$ ranges over all finite subsets of $S$, $m$ ranges over all 
ways of labelling the points in $\P$ by nonzero half-integers, and
$a$ ranges over all ways of labelling these points by nonzero elements
of $\Z_k$ that sum to zero.  We call this subspace the `kinematical'
Hilbert space.  

\subsection{Quantum Einstein Equation}
\label{s5.B}

As explained in Section \ref{s3}, some of the classical Einstein
equations contain no time derivatives and thus act as constraints on the
initial data: the Gauss constraint, the diffeomorphism constraint, and
the Hamiltonian constraint.  In quantum theory, these constraints are to
be incorporated, a la Dirac, as conditions on permissible physical
states.   In the following sections we deal with each of these three
constraints in turn.

\subsubsection{Gauss constraint}
\label{s5.B.1}

The Gauss constraint has in fact been imposed in our definition of the
Hilbert space $\H^{\rm kin}$ in equation (\ref{Hkin}), but let us
explain exactly why this is the case.  At the classical level, the
Gauss constraint generates $\SU(2)$ gauge transformations on the pairs
$(A,\E)$ which reduce to $\U(1)$ gauge transformations preserving the
internal vector field $r$ at the horizon. The group of such gauge
transformations has $\G_V$ as a normal subgroup, and the quotient
group is $\G_Q$.  In the quantum theory, we used this decomposition to
impose the Gauss constraint in two separate stages.  First we
restricted attention to states that are invariant under $\G_V$.  Then
we sought an action of $\G_Q$ on the space of such states and tried to
restrict attention to states that are invariant under $\G_Q$.

The first stage of this plan was carried out in a straightforward way. 
The group $\G_V$ acts trivially on generalized connections on the
surface, but nontrivially on the generalized connections in the volume. 
Thus we only needed to impose invariance under $\G_V$ for the volume
degrees of freedom.  In Section \ref{s4.B} we did this by defining 
the volume Hilbert space $\H_V$ as the subspace of $L^2(\Ab_V)$
consisting of vectors that are invariant under $\G_V$.

The second stage of the plan required more care, and in carrying it out
we encountered an important subtlety.  The volume Hilbert space $\H_V$ is
a direct sum of subspaces $\H_V^{\P,m}$ where $\P,m$ ranges over all finite
subsets of $S$ labelled by nonzero half-integers.  These subspaces have 
the handy property that the action of $g \in \G_Q$ on $\H_V^{\P,m}$ 
is trivial if $g$ is the identity at all
points in $\P$.  Such gauge transformations form a normal subgroup
$\G^\P \subset \G_Q$.  Taking this as our cue, we constructed the
surface Hilbert space $\H_S$ as a similar direct sum of subspaces
$\H_S^{\P,a}$, where in defining $\H_S^\P$ we imposed invariance under all
gauge transformations in $\G^\P$.  Thus, by construction, states in
$\H_V^\P \otimes \H_S^\P$ are automatically invariant under the action 
of the subgroup $\G^\P$.  

To impose invariance under {\it all} of $\G_Q$, we would therefore 
simply need to describe the action of the quotient group $\G_Q/\G^\P =
\U(1)^n$ on $\H_V^\P \otimes \H_S^\P$ and find the subspace of states
invariant under this action.  The action of $\U(1)^n$ on $\H_V^\P$ is
unproblematic, and it is diagonalized by the subspaces $\H_V^{\P,m}$.   
However, it turned out that $\U(1)^n$ does {\it not} act on $\H_S^\P$; 
instead, thanks to the subtleties of quantum Chern-Simons theory, only 
the subgroup $\Z_k^n$ acts on this space.  This action is diagonalized
by the subspaces $\H_S^{\P,a}$.   The best we can do, therefore, is to 
find the subspace of $\H_V^\P \otimes \H_S^\P$ consisting of vectors
invariant under $\Z_k^n$.  Using equations (\ref{V-action}) and
(\ref{S-action}), this subspace turns out to be 
\[  
\bigoplus_{m,a\,\colon\; 2m = -a\, \mod \, k}  
\H_V^{\P,m} \otimes \H_S^{\P,a} 
\]
where $m$ ranges over all ways of labelling points in $\P$ by
half-integers and $a$ ranges over all ways of labelling these points by
elements of $\Z_k$ that sum to zero.

Putting all these subspaces together, we obtain the kinematical 
Hilbert space 
\[ \H^{\rm Kin} = \bigoplus_{\P,m,a\, \colon\; 2m = -a\, \mod \, k}
\H_V^{\P,m} \otimes \H_S^{\P,a}
\]
where $\P$ ranges over all finite subsets of $S$, $m$ ranges over
ways of labelling the points in $\P$ by nonzero half-integers, and
$a$ ranges over ways of labelling them by nonzero elements of $\Z_k$
that sum to zero.  This Hilbert space consists of states that
are invariant under all of $\G_V$ but only the subgroup $\hat \G_Q
\subset \G_Q$ consisting of $\Z_k$-valued gauge transformations on the
horizon.  Thus, while the gauge group in the classical theory is $\G_V
\times \G_Q$, that in the quantum theory is reduced to $\G_V \times
\hat\G_Q$.  Explicitly, the space of states satisfying the Gauss
constraint ---i.e., invariant under under $\G_V \times \hat \G_Q$--- 
is the space $\H^{\rm Kin}$.   

\subsubsection{Diffeomorphism constraint}
\label{s5.B.2}

Denote by $\D$ the group of diffeomorphisms generated by analytic
vector fields on $M$ which are tangential to $S$ and tend to zero at
infinity.  On the classical phase space, the diffeomorphism constraint
generates canonical transformations corresponding to $\D$.  Hence
states mapped to each other by $\D$ are considered physically
equivalent.  In the quantum theory, our task is to construct states
that are invariant under the action of $\D$.  To display the surface
states `explicitly', will use a gauge fixing procedure in which one
chooses exactly one surface state in each orbit of $\D$.  For the
entropy calculation, such an explicit representation of volume states
is not essential.  It is easier to follow the procedure of Section
\ref{s5.B.1} and impose the constraint as an operator equation.  
 
Recall from Section \ref{s4.C.1} that in the construction of Hilbert
spaces $\H_S^{\P ,a}$ and definition of associated operators which
implement $\hat\G_Q$ and measure deficit angles, we had to introduce
an additional structure $\S$ on the horizon surface $S$.  Thus,
strictly, the surface Hilbert spaces should have been labelled
$\H_S^{\S(\P), a}$.  Prior to the present discussion of the
diffeomorphism constraint, we omitted the explicit reference to the
structure $\S$ for the sake of simplicity.  But we can no longer do so
in the present discussion, since this structure is invariant only
under the subgroup $\D^\P$ of $\D$, and we now wish to investigate the
action of the full group $\D$.

Thus, we now begin by considering the entire collection of Hilbert
spaces $\H_S^{\S(\P), a}$, one for each choice of $\S(\P)$. If a
diffeomorphism $\phi \in \D$ maps $\S(\P)$ to $\S'(\P')$, then it
provides a natural unitary mapping
$$  U_\phi:  \H_S^{\S (\P), a} \rightarrow  \H_S^{\S'(\P'), a} $$
between the corresponding Hilbert spaces. The diffeomorphism constraint
implies that these Hilbert spaces are physically equivalent. The
`gauge fixing procedure' we are adopting for surface states says that
we should select only one copy of these physically equivalent Hilbert
spaces.

It is therefore important to examine the action of $\D$ on the space
of structures $\S(\P)$ with a given number $n$ of punctures. The
action is in fact transitive. The proof of this assertion is a bit
technical, but the main steps can be summarized follows.  First one
shows that the analogous result is true in the smooth category.  For
this, the main work involved is to show that smooth diffeomorphisms
fixing each puncture $p_i$ together with a ray in its tangent space
act transitively on the equivalence classes of paths $\gamma_i$ that
look as in Figure 4 with respect to some smooth coordinate system on
an open disc containing all the punctures.  Restricting to a slightly
smaller closed disc, this follows from the fact that smooth
diffeomorphisms act transitively on the set of embeddings of a closed
disc in the 2-sphere \cite{Hirsch}.  Then, having found a smooth
diffeomorphism $f \maps S \to S$ which carries a given structure $\S$
to a given structure $\S'$, one can use approximation methods to find
a real-analytic diffeomorphism $g$ that also does this.  In
particular, by a version of the Weierstrass theorem, for any smooth
diffeomorphism $f \maps S \to S$ one can find a polynomial $p \maps
\R^3 \to \R^3$ which, when restricted to $S$, is arbitrarily close to
$f$ in the $C^1$ topology and has the same values and first
derivatives as $f$ at a chosen finite set of points.  Composing such a
polynomial with radial projection to $S$ gives the desired
real-analytic diffeomorphism $g \maps S \to S$.

Combining these results it follows that, for a given number $n$, all
Hilbert spaces under consideration lie in the orbit of the induced
action of $\D$; they are all physically equivalent.  The quantum
diffeomorphism constraint requires that we consider only one point in
this orbit.  Therefore, for each natural number $n$, we fix a specific
structure $\S(\P)$ on the horizon and construct the corresponding
surface Hilbert space $\H_S^{\S(\P),a}$, which we denote as $\H_S^{n,
a}$.  Each state in this Hilbert space serves as the representative of
all states (belonging to various Hilbert spaces associated with $n$
punctures) in its diffeomorphism equivalence class.

Next, let us consider the volume Hilbert spaces $\H_V^{\P,m}$. In view
of the above `gauge fixing' of surface diffeomorphisms, for any given
integer $n$, we will restrict the set of punctures $\P$ to that used
in the construction of $\H_S^{n,a}$.  Then, it is sufficient to to
consider only the subgroup $\D_S$ of $\D$ consisting of diffeomorphisms
in $\D$ which are identity on $S$.  Our task is construct volume states
which are invariant under the action of this group. For this, one can
essentially repeat the `group averaging procedure' developed and used
in \cite{ALMMT,BS,LT} for the case when $M$ has no internal
boundaries.  Each of these states corresponds to the equivalence class
of spin network states in $\H_V^{\P,m}$ related to one another by a
diffeomorphism in $\D_S$.  Denote by $\H_V^{n,m}$ the resulting volume
Hilbert space.

We can now display the total Hilbert space $\H^{\rm Diff}$ of states
which satisfy the diffeomorphism constraint:
\be 
\H^{\rm Diff} = \bigoplus_{n,m,a\, \colon\; 2m = -a\, \mod \, k} 
\H_V^{n,m} \otimes \H_S^{n,a}    \label{Hdiff} \ee
where the number of punctures $n$ ranges over non-negative integers,
and $m$ and $a$ are as in (\ref{Hkin}).

\subsubsection{Hamiltonian constraint}
\label{s5.B.3}

In this subsection, we will impose the remaining, Hamiltonian
constraint. In the literature, there exist some concrete proposals for
defining this operator on the space of solutions to the diffeomorphism
and Gauss constraints and for finding its kernel, i.e., the physical
states (see, in particular, \cite{T2,MoreHam}.) However, as pointed
out at the end of Section \ref{s2.B}, these proposals are still being
examined in detail. Fortunately, our analysis of the geometry of
quantum horizon and entropy calculation are largely insensitive to
this freedom.

Let us begin by recalling the situation in the classical theory. To
obtain a function on the phase space, the Hamiltonian constraint has
to be smeared by a `lapse field' $N$ on the 3-manifold $M$. The
resulting function is differentiable on the phase space, i.e., defines
a Hamiltonian vector field, \textit{only} if the $N$ tends to zero
both at infinity \textit{and} the horizon. Motions along the resulting
Hamiltonian vector field are `bubble time-evolutions'; in the
space-time picture, the 3-manifold is kept fixed both at infinity and
at the horizon. These motions are to be treated as `gauge' and the
phase space of \textit{physical} states is obtained by quotienting the
full phase space by the orbits of these Hamiltonian vector fields, or,
by picking a cross-section in the phase space which intersects each
orbit once and only once.  (True dynamics corresponds to time evolution
in which the lapse $N$ is not constrained to vanish on the horizon and
infinity and is generated by a genuine Hamiltonian function on the
physical phase space.) Finally, consider the function $A_S$ on the
phase space which measures the horizon area. Because the lapse $N$
smearing the Hamiltonian constraint goes to zero at the horizon, $A_S$
is constant on the gauge orbits and therefore defines an observable on
the physical phase space.

Let us now turn to quantum theory. To impose the Hamiltonian
constraint, we must smear the appropriately constructed constraint
operator by a lapse which goes to zero on $S$. Therefore, the
constraint leaves the surface states untouched and has non-trivial
action only on volume states. Consider the subspace $\H_V^{n,m}$ of
volume states in (\ref{Hdiff}). Imposition of the Hamiltonian
constraint picks out a subspace $\tilde{\H}_V^{n,m}$ of $\H_V^{n,m}$,
which again carries the labels $n,m$ because they refer to punctures
on $S$ and the lapse vanishes on $S$. Thus, the total physical Hilbert
space is now given by:
\be \label{Hphys}       
\H^{\rm Phys} = \bigoplus_{n,m,a\, \colon\; 2m = -a\, \mod \, k} 
\tilde{\H}_V^{n,m} \otimes \H_S^{n,a}   \ee
where $n$ ranges over natural numbers, $m$ ranges over $n$-tuples of 
half-integers, and $a$ ranges over $n$-tuples of nonzero elements of 
$\Z_k$ that sum to zero.  

As in the classical theory, the area observable $\hat{A}_S$ is
well-defined on physical states.  To show this let us proceed step by
step, starting with the kinematical Hilbert space $\H^{\rm Kin}$.
Recall from Section \ref{s4.B} that each spin network state in
$\H^{\rm Kin}$ is an eigenstate of the area operator $\hat{A}_S$ and
the eigenvalue depends only on the spin labels at the
punctures. Therefore, the diffeomorphism-invariant state that results
from group averaging of any spin network state is again an eigenstate
with the same property. Thus $\hat{A}_S$ has a well-defined action on
$\H^{\rm Diff}$.  Finally, the action descends to $\H^{\rm Phys}$
because each $\tilde{\H}_V^{n,m}$ is a subspace of $\H_V^{n,m}$.

The following description of the physical Hilbert space will be useful
in the entropy calculation of the next section.  Define the `physical 
volume Hilbert space' as follows:
\be \label{HphysV}
\H^{\rm Phys}_V = \bigoplus_{n,m}
\tilde{\H}_V^{n,m}   
\ee
where $n$ ranges over natural numbers and $m$ ranges over $n$-tuples
of nonzero half-integers.  Similarly, define the `physical surface 
Hilbert space' by:
\be \label{HphysS}
\H^{\rm Phys}_S = \bigoplus_{n,a}
\H_S^{n,a}   \ee
where $n$ ranges over natural numbers and $a$ ranges over $n$-tuples
of nonzero elements of $\Z_k$ that sum to zero.  By equation (\ref{Hphys})
we have
\be  \label{tensor}
   \H^{\rm Phys} \subset \H^{\rm Phys}_V \otimes \H^{\rm Phys}_S.  
\ee
This decomposition allows us to isolate the volume and surface degrees
of freedom of the black hole system.   In the next section we will
use this decomposition to compute the entropy of a non-rotating black
hole whose event horizon has its area lying in a given range.  

\section{Entropy}
\label{s6}

We now have all the machinery needed for the computation of entropy of
isolated horizons.  Recall first that our fundamental macroscopic
parameters are defined intrinsically at the horizon, without reference
to infinity: they are area, angular momentum and charges associated
with matter fields.  (In this framework, mass is a secondary quantity,
expressed as a specific function of the fundamental ones.)  In the
non-rotating case now under consideration, it is then natural to begin
with a microcanonical ensemble consisting of physical quantum states
which endow the horizon $S$ with an area lying in a small interval
containing a fixed value $a_0$ and charges lying in a small interval
containing fixed values $Q_0$ and count the independent
\textit{surface states} in the ensemble.  We will carry out this task
in three steps.  The first will focus on pure general relativity where
the only relevant horizon parameter is the area $a_0$.  In Section
\ref{s6.A} we outline the strategy and in Section \ref{s6.B} we carry
out the desired counting in detail.  Matter fields and charges are
introduced in the third step.  This step is carried out in Section
\ref{s6.C}, which also summarizes the current viewpoint towards the
final result.  Further remarks on the physical meaning of this entropy
and the relation of our calculation to those carried out in other
approaches can be found in Section \ref{s7}.

\subsection{Strategy}
\label{s6.A}

Consider isolated horizons whose area $a$ lies in the range $a_0 -
\delta \le a \le a_0 + \delta$.  Since we are only interested in the
entropy of the horizon itself, not the surrounding spacetime, we start
by considering all physical states for which the horizon area lies in
this range, and then trace out over the volume states to obtain a
density matrix $\rho_\bh$ on $\H^\phys_S$ describing a maximal-entropy
mixture of surfaces states for which the horizon area lies in this
range.  The statistical mechanical entropy of this mixture will then
be given by
\[ S_\bh = -\Tr(\rho_\bh \ln \rho_\bh) .\]
In more detail, first recall from Section \ref{s5.B.3} that we can
define the area operator $\hat A_S$ as an operator on the physical
Hilbert space $\H^\phys$.  Let $\H^\bh \subset \H^\phys$ be the
subspace spanned by eigenstates of $\hat A_S$ with eigenvalues $a$
lying in the range $a_0 - \delta \le a \le a_0 + \delta$. Because we
allow for the presence of arbitrary radiation in the bulk, the space
$\H^\bh$ will  be infinite-dimensional, so it will be impossible
to normalize the projection onto this subspace to obtain a density
matrix on $\H^\phys$.  However, using the decomposition given in
equation (\ref{tensor}), we can write any vector $\psi \in \H^\bh$ in
the form
\[           \psi = \sum_i \psi^i_V \otimes \psi^i_S  \]
where $\psi^i_V$ lie in the physical volume Hilbert space and
$\psi^i_S$ lie in the physical surface Hilbert space.  In particular,
there is a smallest subspace of $\H^\phys_S$, which we call
$\H^\bh_S$, with the property that any vector in $\H^\bh$ can be written 
as above with $\psi^i_S \in \H^\phys_S$.  States in $\H^\bh_S$ describe 
the surface degrees of freedom of states in $\H^\bh$.  The space
$\H^\bh_S$ is finite-dimensional (as we shall soon see), 
so we may normalize the projection from $\H_S$ onto $\H^\bh_S$ to
to obtain a density matrix $\rho_\bh$.  This density matrix
describes the maximal-entropy mixed state of the surface geometry
compatible with the constraint $a_0 - \delta \le a \le a_0 + \delta$.  
The entropy of this state is
\[      S_\bh = \ln N_\bh  \]
where $N_\bh$ is the dimension of the space $\H^\bh_S$, i.e., the 
number of physical surface states compatible with the above constraint
on the horizon area.   To compute this entropy, we thus need to count
states forming a basis of $\H^\bh_S$.   

To do this count we first need some definitions.  Given an ordered
list of positive half-integers $j = (j_1,\dots,j_n)$, let $A(j)$ be the
corresponding eigenvalue of the area operator:
\[     A(j) = 8 \pi \gamma \l^2 \sum_i \sqrt{j_i(j_i+1)} . \]
We say the list $j$ is `permissible' if it satisfies
\begin{equation}
a_0 - \delta \le A(j) \le a_0 + \delta .
\label{counting.ineq}
\end{equation}
Given a list of half-integers $(m_1,\dots,m_n)$, we say it is
`permissible' if for some permissible list of positive half-integers
$j = (j_1,\dots,j_n)$ we have $m_i \in \{-j_i, -j_i + 1, \dots,
j_i\}$.  Finally, given a list $(a_1,\dots,a_n)$ of elements of
$\Z_k$, we say it is `permissible' if $a_1 + \cdots + a_n = 0$ mod $k$
and $a_i = -2m_i$ mod $k$ for some permissible list of half-integers
$(m_1,\dots,m_n)$.  

Given these definitions, we may determine the dimension
of $\H^\bh_S$ with the help of a weak
assumption on the nature of the Hamiltonian constraint.  We assume
that for any permissible lists $j$, $m$ there exists at least one
state $\psi_V \in \H_V^{\P,j} \cap \H_V^{\P,m}$ that is 
annihilated by the Hamiltonian constraint.  Given this assumption, 
the dimension of $\H^\bh_S$ is exactly the number of permissible lists
$a$ of elements of $\Z_k$.   (If this assumption fails, the dimension
of $\H^\bh_S$ will be smaller, so it is finite in any event.)

In the following section we use this description of the dimension of
$\H^\bh_S$ to establish upper and lower bounds on it which enable us
to show that for $\delta$ sufficiently large, but still on the order
of $\l^2$,
\[ S_\bh = {\ln 2 \over 4 \pi \sqrt{3} \gamma \l^2} a_0 + o(a_0) 
\]
Here `$o(a_0)$' refers to a quantity which, divided by $a_0$,
approaches zero in the limit $a_0 \to \infty$.  Thus, our result
agrees with Hawking's semi-classical calculation \cite{H} in the
sector of the quantum theory on which the Barbero-Immirzi
parameter $\gamma$ equals $\gamma_0$ with
\[     \gamma_0 = {\ln 2\over \pi \sqrt{3}}\,  .\]
We will return to this point in Section \ref{s6.C}.

A quick sketch of the calculation in the following section reveals its
physical significance.  The count of black hole horizon states is
dominated by states in which all the spin network edges piercing the
horizon are labelled by spins $j_i = 1/2$.  Each such edge contributes
$8 \pi \gamma \l^2 \sqrt{j_i(j_i + 1)} = 4 \pi \sqrt{3} \gamma \l^2$
to the area of the event horizon, so the total number of spin network
edges puncturing the horizon is approximately $a_0 /4 \pi \sqrt{3}
\gamma \l^2$.  Moreover, when $j_i = 1/2$ there are 2 allowed values
of $a_i$, namely $\pm 1$, corresponding to angle deficits of $\pm
4\pi/k$ for the Levi-Civita connection at the puncture $p_i$.  It
follows that each such puncture contributes $\ln 2$ to the black hole
entropy.  The total entropy of the black hole is therefore asymptotic
to $(\ln 2 / 4 \pi \sqrt{3} \gamma \l^2) a_0$.

\subsection{Counting}
\label{s6.B}

A lower bound on $N_\bh$ can be obtained as follows.  Let us consider
lists of positive half-integers $j = (j_1,\dots,j_n)$ in which all the
$j_i$ are equal to $1/2$.  The area corresponding to such a list is
\[       A(j) =  4 \pi \sqrt{3}\gamma \l^2 n .\]
Thus if 
\be 
\label{delta}
\delta > 8 \pi \sqrt{3} \gamma \l^2 
\ee
one can always find an {\it even} integer $n$ such that
(\ref{counting.ineq}) is satisfied, giving us a permissible list $j$.  
{}From this choice of $j$ we obtain $2^n$ permissible lists $m =
(m_1,\dots,m_n)$, since each $m_i$ can be either $1/2$ or $-1/2$.  
Since $n$ is even, ${n \choose n/2}$ of these lists satisfy $m_1 + \cdots +
m_n = 0$, and thus give rise to admissible lists $a$ of elements of
$\Z_k$ via $a_i = -2m_i$ mod $k$.  When $k > 2$, which is certainly true
when the black hole area is large in Planck units, all of these
admissible lists $a$ are distinct.  

In short, given that $\delta$ satisfies (\ref{delta}) and $a_0$
is large enough, we have the lower bound
\[     N_\bh \ge {n \choose n/2}  \]
where 
\[   a_0 - \delta \le  
4 \pi \sqrt{3} \gamma \l^2 n \le 
a_0 + \delta.  \]
Using Stirling's formula, one can show that for large $n$ we have
\[      {n \choose n/2} \sim 2^{n+1/2}/\sqrt{\pi n}   \]  
We thus have
\[      S_\bh = \ln N_\bh \ge \ln(2^n) - o(n)  \]
for large $n$, where `$o(n)$' refers to a quantity which, divided
by $n$, approaches zero as $n \to \infty$.  It follows that 
\be
\label{lower}
S_\bh \ge {\ln 2 \over 4 \pi \sqrt{3} \gamma \l^2} a_0 - o(a_0) .
\ee

Next we obtain an upper bound on $N_\bh$.  We certainly have
\be
\label{upper.1}
       N_\bh \le \sum_j  d(j)
\ee
where we sum over all permissible lists $j = (j_1,\dots,j_n)$ of positive
half-integers, and $d(j)$ is the number of permissible lists $m$ of
half-integers with $m_i \in \{-j_i,-j_i+1,\dots,j_i\}$, namely:
\[      d(j) = \prod_{i = 1}^n (2j_i + 1)    \]
To get a more explicit upper bound, we introduce the `density of 
states' 
\[        g(A) = \sum_j d(j) \delta(A - A(j))  \]
where we sum over {\it all} lists $j$ of positive half-integers.
By equation (\ref{upper.1}) we clearly have
\be
\label{upper.2}
     N_\bh \le \int_{a_0 - \delta}^{a_0 + \delta} g(A) dA  
\ee
To make further progress, we need an upper bound on the rate of growth of
the density of states as $A \to \infty$.  To do this, we introduce a
`partition function', which is the Laplace transform of the density of
states:
\[        Z(\alpha) = \int_0^\infty e^{-\alpha A} g(A) dA . \]
Here $\alpha$ represents an intensive variable conjugate to the
horizon area.   This integral converges when the real part of $\alpha$ 
is sufficiently large, but blows up when the damping factor $e^{-\alpha
A}$ is insufficient to counteract the growth of $g(A)$, so the
function $Z$ is analytic on a half-plane of the form $\Re(\alpha) > \alpha_0$.  
As we shall see, it extends to a meromorphic function on the 
half-plane $\Re(\alpha) > 0$.  The poles with the largest real part 
occur when $\Re(\alpha) = \alpha_0$, i.e., exactly at the point where the 
integral defining $Z(\alpha)$ ceases to converge.   In what follows, 
we use this to put an upper bound on the growth of the density of states.

Since $g$ is positive, for $\alpha > 0$ we have
\ban    \int_{a_0 - \delta}^{a_0 + \delta} g(A) dA
     &\le& \int_0^{a_0 + \delta} g(A) dA  \\
&\le& \int_0^{a_0 + \delta} e^{a_0 + \delta - A} g(A) dA \\
&\le& e^{\alpha(a_0 + \delta)} Z(\alpha)  \ean
so by equation (\ref{upper.2}) we have
\be
\label{upper.3}
 N_\bh \le e^{\alpha(a_0 + \delta)} Z(\alpha) 
\ee
whenever the integral defining $Z(\alpha)$ converges.

A simple calculation shows that
\[    Z(\alpha)    = \sum_j e^{-\alpha A(j)} d(j) \]
where we sum over all lists $j$ of positive half-integers.  
By the definition of $d(j)$ we have
\[   Z(\alpha) = \sum_n \sum_{j_1} \ldots \sum_{j_n}
e^{-\alpha \left[ 8 \pi \gamma \l^2 \sum_{i=1}^n \sqrt{j_i(j_i+1)}\right]}
\prod_{i=1}^n (2j_i+1)   \]
and this in turn implies
\[   
Z(\alpha) = \prod_l {1 \over 1 - 
(2l +1)e^{-\alpha \left[ 8 \pi \gamma \l^2 \sqrt{l(l+1)}\right] }}\,.
\]
where we take the product over positive half-integers $l$.
Note that each factor in this infinite product is analytic as a function
of $\alpha$ everywhere except for a simple pole where the denominator
vanishes, namely at the points
\[    \alpha = {\ln(2l+1) + 2\pi i n \over 8\pi \gamma \l^2 \sqrt{l(l+1)}} .\]
Note also that for any fixed choice of $\alpha$ with $\Re(\alpha) > 0$, 
the terms in the infinite product rapidly approach $1$ as $j\to\infty$.  
It follows that $Z(\alpha)$ is analytic on the right half-plane except for
simple poles at the above points.  The poles 
of $Z(\alpha)$ with the largest real part occur when $l = 1/2$,
so $Z(\alpha)$ is analytic for 
\[           {\rm Re}(\alpha) > \alpha_0 = 
{\ln 2 \over 4\pi \sqrt{3} \gamma \l^2 } \]
and $Z(\alpha)$ has a simple pole at $\alpha_0$.   

It follows that for $\alpha > \alpha_0$ the integral defining 
$Z(\alpha)$ converges, and we have 
\[     Z(\alpha) \le {K\over \alpha - \alpha_0}  \]
for some constant $K$ when $\alpha$ is close to $\alpha_0$.  
Equation (\ref{upper.3}) thus gives
\[     N_\bh \le {K \over \alpha - \alpha_0} e^{\alpha(a_0 + \delta)}. \]
Taking the logarithm of both sides, we obtain
\[     S_\bh \le \alpha(a_0 + \delta) + \ln({K \over \alpha - \alpha_0}) \]
so that as $a_0 \to \infty$ we have
\be
\label{upper}
S_\bh \le {\ln 2 \over 4 \pi \sqrt{3} \gamma \l^2} a_0 + o(a_0) .
\ee

Combining the upper bound (\ref{upper}) with the lower bound 
(\ref{lower}), we obtain 
\ba
\label{asymptotic}
 S_\bh = {\ln 2 \over 4 \pi \sqrt{3} \gamma \l^2 } a_0 + o(a_0).  
\ea
whenever $\delta$ is sufficiently large, as in (\ref{delta}).
This formula asymptotically matches the Bekenstein-Hawking formula: 
\[  S_\bh = {1\over 4} {a_0\over \l^2}  \]
if we take the Barbero-Immirzi parameter $\gamma$ to equal
\be    \gamma_0 = {\ln 2 \over \pi \sqrt{3}} .\ee
This concludes the mathematics of counting states. We comment further on the 
final result at the end of Section \ref{s6.C} and also in Section \ref{s7}.

\subsection{Charged Black Holes}
\label{s6.C}

Using Hamiltonian methods, laws of black hole mechanics have been
extended to isolated horizons for general relativity coupled to
Maxwell, dilaton, and Yang-Mills fields \cite{abf,other}.  In all
cases, the surface gravity $\kappa$ is constant on isolated horizons
and the change in the horizon energy $E_S$ is related to the changes
in its parameters via: $\delta E_S = (\kappa/8\pi G) \delta a_S +
\hbox{{\rm work terms}}$.  Comparison with the familiar zeroth and
first laws of thermodynamics suggests that, as in standard black hole
mechanics, one should assign an isolated horizon a temperature $T =
(\kappa\hbar/2\pi)$ and an entropy $S= (a_S/4G\hbar) \equiv
(a_S/4\l^2)$.  In particular, these semi-classical considerations
suggest that the entropy is purely geometrical: it is a multiple of
the area of the horizon, irrespective of the values of matter fields
and their charges \textit{at} the horizon.  A statistical mechanical
derivation of entropy has to account for this key fact.

For simplicity, in this section, we will restrict ourselves to Maxwell
and dilaton fields: the inclusion of Yang-Mills fields only adds
certain technical complications without altering the essence of our
arguments or the final results.  (If the dilaton field vanishes, we
are left with the Einstein-Maxwell system.)  Let us then begin by
briefly recalling the relevant aspects of the classical theory of the
Einstein-Maxwell-dilaton system from \cite{ack,abf,other}.  In this
theory, the isolated horizon carries four parameters which can be
taken to be the area $a_0$, electric charge $Q_0$, the magnetic charge
$P_0$ and the value of the dilaton field $\phi_0$, all defined
intrinsically on the horizon.  With suitable boundary conditions, the
action principle for the total system is well-defined if we simply add
the standard matter action terms to the gravitational action of
\cite{ack}.  In particular, there is no need to introduce any new
surface terms at the horizon in the matter sector.  One can then pass
to the phase space through the standard Legendre transform.  The total
phase space $\X_{\rm total}$ has three striking features which are
important for quantization.

First, the gravitational part $\X_{\rm grav}$ of the phase space is
exactly the same as $\X$ of Section \ref{s2}.  In particular, the key
boundary condition (\ref{bc}) constraining the gravitational phase
space variables $(A, \E)$ at the horizon is left unaltered in spite of
the presence of matter fields on the horizon itself.  This is
remarkable because the presence of matter \textit{does} modify the
curvature $F$ of $A$: while in absence of matter, $F$ is determined by
the Weyl tensor (technically the Newman-Penrose component $\Psi_2$ at
the horizon), in presence of matter it depends also on the Ricci
curvature (the $\Phi_{11}$ component and the scalar curvature).
However, because of the isolated horizon boundary conditions, the
pullback of $F$ to $S$ is always proportional to the pullback of $\E$,
and the Gauss-Bonnet theorem implies that the proportionality factor
is always $ -2\pi/a_0$ \textit{irrespective of the value of the Ricci
curvature at the horizon}.  Furthermore, as one might expect, the
gravitational part of the symplectic structure is unaffected by the
presence of matter.  These properties are important to our analysis
because they imply that the description of the quantum horizon geometry
developed in Sections \ref{s4} and \ref{s5} continues to be valid,
without any change whatsoever, in the presence of matter fields.

The matter sector of the phase space consists of fields $(\phi, \pi;
\bfA, \bfpi)$ on the spatial 3-manifold $M$, satisfying appropriate
boundary conditions.  Here $\phi$ is the dilaton scalar field, the
3-form $\pi$ is its conjugate momentum, $\bfA$ is the electromagnetic
vector potential, and $\bfpi$ is its conjugate momentum, related to
the electric field 2-form $\bfE$ via $\bfpi = \exp(-2\alpha\phi)
\bfE$.  (Here $\alpha$ is the dilaton coupling constant.)  We denote
the curvature of $\bfA$ ---the magnetic field 2-form--- by $\bfF$.  The
boundary conditions at infinity are the standard ones.  The
gravitational boundary conditions at the horizon imply that the
dilaton field $\phi$ is constant on $S$.  We will denote its value by
$\phi_0$.  The remaining horizon parameters can be taken to be the
electric and magnetic charges,
\be 
\label{charges} 
P_0 := \frac{1}{4\pi}\, \oint_S \bfF\,  \quad {\rm and} \quad 
Q_0 := \frac{1}{4\pi}\, \oint_S \bfE\, .
\ee 
All three quantities are conserved in time.  The remaining horizon
boundary conditions, relevant for quantization and entropy, are: 
\be
\label{matterbc} 
\phi = \phi_0, \quad \underline{\bfF} = 2 P_0 dW, \quad {\rm and}
\quad \underline\bfE = 2Q_0 dW 
\ee 
where, as before, underbars denote pullbacks to $S$ and $W$ denotes
the $U(1)$ gravitational connection.  Thus, given the horizon
parameters, the \textit{local} values of matter fields on the horizon
are completely determined by the \textit{gravitational} connection
already at the kinematical level, without reference to the bulk
equations of motion.This is the second important feature. It already
provides a partial explanation of why the entropy is purely
geometrical.

The third important feature of the Hamiltonian framework is the form
of the matter symplectic structure.  It is given by:
\be 
\label{matterss}
\Omega_{\rm matter} (\phi,\pi; \delta\bfA, \delta\bfpi) =
\int_M (\phi\pi + \bfA \wedge \bfpi )\, .
\ee
The symplectic structure on the full phase space is simply
$\Omega_{\rm total} = \Omega_{\rm grav} + \Omega_{\rm matter}$, the
gravitational symplectic structure being given by (\ref{ss}).  Note
that, unlike the $\Omega_{\rm grav}$, the matter contribution
$\Omega_{\rm matter}$ does not contain a surface term.  This
difference has major ramifications for quantization.

For, in the gravitational sector, the presence of a surface term in
$\Omega_{\rm grav}$ led us naturally to construct a bulk phase space
$\X_V$ and a surface phase space $\X_S$ with projections $p_V$ and
$p_S$ from the gravitational phase space $\X_{\rm grav}$ to $\X_V$ and
$\X_S$ respectively.  (See the remark at the end of Section \ref{s2}).
This in turn suggested that in quantum theory we should begin with a
total Hilbert space $\H$ of the form $\H_V \otimes H_S$.  The volume
states were obtained by quantization of the volume phase space $\X_V$
and the surface states by quantization of $\X_S$.  In the matter
sector, on the other hand, there is no surface term in $\Omega_{\rm
matter}$.  Therefore, the matter sector has to be quantized
differently: now only the volume Hilbert space can be constructed
naturally.  Furthermore, as argued in Section \ref{s6.A}, the details
of even this construction are not needed in the entropy calculation
which only involves counting of \textit{surface} states in the
microcanonical ensemble.  The fact that the only independent surface
states come from gravity explains, in this approach, the purely
geometric origin of entropy.

What happens then to the matter boundary conditions (\ref{matterbc})
in the quantization procedure?  These are incorporated by eliminating
the surface Maxwell fields $\underline\bfF$ and $\underline\bfE$ in
favor of $(1/P_0)\, dW$ and $(1/Q_0)\, dW$ respectively \textit{prior}
to quantization.  Thus, any classical observable involving the surface
values of the Maxwell fields becomes, in quantum theory, an operator
on the \textit{gravitational} surface Hilbert space.  \textit{All}
horizon surface physics in the Einstein-Maxwell-dilaton system occurs
only on the gravitational Hilbert space $\H_S$ constructed in Sections
\ref{s4}.  Hence, the counting argument of Section \ref{s6.B} is
unaltered and entropy of an isolated horizon is independent of the
values of the charges $P_0, Q_0$ and $\phi_0$.  For large values of
$a_0/\l^2$ the leading term in the formula for the entropy is always
given by (\ref{asymptotic}).  Thus, all the surface degrees of freedom
associated with a quantum isolated horizon belong to the gravitational
sector alone.  In more physical terms, the quantum fluctuations of
matter fields on the horizon manifest themselves only through the
quantum fluctuations of the horizon geometry.%
\footnote{It is interesting to consider an alternate, hypothetical
strategy.  Although there is no natural way to construct the surface
Hilbert space for the Maxwell theory, suppose one just chose, in an ad
hoc manner, a quantization procedure without first solving for the
surface Maxwell fields in terms of the gravitational $dW$.  Then, one
would be led to impose the boundary conditions (\ref{matterbc}) as
operator restrictions on permissible quantum states.  How would one
then see that there are no independent Maxwell surface states?
Setting $F= dW$ as in Section \ref{s3}, the equation constraining
Maxwell \textit{surface} states would then read $(\Psi_{{\rm grav},V}
\otimes \Psi_{{\rm grav},S}) \otimes(\Psi_{{\rm Max},V}\, \otimes\,
\exp i\hat{\underline{\bfF}}\, \Psi_{{\rm Max},S}) = (\Psi_{{\rm
grav},V} \otimes \exp 2iP_0\hat{F}\, \Psi_{{\rm grav},S})
\otimes(\Psi_{{\rm Max},V} \otimes \Psi_{{\rm Max},S})$. Because of
this tight correlation, the entropy calculation would again reduce to
the counting only of independent gravitational surface states.}

The fact that the operators corresponding to surface matter fields act
on the gravitational Hilbert space $\H_S$ may seem somewhat
counter-intuitive at first.  This arose because, already at the
kinematical level, we were naturally led to put all the degrees of
freedom of the coupled system in the gravitational sector.  A
qualitatively similar situation occurs in quantum gravity coupled to
the Maxwell (or scalar) field in 2+1 dimensions.  The rotationally
symmetric sector of this system provides an exactly soluble quantum
field theory with an infinite number of degrees of freedom \cite{ap}.
In this theory, it is natural to arrange matters such that the true
degrees of freedom reside in the Maxwell field which, furthermore, can
be quantized readily.  The gravitational observables, such as the
fully gauge fixed metric, are then operators on the Maxwell Hilbert
space.  If one constructed a microcanonical ensemble in this theory
using some Maxwell \textit{and} gravitational ---or, indeed, even
purely gravitational--- observables, the state counting would still be
carried out on the appropriate sector of the pure Maxwell theory.
(Unfortunately, however, this theory does not admit black holes and so
one can not use this procedure to shed light on black hole entropy.)
In the present case, the situation is analogous, although the roles of
the gravitational and matter fields are now reversed and the reasoning
is not tied to the details of dynamics.  (For further remarks on the
issue of dynamics, see Section \ref{s7}.)

We conclude with a few remarks on the results obtained in Section 
\ref{s6} as a whole.   \medskip

\textsl{1. Extremal black holes.}  For simplicity, let us consider 
the Einstein-Maxwell system without dilatons.%
\footnote{In this remark we are interested in extremal black holes and
the horizon geometry is non-singular in extremal static solutions only
when the dilatonic coupling constant is zero, i.e., only in the
Einstein-Maxwell sector of the theory.  Inclusion of a dilaton
therefore adds unnecessary complications to this discussion.}
In this case, we were led to eliminate the surface matter fields in
favor of the surface gravitational field because of three key factors.
First, the gravitational boundary condition (\ref{bc}) could be
expressed entirely in terms of the parameter $a_0$ without any
reference to charges $Q_0$ and $P_0$.  Second, the surface term in the
total symplectic structure $\Omega_{\rm grav}$ is also purely
gravitational.  Finally, while the gravitational boundary conditions
(\ref{bc}) made no reference to matter fields, the gravitational
curvature $dW$ plays a key role in the matter boundary conditions
(\ref{matterbc}).  Therefore, at the kinematical level, while it was
possible to first quantize the gravitational sector without any
reference to the values of electromagnetic charges, it was not
possible to reverse the strategy and first quantize the matter fields.
However, if we restrict ourselves only to the extremal, static black
holes, we can express the area $a_0$ in terms of the electromagnetic
charges.  Furthermore, in this case, the Weyl tensor vanishes
identically on the horizon whence the gravitational curvature $dW$ is
completely expressible in terms of the electromagnetic field.
Therefore, it is quite possible that the surface term in the
symplectic structure could also be recast in terms only of the
electromagnetic field.  If this can be done, it should be possible to
reverse the strategy, transfer the gravitational surface degrees of
freedom to the electromagnetic ones and compute the entropy purely in
terms of the electromagnetic surface states.  This calculation would
be instructive and may well enable one to relate the present approach
to that based on string theory.  \medskip

\textsl{2. The Barbero-Immirzi parameter.} We have shown that,
irrespective of the values of matter charges, entropy of a
non-rotating isolated horizon is proportional to its area $a_0$.  In
this calculation, the proportionality factor depends on the value of
the Barbero-Immirzi parameter $\gamma$ simply because the elementary
quantum of area depends on this value.  This is an inherent feature of
the approach: just as there is the $\theta$-ambiguity in QCD, we have
a 1-parameter family of quantization ambiguities, labelled by
$\gamma$, and in general physical predictions depend on which
$\gamma$-sector of the quantum theory one chooses.  In QCD, one can
hope to make a single judicious experiment to fix the value of
$\theta$; then all further experiments can test the theory. The
situation here is rather similar.  If we could somehow make a
judicious experiment to measure the value of, e.g., the area quantum,
the value of $\gamma$ ---and hence the quantum theory of geometry---
would be fixed. Although such a direct measurement seems completely
out of reach at present, one can regard the entropy calculation as
giving an indirect handle on the elementary quantum of area, thereby
selecting the value of $\gamma$ realized in Nature.

More precisely, one can adopt the following `phenomenological
approach.' In the classical theory, all $\gamma$ sectors are related
simply by a canonical transformation and hence completely equivalent
to one another.  However, as with the $\theta$-sectors of QCD, they
lead to different physical predictions as soon as $\hbar$ is non-zero,
i.e. already at the semi-classical level.  In particular, although the
non-perturbative quantum theories corresponding to various $\gamma$
values are internally consistent, they would not all agree with the
standard semi-classical calculations performed in the framework of
quantum field theory in curved spacetimes.  One can thus hope to
constrain the value of $\gamma$ by demanding that for large black
holes, predictions of the full, non-perturbative quantum theory should
reduce to those of the standard quantum field theory in curved
spacetimes. 

Since the semi-classical calculations have been performed for general
relativity coupled to a variety of matter fields, a very large family
of consistency checks is thus available. We could consider the
isolated horizon in a de Sitter spacetime, or of an uncharged
Schwarzschild black hole, or of a black hole with specific electric,
magnetic and dilatonic charges.  Now, it turned out that the
requirement that the leading term in full quantum answer agree with
the semi-classical result for \textit{any one} of these horizons
already fixes the value of $\gamma$ uniquely; $\gamma = \gamma_0 = \ln
2/\pi\sqrt{3}$. This in turn fixes the theory. Then one can ask: Does
one obtain the correct answer in all other contexts?  We saw that the
answer is in the affirmative.  There is an infinite-dimensional space
of spacetimes admitting non-rotating isolated horizons and in all
these cases, the statistical mechanical entropy agrees to leading
order with the Hawking-Bekenstein formula in the $\gamma_0$ sector.
Thus, a very large number of consistency checks are met within the
current quantum geometry approach.

We should emphasize, however, that what we have summarized here is
only the present `working viewpoint'.  A more definitive stand would
emerge only after a number of other semi-classical checks are made and
the relation between the non-perturbative theory based on quantum
geometry and more conventional perturbative quantum field theories in
the continuum is better understood. \medskip

\textsl{3. Wheeler's `It from Bit'.} The detailed calculation of
Section \ref{s6.B} shows that the dominant contribution to entropy
comes from states in which there is a \textit{very large} number of
punctures, each labelled by $j = \textstyle{1\over 2}$ and $a= \pm 1$.
Thus, there is a precise sense in which the quantum geometry
calculation realizes John Wheeler's `It from Bit' picture of the
origin of black hole entropy \cite{jw} based on qualitative
considerations from information theory. \medskip

\textsl{4. Robustness of the calculation.} 
Isolated horizon boundary
conditions imply that the independent information in the gravitational
connection $A$ at the boundary $S$ is coded in the $\U(1)$ connection
$W$ (in the sense made explicit in Section 3.C of
\cite{ack}). Therefore, in our analysis, surface states were described
by the $\U(1)$ Chern-Simons theory.  However, even if one ignores this
fact and just uses $\SU(2)$ Chern-Simons theory to describe surface
states, the leading order contribution to the entropy again turns out
to be given by equation (\ref{asymptotic}) \cite{KM1,Smolin}. This 
precise agreement seems mysterious and it is important to understand 
systematically if this robustness could have been predicted on general 
grounds. If so, one may be able to relate our framework to conformal field 
theories that feature in other approaches \cite{sc}. Furthermore, there is
already an elegant derivation of the sub-leading correction to entropy
in the $\SU(2)$ framework \cite{KM2} and the resulting logarithmic
correction is supported by a closer analysis of the Cardy formula
\cite{sc2}. 

As another example of the robustness of the entropy calculation, note that
the $j$ values that determine the black hole area via equation (\ref{area2}) 
are considered as volume rather than surface degrees of freedom in our 
approach.  This arises naturally from the fact that the
area operator is built using the $\E$ field, which acts as an operator
on the volume Hilbert space.  But one might still wonder: if
we decided to treat these $j$ values as surface degrees of freedom for
the purposes of computing black hole entropy, how would the entropy
formula be affected?  The answer is that this would \textit{not}
affect the leading-order term in the entropy calculation.  More
precisely, suppose we redefined $N_\bh$ to be the number, not of permissible 
lists $(a_1,\dots,a_n)$, but of permissible lists $(j_1,\dots,j_n)$ 
together with choices for each spin $j_i$ of an element 
$a_i \in \Z_\k$ satisfying the conditions described in Section \ref{s6.A}.
Then we would still obtain the asymptotic formula (\ref{asymptotic})
for the entropy $S_\bh = \ln N_\bh$.  

\section{Discussion}
\label{s7}

This is the second paper in a series. The first paper \cite{ack}
developed the classical Hamiltonian framework for the sector of
general relativity consisting of spacetimes admitting a non-rotating
isolated horizon as their inner boundary. Using this framework as the
point of departure, in the first part of this paper we extended the
quantum theory of geometry to describe the geometry of quantum
horizons. The approach is rather general: we could use a single
framework to incorporate isolated black hole horizons with no
restrictions on values of electromagnetic charges or on the ratio of
the horizon area to the cosmological constant. Furthermore, the
cosmological horizons of the type discussed in \cite{gh} are
incorporated from the beginning \cite{abf}. In the second part of the
paper, we constructed a microcanonical ensemble consisting of
spacetimes whose isolated horizons have areas and charges lying in
small intervals around classical values, $a_0, Q_0, \ldots$ (with $a_0
\gg \l^2$), and calculated the number of surface states to obtain the
statistical mechanical entropy associated with these horizons. We
showed that, irrespective of the values of charges, to leading order
the entropy is always proportional to the area $a_0$.  That is, for
large black holes, the number of micro-states of the horizon geometry
completely determines the state-counting; there are no independent
surface states from the matter sector. This explains the geometrical
nature of the Bekenstein-Hawking entropy.

Perhaps the most intriguing feature of this analysis is that it rests
on a delicate and subtle interplay between general relativity which
provides the isolated horizon boundary conditions, quantum theory of
geometry which leads to a `polymer picture' of geometry in the bulk
and quantum Chern-Simons theory which describes the geometry of the
quantum-horizon. First, the ingredients needed for quantization of the
Chern-Simons theory ---the symplectic structure and the punctures
labelled by spins--- are provided, respectively, by general relativity
and quantum theory of geometry. Second, the quantum-horizon boundary
condition can be meaningfully imposed \textit{only} because the
spectra of certain quantum geometry operators on the volume Hilbert
space and those of certain other operators on the Chern-Simons Hilbert
space match exactly eventhough the two calculations are completely
independent. This matching is in turn possible because the symplectic
structure dictated by classical general relativity provides a specific
value of the `level' $k$ of the Chern-Simons theory. Finally, the bulk
quantum theory provided by quantum geometry and the surface quantum
theory provided the Chern-Simons theory match seamlessly to provide a
full, coherent theory. In particular, in the final picture, the
quantum-horizon boundary condition ---which came from the structure of
isolated horizons in general relativity--- can be interpreted as
requiring that the bulk and the surface states be so intertwined that
the total state is gauge invariant.

There are three outstanding open issues:\\ 
i) First, the classical framework developed in \cite{ack} only allowed
undistorted, non-rotating black holes. However, these restrictions
were removed recently \cite{other}. Furthermore, as discussed below,
there is a well-defined sense in which the quantum framework developed
here already incorporates distortion. The first outstanding problem is
to extend it further to incorporate rotation. Away from extremality,
this is also an open issue in approaches based on string theory.\\
ii) The second major issue is to gain a better understanding of the
Barbero-Immirzi parameter $\gamma$.  While this parameter plays no
physical role in the classical theory, quantum sectors corresponding
to different values of $\gamma$ are unitarily inequivalent.  Just as
one needs an external input to resolve the $\theta$-ambiguity in QCD,
here we need experimental data (e.g., on values of the quanta of the
geometric operators), or, a theoretical principle to resolve the
$\gamma$-ambiguity and arrive at an unique quantum gravity theory.
The requirement that the non-perturbative framework should yield the
Bekenstein-Hawking entropy for one large isolated horizon suffices to
select a specific value of $\gamma$.  Then agreement continues to hold
for \textit{all} non-rotating isolated horizons, including the
cosmological ones, irrespective of the values of charges and the
cosmological constant.  Further considerations \cite{K3} show that the
framework also yields the correct Hawking temperature for the same
value of $\gamma$.  Thus, the `external' requirement that the
non-perturbative framework agree with quantum field theory in curved
spacetimes with large isolated horizons suffices to remove the
ambiguity and fix the framework.  We now need to develop further
semi-classical tests to check the robustness of this conclusion.
Another avenue that might shed light on the preferred value of
$\gamma$ is suggested by Carlip's approach to entropy \cite{sc}, based
on symmetries.  A detailed examination of symmetries associated with
the isolated horizon and their action on quantum states in various
$\gamma$-sectors may provide new insights.  However, in their present
form, these symmetry ideas are not well-adapted to the non-rotating
case and so their implementation may well have to await the
incorporation of rotation.\\
iii) Finally, there is the issue of a systematic derivation of the
Hawking radiation.  First steps in this direction have already been
taken \cite{K3}.  Using properties of our surface states, one can
refer to the rather general arguments given by Bekenstein \cite{Bek}
to conclude that the spectrum should be thermal.  However, as usual
\cite{H}, the intensity distribution contains a multiplicative factor
involving the absorption cross-section of the horizon.  A derivation
of this cross-section from full quantum gravity is still missing.  A
systematic study of the semi-classical approximation may be necessary
before one can carry out this calculation.

This completes the summary of major open issues.  Since our analysis
uses ideas from several distinct areas of mathematical physics, we
conclude by making a number of remarks to clarify the role of various
structures and the relation of this approach to some others.\medskip

\textsl{1.  Degrees of freedom: this approach.} We began our analysis
with a classical system: the sector of general relativity admitting a
non-rotating isolated horizon with given area (and charges).  We used
the boundary conditions to eliminate all components of the $\SU(2)$
connection at the horizon other than the $\U(1)$ connection $W$, which
alone appears in the surface term of the symplectic structure
(\ref{ss}).  Now, the boundary conditions (\ref{bc}) imply that, given
the value $a_0$ of area, $W$ is unique up to gauge transformations and
diffeomorphisms \cite{ack}.  Thus, in the classical theory, there are
no true (`configuration space') degrees of freedom on the horizon.
Indeed, if there were, the quantum Hilbert space would have been
infinite dimensional even for fixed values of horizon parameters.
(Recall that if the classical configuration space is $\R^n$, the
quantum Hilbert space is $L^2(\R^n)$.)

However, in the passage to quantum theory, we did not first `solve'
the boundary condition (\ref{bc}) on $W$ and $\underline{\E}$
classically.  Rather, we imposed the horizon boundary condition as an
operator restriction (\ref{qbc2}) on allowed quantum states.  This
procedure was motivated by the physical expectation that the horizon
`should be allowed to fluctuate' in the quantum theory.  Thus, we did
not freeze either the connection $W$ nor the pullback $\underline\E$
of the triad 2-form to the horizon.  Both are allowed to fluctuate,
but must do so `in tandem', respecting (\ref{qbc2}).  The quantum
boundary condition (\ref{qbc2}) then led us to the Chern-Simons theory
with punctures.  This theory has a \textit{finite} number of
states because the phase space of the classical Chern-Simons theory
with a given set of punctures is \textit{compact}, reflecting the fact
that the classical theory does not have `true' configuration degrees
of freedom.  These states describe the geometry of the quantum horizon
and account for entropy.  Thus, thanks to the strategy of
incorporating the horizon boundary condition through (\ref{qbc2}), we
could fulfill the statistical mechanical expectation that a single,
physical macrostate should correspond to a large number of
micro-states and, \textit{furthermore,} arrive at these micro-states
through quantization.\medskip

\textsl{2. Degrees of freedom: comparison.} In string theory, as well
as in the analysis based on symmetries a la Carlip \cite{sc}, one
first quantizes a \textit{larger} system with true degrees of freedom
and then associates the given black hole (with specific values of
charges) with a finite dimensional subspace of the infinite
dimensional Hilbert space of that larger system.  Thus, rather than
first restricting and then quantizing, one first quantizes and then
restricts.  Note that this is qualitatively analogous to our strategy
with the quantum boundary conditions: We first ignored the boundary
conditions and quantized a larger system and then projected to its
subspace on which (\ref{qbc2}) holds.  Of course, the larger systems
are quite different in the three cases.

`Edge states' introduced in other contexts in the literature have the
flavor of the surface states we introduced in this paper.  However,
there are some important differences.  In \cite{B}, for example,
techniques used in the quantum Hall effect were carried over to
canonical gravity using `geometrodynamics'.  However, in that
discussion, the boundary conditions at internal boundaries are not
explicitly stated and so the discussion is not specific to horizons.
Furthermore, it is assumed that the conditions should be such that
only those (spatial) diffeomorphisms which are identity on the
internal boundary are regarded as gauge.  This strategy plays a key
role in their construction of an infinite set of boundary observables.
In our treatment, on the other hand, the specific horizon boundary
conditions dictated what is `gauge'.  Furthermore, we were led to
regard all spatial diffeomorphisms which map the horizon $S$ to itself
as gauge.  Indeed, without this, our entropy would have been infinite!
Thus the `origin' and physical meaning of surface states is quite
different in the two cases.  This point also serves as a key
distinction between our approach and that of \cite{M}.  Indeed, in
contrast to these analyses \cite{B,M}, in our approach there are no
independent surface degrees of freedom at the classical level.
Independent surface states arise in quantum theory only because the
quantum configuration space is larger than the classical, admitting
distributional connections.  By contrast, in obtaining the `edge
states' and `surface degrees of freedom' in \cite{B,M} these
functional analytical subtleties play no role.  Finally, there is some
similarity between our surface states and those considered in
\cite{Smolin}.  However, that analysis refers to Euclidean gravity and
uses boundary conditions which are unrelated to the presence of
horizons.  There is no $\U(1)$ reduction on the boundary so that the
Chern-Simons theory considered there refers to $\SU(2)$.  Finally, the
boundary in \cite{Smolin} is the outer boundary, not inner.  Thus,
while these three treatments share some features of our analysis, they
consider quite different physical situations and their full
mathematical frameworks are also different from ours.\medskip

\textsl{3. Horizon boundary conditions.} Note that the isolated
horizon boundary conditions play several distinct roles in our
classical analysis \cite{ack,abf}. First, they enable us to show that
the full $\SU(2)$ connection $\underline{A}$ at the horizon can be
recovered from the $\U(1)$ connection $W$. Second, they lead us to the
surface term in the symplectic structure involving $W$. This structure
brings out the relation between the `total', the `volume' and the
`surface' phase spaces, described at the end of Section \ref{s2.A},
suggesting that the space of quantum states should have an analogous
structure. We were therefore led to begin the quantum analysis on the
Hilbert space $\H = \H_V \otimes \H_S$. Finally, the residual relation
(\ref{bc}) led to the quantum boundary condition (\ref{qbc2}) which
played a key role in the present paper.

Because equation (\ref{qbc2}) played such an important role in this paper,
one might be tempted to the ignore other roles of the boundary
conditions.  Indeed, this was done in \cite{vh} where the author noted
that (\ref{bc}) does not suffice to capture the idea that $S$
represents a horizon and went on to replace that condition with
another.  However, as is clear from the detailed structure at the
horizon discussed in \cite{ack,abf}, the fact that the $\SU(2)$
connection $\underline{A}$ has a specific form ---the first role of
the boundary condition noted above--- does suffice to ensure that $S$
is a marginally trapped surface.%
\footnote{Furthermore, the proposed modification of \cite{vh} does not
appear to be viable because the corresponding action (equation (10) of
\cite{vh}) fails to be functionally differentiable with respect to $e$
unless $\underline{F}$ vanishes on the boundary, a condition which is
violated already in the Schwarzschild family. By contrast, (\ref{bc})
holds on all non-rotating isolated horizons. However, we should add
that the main confusion in \cite{vh} probably came about because, due
to a space limitation, the boundary conditions were not spelled out in
detail in the brief report \cite{abck}.}

Nonetheless, since our quantum considerations depend primarily on
(\ref{qbc2}), one might ask if there there exist contexts, other than
isolated horizons, where our ideas are applicable.  As pointed out in
the first part of \cite{vh}, the answer is in the affirmative: One can
indeed construct mathematical models, even unrelated to general
relativity, to which this analysis is applicable.  This is not
surprising because the ideas underlying polymer geometry are robust
and applicable to a wide variety of situations.  Of course, as noted
in \cite{vh}, the physical interpretation of surface quantum states
and of the resulting entropy would depend on the specific context to
which these ideas are applied.  \medskip

\textsl{4. Examples.} Our isolated horizon boundary conditions are
satisfied at the event horizons of the Reissner-Nordstr\"om family as
well as cosmological horizons of de-Sitter space times.  As discussed
in detail in \cite{abf}, one can construct a large family of examples
by appropriately `adding radiation' to the exterior regions of these
spacetimes.  In all these cases, our analysis of the geometry of
quantum horizons goes through provided $M$ can be chosen as a partial
Cauchy slice in an appropriate piece of these spacetimes (see Figures
5 and 6 in the second paper in Ref.\ \cite{abf}).  Furthermore, since
we trace over the volume degrees of freedom in our entropy
calculation, the presence of radiation in the exterior region does not
affect that analysis.

Finally, although we began the classical analysis in \cite{ack} with
undistorted horizons, there is a precise sense in which distortion is
automatically included in the final theory. In the classical theory,
we did not explicitly restrict ourselves to spherical horizons. Rather
our boundary conditions, together with standard differential geometric
identities, imply that the scalar curvature of the horizon 2-metric is
constant, i.e., the 2-metric is spherically symmetric with respect to
\textit{some} 3-dimensional $\SO(3)$ subgroup of $\Diff(S)$. Now, in
quantum geometry, these identities are not available, whence an
analogous conclusion can not be drawn. Indeed, there is a precise
sense in which the quantum geometry of $S$ is flat everywhere except
at the punctures. At the $i$th puncture $p_i$, there is a quantized
angle deficit of $4\pi a_i/k$ for some integer $a_i$. {}From the
classical theory of distorted horizons now available \cite{other}, one
can interpret $a_i$ as giving us the `strength' of the scalar
curvature at $p_i$. Since the conditions which the integers $a_i$ are
subject to allow them to vary from puncture to puncture, it is clear
that the quantum geometry is allowed to be `non-spherical'. More
precisely, given any distorted, classical 2-geometry on $S$, the
Hilbert space $\H^{\rm Kin}$ includes semi-classical states which,
upon coarse-graining, approximate that 2-geometry. Thus, the classical
limit of our quantum theory already includes all non-rotating
distorted horizons. In view of the fact that the quantum geometry has
discrete, distributional character, this result is not
surprising. Indeed, none of our quantum states can have an exact
spherical symmetry. More generally, as in the quantum theory of
solids, because of the `atomic' nature of geometry, quantum states can
not have any continuous spatial symmetries.\medskip

\textsl{5. Dynamics and related issues.}  The fact that we are dealing
with general relativity dictated our choice of the phase space and the
symplectic structure. Therefore, our analysis is not valid for higher
derivative theories where the phase space is larger and the symplectic
structure and the constraints take on a different form.  Indeed, the
first law of classical black hole mechanics suggests that the entropy
should not be proportional to area in such theories \cite{W}.

In quantum general relativity, physical states must solve quantum
constraints. In our calculations, the Gauss and the diffeomorphism
constraints play an important role; in particular, without their
imposition, the entropy would have been infinite. The Hamiltonian
constraint, on the other hand, plays a minor role. This is because, in
the classical theory, it leads to a well-defined Hamiltonian vector
field on the phase space \textit{only if} it is smeared with a lapse
function which goes to zero on $S$ (and at infinity). `Time evolution'
along the horizon is generated not by constraints but by a `true
Hamiltonian' with a non-vanishing surface term at $S$
\cite{ack,abf,other}.  Consequently, physical states have to be
annihilated only by the constraint smeared with a lapse which vanishes
at $S$, whence this constraint does not play a direct role in the
determination of the geometry of the quantum horizon.  What we need is
only that, for generic, permissible lists $(j_1, \dots,  j_n)$ and
$(m_1,\dots, m_n)$, there exists a solution to the Hamiltonian constraint
in the bulk (with lapse going to zero on $S$) constructed out of a
spin network $n$ of whose edges intersect $S$ and carry labels $j$
and $m$.  
Thus, our construction does not depend on full dynamics.%
\footnote{Note however that \textit{given the phase space},
considerations motivated by spacetime covariance essentially determine
gravitational terms in the Hamiltonian constraint \cite{hkt}. In this
sense, within our setting, the remaining freedom in dynamics comes
essentially from matter couplings and the value of the cosmological
constant.}
This feature seems to be rather general. Carlip's approach
\cite{sc}, for example, is based on symmetries rather than details of
dynamics.  Similarly, in string theory, one does not yet have control
on full dynamics (e.g., the interaction between branes and
anti-branes) and entropy calculations are possible precisely in those
cases where they do not depend on these unresolved aspects of
dynamics.

We began our analysis by restricting ourselves to spacetimes which
admit isolated horizons as their inner boundaries.  To ensure that no
radiation falls across the portion of horizon under consideration, the
allowed initial data on a spatial surface must satisfy some implicit
conditions (in addition to the initial value constraints of Einstein's
theory).  Therefore, strictly speaking, quantization of the bulk
theory is more subtle than in cases without internal boundaries.
However, these subtleties should not affect the analysis of surface
states which are sensitive only to the boundary conditions \textit{at}
the horizon.  In the calculation of entropy, one traces over the bulk
degrees and hence these subtleties get `washed away'; in the detailed
calculation of Section \ref{s6}, we counted only surface states.  This
is also the reason why details of bulk spin networks such as their
knotting in the bulk and the precise nature of intersection of edges
with $S$ do not affect the quantum geometry and entropy of the
horizon. \medskip

\textsl{6. Entropy: physical considerations.} The entropy we calculated
is not an intrinsic attribute of the spacetime as a whole but depends
on the division of the spacetime in to exterior and interior
regions. Operationally, it is tied to the class of observers who live
in the exterior region, for whom the isolated horizon is a {\it
physical} boundary that separates the part of the spacetime they can
access from the part they can not.  (This is in sharp contrast to
early work \cite{K,Rov} which focussed only on the interior.)  This
point is especially transparent in the case of cosmological horizons
in de Sitter spacetime since that spacetime does not even admit an
invariantly defined division. Note however that, while there is an
`observer dependence' in this sense, our entropy does {\it not} refer
to the number of interior degrees of freedom which are inaccessible to
the observers under consideration. Indeed, on general grounds, it
would seem unreasonable to associate entropy with interior states:
since one can imagine multiple universes inside the horizon which do
not communicate to the exterior region, the number of potential
interior states compatible with the data accessible to the exterior
observers is uncontrollably large. In our analysis, interior states
were never mentioned. In particular, our `tracing' was done with
respect to the bulk states in the \textit{exterior}. Our entropy
refers to the micro-states of the boundary itself which are compatible
with the macroscopic constraints on the area and charges of the
horizon; it counts the physical micro-states which can interact with
the outside world, and are not disconnected from it.

The goal of our framework is to answer the following question: Given
that there is an isolated horizon, what is the entropy associated with
it? In light of the conditional nature of this question, it was
appropriate to begin with a suitably restricted sector of general
relativity and \textit{then} carry out quantization. However, as a
result, our description is an effective one. In a fundamental
description, one would first quantize the whole theory and then
isolate states which admit `quantum horizons' with given area and
charges. However, since the notion of a horizon is deeply tied to
classical geometry, at present it seems difficult to state precisely
what one would mean by quantum horizons in a full quantum theory,
irrespective of the approach one wants to use. Fortunately, for
thermodynamic considerations involving large black holes, effective
descriptions are adequate.\medskip

\textsl{7. Comparison with results from string theory.} Because one
begins with classical general relativity and uses non-perturbative
quantization methods, in the present approach one can keep track of
the physical, curved geometry.  In particular, as required by physical
considerations, the micro-states which account for entropy can
interact with the physical exterior of the black hole.  In string
theory, by contrast, detailed calculations \cite{SV,MS} are generally
performed in flat space and non-renormalization arguments and/or
duality conjectures are then invoked to argue that the results so
obtained refer to macroscopic black holes.  Therefore, relation to the
curved space geometry and physical meaning of the degrees of freedom
which account for entropy is rather obscure.  More generally, lack of
direct contact with physical spacetime can also lead to practical
difficulties while dealing with other macroscopic situations in string
theory.  For example, while we could easily account for the entropy
normally associated with de Sitter horizons in four spacetime
dimensions, this task appears to be rather difficult in string
theory.%
\footnote{Recently, 2-dimensional de Sitter spacetimes were were
discussed in the context of AdS/CFT duality (modulo a caveat on the
existence of configurations of the required type in full string
theory). However, in higher dimensions it is difficult to arrive at
precise conclusions because of certain divergences \cite{hms}.}
On the other hand, in the study of genuinely quantum, Planck size
black holes, this `distance' from the curved spacetime geometry may
turn out to be a blessing, as classical curved geometry will not be an
appropriate tool to discuss physics in these situations.  In
particular, a description which is far removed from spacetime
pictures may be better suited in the discussion of the last stages of
Hawking evaporation and the associated issue of `information loss'.

The calculations based on string theory have been carried out in a
number of spacetime dimensions while the approach presented here is
directly applicable only to four dimensions.  An extension of the
underlying non-perturbative framework to higher dimensions was
recently proposed \cite{fkp} but a systematic development of quantum
geometry has not yet been undertaken.  Also, our quantization
procedure has an inherent $\gamma$-ambiguity which trickles down to
the entropy calculation.  By contrast, calculations in string theory
are free of this problem.  On the other hand, within string theory,
detailed calculations (based on D-branes) have been carried out only
for (a sub-class of) extremal or near-extremal black holes.  While
these black holes are especially simple to deal with mathematically,
unfortunately, they are not of direct relevance to astrophysics, i.e.,
to the physical world we live in. 

Away from extremality, there is another argument \cite{hpd} in which
the Schwarzschild black hole is regarded as a highly excited bosonic
string. But it is more of a semi-qualitative estimate rather than a
systematic, ab-initio calculation. In particular, while the estimated
entropy \textit{does} turn out to be proportional to the area, there
is no control on the numerical coefficient and, moreover, when
extended to include charged black holes, the coefficient seems to
depend on the charge.  More recently, the Maldacena conjecture has
been used to calculate the entropy of non-extremal black holes.
However, the numerical coefficient in front of the entropy 
can only be calculated in the free field approximation and turns out to 
be incorrect (see, e.g., \cite{review}). This is not a 
discrepancy since the (super) gravity approximation can be trusted
only for large couplings, but in this regime the entropy calculation on 
the field theory side seems practically impossible. Moreover,
as is generally recognized, the boundary conditions used in the
Maldacena duality are quite unphysical since the radius of the
compactified dimensions is required to equal the cosmological radius
even near infinity.  Hence the relevance of these mathematically
striking results to our physical world remains unclear.  In the
current approach, by contrast, ordinary, astrophysical black holes in
the physical four spacetime dimensions are included from the
beginning.

In spite of these contrasts, there are some striking similarities.
Our polymer excitations resemble strings.  Our horizon looks like a
`gravitational 2-brane'.  Our polymer excitations ending on the
horizon, depicted in Figure \ref{fig2}, closely resemble strings with
end points on a membrane.  As in string theory, our `2-brane' carries
a natural gauge field.  Furthermore, the horizon degrees of freedom
arise from this gauge field. These similarities seem astonishing.
A closer examination brings out some differences.  In
particular, being the horizon, our `2-brane' has a direct interpretation
in terms of the curved \textit{spacetime geometry} and our $\U(1)$
connection is the \textit{gravitational} spin-connection on the
horizon.  Nonetheless, it is not impossible that, when quantum gravity
is understood at a deeper level, it will reveal that the striking
similarities are not accidental, i.e., that the two descriptions are
in fact related.

\section{Acknowledgements} 
\label{s8}

The authors thank Maarten Bergvelt, 
Alejandro Corichi, Gary Horowitz, Minhyong Kim, Jerzy Lewandowski, Robert 
Israel, Ted Jacobson, Greg Kuperberg, Don Marolf, Carlo Rovelli, Daniel 
Sudarski and Alan Weinstein for useful discussions.  The authors were 
supported in part by the NSF grants 
PHYS94-07194, PHY95-14240, INT97-22514 and by the Eberly research funds of 
Penn State.  In addition, KK was supported by the Braddock fellowship of 
Penn State.  JB thanks the Center for Gravitational Physics and Geometry 
for their hospitality while this paper was being written.  All authors 
acknowledge the support from the Erwin Schr\"odinger Institute, Vienna 
where part of this work was carried out.

\section{Index of Notation}
\label{s9}

\noindent 
For each symbol we list the section or
sections where it is defined.

\medskip
\noindent 
$a$ --- list $\{a_1, \dots, a_n\}$ of elements of $\Z_k$ labelling
punctures.  IVC2.

\noindent 
$\A$ --- classical configuration space, space of $\SU(2)$ connections
on $P$.   IIB, IVA.

\noindent 
$\A^\P$ --- generalized $\U(1)$ connections on $Q$, flat except at
punctures in $\P$.  IVC1.

\noindent 
$\Ab$ --- quantum configuration space, space of generalized $\SU(2)$
connections on $P$.   IIB, IVA.

\noindent 
$\Ab_S$ --- space of generalized connections on the surface.  IVA.

\noindent 
$\Ab_V$ --- space of generalized connections in the volume.  IVA.

\noindent 
$\A_\g$ --- space of $\SU(2)$ connections on the graph $\g$.   IIB.

\noindent 
$A(j)$ --- area associated to list of spins $j$.  VIA.

\noindent 
$\hat A_S$ --- operator corresponding to area of horizon surface.  IVB.

\noindent 
$a_0$ --- horizon area.   IIA.

\noindent 
$A_a^i$ --- $\SU(2)$ connection on the bundle $P$ over $M$.  IIA.

\noindent 
$\D$ --- group of diffeomorphisms of $M$ generated by analytic vector
fields tangential to $S$ and going to zero at infinity.  VB2.

\noindent 
$\D^\P$ --- group of real-analytic diffeomorphisms of $S$ respecting 
certain structure associated with punctures in $\P$.   IVC1.

\noindent 
$E_i^a$ --- orthonormal triad of density weight one on $M$.  IIA.

\noindent 
$F_{ab}$ --- curvature of the connection $W$.  III.

\noindent 
$g_i$ --- holonomy of $W$ along $\gamma_i$.   IVC1.

\noindent 
$\G$ --- group of not necessarily continuous gauge transformations
of $P$ that equal the identity on $S$.   IVB.

\noindent 
$\G^\P$ --- group of not necessarily continuous $\U(1)$ gauge 
transformations of $Q$ that equal identity at punctures in $\P$.  IVC1.

\noindent 
$\G_Q$ --- group of not necessarily continuous gauge transformations
of the bundle $Q$.  VA1.

\noindent 
$\hat \G_Q$ --- group of not necessarily continuous gauge transformations
of the bundle $Q$ taking values in $\Z_k$.    VA1.

\noindent 
$\G_S$ --- group of not necessarily continuous gauge transformations
of the bundle $P$ restricted to $S$.  IVB.

\noindent 
$\g$ --- graph embedded in $M$.   IIB, IVA.

\noindent 
$h_i$ --- holonomy of $W$ along $\eta_i$.   IVC1.

\noindent 
$\hat h_i$ --- quantum operator corresponding to $h_i$.   VA2.

\noindent 
$\H_\g$ --- $L^2(\Ab_\g)$.   IIB.

\noindent 
$\H^{\rm Kin}$ --- kinematical Hilbert space.  VA3.

\noindent 
$\H^{\rm Phys}$ --- physical Hilbert space.  VB3.

\noindent 
$\H^{\rm bh}$ --- Hilbert space of states satisfying area constraint.  VIA.

\noindent 
$\H_S^{\rm bh}$ --- Hilbert space of surface states satisfying area
constraint.  VIA.

\noindent 
$\H_S$ --- surface Hilbert space.  IVC2.

\noindent 
$\H_S^{\rm Phys}$ --- physical surface Hilbert space.  VB3.

\noindent 
$\H^\P_S$ --- subspace of surface Hilbert space associated to
set of punctures $\P$.   IVC2.

\noindent 
$\H^{\P,a}_S$ --- subspace of surface Hilbert space associated to
set of punctures $\P$ labelled by $\Z_k$ elements $a$.  VA2.

\noindent 
$\H^{\S(\P),a}_S$ --- subspace of surface Hilbert space associated to
set of punctures $\P$ labelled by $\Z_k$ elements $a$ and extra structure
$\S$.  VB2.

\noindent 
$\H^{n,a}_S$ --- Hilbert space of surface states associated to $n$ punctures
labelled by $\Z_k$ elements $a$.  VB2.

\noindent 
$\H_V$ --- volume Hilbert space.  IVB.

\noindent 
$\H_V^{\rm Phys}$ --- physical volume Hilbert space.  VB3.

\noindent 
$\H^\P_V$ --- subspace of volume Hilbert space associated to set 
of punctures $\P$.  IIIB.

\noindent 
$\H_V^{\P,j}$ --- subspace of volume Hilbert space associated to
set of punctures $\P$ labelled by spins $j$.  IVB.

\noindent 
$\H_V^{\P,m}$ --- subspace of volume Hilbert space associated to
set of punctures $\P$ labelled by nonzero half-integers $m$.  IVB.

\noindent 
$\H_V^{n,m}$ --- Hilbert space of volume states associated to $n$
punctures labelled by nonzero half-integers $m$.  VB3.

\noindent 
$\H_V^{n,m}$ --- space of volume states satisfying Hamiltonian
constraint associated to $n$ punctures labelled by nonzero 
half-integers $m$.  VB3.

\noindent 
$j$ --- list of spins $\{j_1,\dots,j_n\}$ labelling punctures.  IVB.

\noindent 
$J^i(p)$ --- generator of $\SU(2)$ gauge rotations at point $p$.   
IIB, IVB.

\noindent 
$K_a^i$ --- $\Ad P$-valued 1-form constructed from the extrinsic 
curvature of $M$.  IIA.

\noindent 
$k$ --- `level' in $\U(1)$ Chern-Simons theory, equal to 
$a_0/4 \pi \gamma \l^2$.   III.
 
\noindent 
$k_{ij}$ --- Cartan-Killing form on $\su(2)$.  IIA.

\noindent 
$L$ --- complex line bundle over $\X^\P$.   IVC2.

\noindent 
$\Lambda$ --- the lattice $(2\pi \Z)^{2(n-1)}$.   IVC1.

\noindent 
$m$ --- list of nonzero half-integers $\{m_1,\dots,m_n\}$ labelling
punctures.  VA1.

\noindent 
$M$ --- the spatial 3-manifold, the complement of the open unit ball in $\R^3$.
IIA.  

\noindent 
$N_\bh$ --- number of black hole surface states.  VIA.

\noindent 
$P$ --- trivial $\SU(2)$ bundle over $S$.   IIA.

\noindent 
$\P$ --- finite set $\{p_1,\dots,p_n\}$ consisting of `punctures' $p_i$,
namely points in $S$.   IIIB, IVC1.

\noindent 
$q_{ab}$ --- the 3-metric, Riemannian metric on $M$.  IIA.

\noindent 
$Q$ --- spin bundle of horizon surface, a $\U(1)$ sub-bundle of $P$ 
restricted to $S$.  IIA.

\noindent 
$r$ --- unit internal vector field on horizon surface, a smooth 
function from $S$ to $\su(2)$.  IIA.

\noindent 
$R$ --- projective representation of translation group on holomorphic
functions on $\C^{n-1}$, giving representation of $\Z^n_k$ on $\H^\P_S$.   
IVC2, VA2.

\noindent 
$S$ --- the horizon surface, unit sphere in $\R^3$.  IIA. 

\noindent 
$\S$ --- extra structure on horizon surface.   IVC1.

\noindent 
$S_\bh$ --- black hole entropy.  VIA.

\noindent 
$W$ --- $\U(1)$ connection on $Q$ constructed from the Levi-Civita
connection on $S$.  IIA.

\noindent 
$x_i$ --- coordinates on $\X^\P$.   IVC1.

\noindent 
$X_i$ --- 1-forms on $S - \P$.   IVC1.

\noindent 
$\X$ --- kinematical phase space.  IIA.

\noindent 
$\tilde \X$ --- physical phase space.  IIA.

\noindent 
$\X^\P$ --- surface phase space associated with set of punctures $\P$.  IVC1.

\noindent 
$\X_S$ --- surface phase space.  IIA.

\noindent 
$\X_V$ --- volume phase space.  IIA.

\noindent 
$Y_i$ --- 1-forms on $S - \P$.  IVC1.

\noindent 
$y_i$ --- coordinates on $\X^\P$.   IVC1.

\noindent 
$\gamma$ --- the Barbero-Immirzi parameter.  IIA.

\noindent 
$\gamma_0$ --- special value of Barbero-Immirzi parameter, equal to
$\ln 2/ \pi \sqrt{3}$.   VIA, VIB.

\noindent 
$\gamma_i$ --- paths in $S$.   IVC1.

\noindent
$\Gamma_a^i$ --- $\SU(2)$ connection on $P$ constructed from the 
Levi-Civita connection on $M$.  IIA.

\noindent 
$\eta_i$ --- loops in $S$.   IVC1.

\noindent 
$\mu$ --- uniform measure on $\Ab$.   IIB, IVA.

\noindent 
$\mu_\g$ --- Haar measure on $\A_\g$.   IIB.

\noindent 
$\Phi$ --- diffeomorphism from $\X^\P$ to torus.  IVC1.

\noindent 
$\psi^\P_a$ --- basis of $\H^\P_S$.  IVC2.

\noindent 
$\rho_\bh$ --- black hole density matrix.   VIA.

\noindent 
$\Sigma^i_{ab}$ --- $\Ad P$-valued 2-form on $M$, canonically conjugate to $A$.
IIA.

\noindent 
$\underline{\Sigma}^i_{ab}$ --- pullback of $\Sigma^i_{ab}$ to horizon
surface $S$.   IIA.

\noindent 
$\underline{\hat\Sigma}^i_{ab}$ --- quantum operator corresponding to
$\underline{\Sigma}^i_{ab}$.   III.

\noindent 
$\omega$ --- symplectic structure on $\X^\P$.   IVC1.

\noindent 
$\Omega_{\rm grav}$ --- symplectic structure on $\X$.   IIA.

\bigskip
\noindent{\sl E-mail addresses:} 
\bigskip

\noindent{\tt ashtekar@gravity.phys.psu.edu} \\
\noindent{\tt baez@math.ucr.edu} \\
\noindent{\tt krasnov@cosmic.physics.ucsb.edu}

\end{document}